\documentclass[aip,floatfix,
 amsmath,amssymb,
 reprint,%
]{revtex4-1}

\usepackage{graphicx}% Include figure files
\usepackage{dcolumn}% Align table columns on decimal point
\usepackage{chemfig}
\usepackage{amsmath}
\usepackage{amssymb}

\usepackage[utf8]{inputenc}
\usepackage[T1]{fontenc}
\usepackage{mathptmx}
\usepackage{etoolbox}
\usepackage{float}

\newcommand{\Sec}[1]{Sec.\,\ref{#1}}
\newcommand{\Eq}[1]{Eq.\,(\ref{#1})}
\newcommand{\Fig}[1]{Fig.\,\ref{#1}}

\newcommand{\RNum}[1]{\uppercase\expandafter{\romannumeral #1\relax}}
\usepackage{tikz}

\makeatletter
\def\@email#1#2{%
 \endgroup
 \patchcmd{\titleblock@produce}
  {\frontmatter@RRAPformat}
  {\frontmatter@RRAPformat{\produce@RRAP{*#1\href{mailto:#2}{#2}}}\frontmatter@RRAPformat}
  {}{}
}%
\makeatother
\begin{document}

\preprint{AIP/123-QED}

\title{Cooperative Chemical Reactions in Optical Cavities: A Complex Interplay of Mode Hybridization, Timescale Balance, and Pathway Interference}
% Force line breaks with \\
\author{Yaling Ke}
\email{yaling.ke@phys.chem.ethz.ch}
\affiliation{ 
Department of Chemistry and Applied Biosciences, ETH Zürich, 8093 Zürich, Switzerland%\\This line break forced with \textbackslash\textbackslash
}%

\begin{abstract}
Harnessing strong light–matter interactions to control chemical reactions in confined electromagnetic fields offers a promising route toward deepening our understanding of chemical dynamics at the collective quantum-mechanical level, with potential implications for future chemical synthesis paradigms. Achieving this goal, however, requires an in-depth mechanistic understanding of the underlying dynamical processes. As a step in this direction, we present a systematic and numerically exact quantum dynamical study of cooperative reaction dynamics inside an optical microcavity. Using a hierarchy of model systems with increasing complexity, we elucidate how cavity-modified reactivity emerges from--and is highly sensitive to--subtle structural and environmental variations. Our models consist of optically dark reactive molecules, each represented by a symmetric double well potential, coupled to infrared-active non-reactive intramolecular or solvent vibrational modes, as well as their respective dissipative environments. Our results demonstrate that cavity-induced rate modifications arise from a delicate interplay among mode hybridization in strong-coupling regimes, the dynamical balance of all participating energy exchange processes, and quantum interference between multiple fluctuation–dissipation-mediated reaction pathways enabled by collective cavity coupling. By continuously tuning a single system parameter or introducing molecular collectivity, we observe qualitatively distinct rate modification profiles as functions of the cavity frequency, including resonant rate enhancement, resonant rate suppression, hybridization-induced peak splitting, and, notably, asymmetric Fano-type line shapes in which enhancement peaks and suppression dips coexist within a narrow resonance window, highlighting the important role of quantum interference in cavity-modified chemical reactivity. 
\end{abstract}
\maketitle

\section{Introduction}
The synergy between cavity quantum electrodynamics and organic chemistry,\cite{Ebbesen_2016_ACR_p2403,Kasprzak_Nat_2006_p409,AberraGuebrou_2012_PRL_p66401,Coles_Nat.Mater._2014_p712,Orgiu_2015_NM_p1123,Hagenmueller_2017_PRL_p223601,Hsu_2017_JPCL_p2357,Baieva_2017_AP_p28,Mony_2021_AFM_p2010737,Hayashi_2024_JCP_p181101,Sandik_2025_NM_p344} in which the vacuum electromagnetic field is confined within Fabry–P\'erot cavities or plasmonic nanostructures, offers a fundamentally new platform for controlling molecular interactions and chemical reaction dynamics. This opportunity arises, on the one hand, from the intrinsically quantum nature of confined light fields, which necessitates a discrete-state description, and on the other hand from their spatial delocalization, which can coherently couple an ensemble of molecules into an effectively interconnected network, thereby enabling collective effects. When combined with the diversity and tunability of organic chemical reactions, polariton chemistry--formed under strong light–matter coupling--holds great promise for uncovering new mechanisms of chemical control and enabling unprecedented technological applications. 

This vision has been partly realized through a growing body of experiments reporting pronounced resonant modifications of chemical reactivity under vibrational strong coupling, achieved by tuning the cavity frequency into resonance with specific molecular absorption bands.\cite{Thomas_2016_ACE_p11634,Vergauwe_2019_ACIE_p15324,Thomas_2019_S_p615,Hirai_2020_ACE_p5370,Pang_2020_ACIE_p10436,Sau_2021_ACIE_p5712,Ahn_Sci_2023_p1165,Patrahau_Angew.Chem.Int.Ed._2024_p202401368,Verdelli_2024_ACIE_p202409528,Mahato_2025_ACIE_p202424247} Despite these advances, experimental observations have revealed striking variability in cavity-modified chemical kinetics, even for nominally identical or closely related reactions.\cite{Lather_2019_ACIE_p10635,Lian_2025_AP_p3557,Wiesehan_2021_JCP_p241103,Hiura_2019__p,Imperatore_2021_JCP_p191103,Wang_2024_Ap_p263} For instance, in the solvolysis of para-nitrophenyl acetate, Lather et al. reported an order-of-magnitude rate enhancement,\cite{Lather_2019_ACIE_p10635} whereas a subsequent study by Lian et al. observed only a 2.7-fold enhancement,\cite{Lian_2025_AP_p3557} and Wiesehan and Xiong reported negligible catalytic effects.\cite{Wiesehan_2021_JCP_p241103} Such disparate outcomes strongly suggest that cavity-induced chemical responses are highly sensitive to subtle environmental and structural variations.

Indeed, polaritonic chemical reactions involve complex interactions among many distinct components, including intramolecular and intermolecular couplings, light–matter interactions, cavity losses, and solvent-induced dissipation. Together, these ingredients define a high-dimensional parameter space in which small perturbations may lead to qualitatively different dynamical outcomes. Moreover, when molecules are collectively coupled to a cavity mode, multiple energy transfer pathways may coexist and coherently interfere, giving rise to nontrivial quantum effects, such as constructive or destructive interference, that can further complicate the reaction dynamics.

The goal of this work is to elucidate, from a theoretical perspective, how such sensitivity arises in cavity-modified chemical reaction dynamics. To this end, we investigate a series of model systems with progressively increasing complexity, focusing on cooperative reaction dynamics in condensed phase involving optically dark reactive vibrational coordinates coupled to optically active, non-reactive modes. These spectator modes represent either intramolecular vibrations or characteristic solvent motions and serve as crucial mediators between the cavity field and reactive coordinates, both outside and inside an infrared microcavity.\cite{Lindoy_2023_NC_p2733,Vega_2025_JACS_p19727} All degrees of freedom (DoFs) are treated fully quantum mechanically using a numerically exact hierarchical equations of motion (HEOM) approach combined with an efficient tree tensor network state (TTNS) solver.\cite{Ke_2023_JCP_p211102}

Our results reveal that cavity-induced rate modifications emerge from a delicate interplay of multiple factors, including mode hybridization in strong-coupling regimes, the balance of characteristic energy transfer timescales, and quantum interference between cavity-assisted and intrinsic reaction pathways. By continuously tuning a single coupling parameter or introducing collective effects, we demonstrate that the reaction rate as a function of cavity frequency can display qualitatively distinct behaviors, ranging from resonant enhancement, to resonant suppression, or even asymmetric Fano-type line shapes in which enhancement peaks and suppression dips coexist within a narrow frequency window. These findings establish quantum interference as a unifying framework for understanding the diversity of cavity-induced reactivity trends.

The remainder of this paper is organized as follows. In \Sec{sec:theory}, we introduce the model systems and briefly describe the theoretical methods employed. In \Sec{sec:results}, we present reaction rates for scenarios of increasing complexity, both outside and inside a single-mode cavity. Finally, \Sec{sec:conclusion} summarizes our key findings and outlines directions for future research.

\section{\label{sec:theory}Theory}
We consider a system composed of $N_{\rm mol}$ molecules, each characterized by a reactive coordinate $x_i$, coupled to $N_{\rm nor}$ interacting non-reactive coordinates $x_j$. Either the reactive or the non-reactive modes can be coupled to a single cavity mode. The light--matter interaction is described by the Pauli--Fierz quantum electrodynamics Hamiltonian in the dipole gauge and under the long-wavelength approximation. Throughout this work, we set $\hbar = 1$. The system Hamiltonian reads explicitly\cite{Flick_2017_PotNAoS_p3026,Rokaj_2018_JPBAMOP_p34005,Mandal_2023_CR_p9786,Lindoy_2023_NC_p2733}
\begin{equation}
\label{systemHamiltonian}
\begin{split}
H_{\mathrm{S}} =& \sum_{i=1}^{N_{\rm mol}}\frac{p_{i}^2}{2}+ E_{\mathrm{b}}\left(\frac{x_{i}^2}{a^2}-1\right)^2 + 
\sum_{j=1}^{N_{\rm nor}}\frac{p_{j}^2}{2}+ \frac{1}{2}\omega_{j}^2\left(x_{j}-\sum_i\frac{\eta_{ij} x_i}{\omega_{j}^2}\right)^2  \\
&+ \frac{p_{\mathrm{c}}^2}{2} + \frac{1}{2} \omega_{\mathrm{c}}^2\left(x_{\mathrm{c}} - \sqrt{\frac{2}{\omega_{\mathrm{c}}}} \eta_{\mathrm{c}} \vec{D}\cdot \vec{e} \right)^2.  
\end{split}
\end{equation}
Here, $p_i$ and $p_j$ denote the mass-scaled momenta of the $i$th reactive molecule and the $j$th non-reactive mode, respectively. The reactive potential energy surface is modeled as a symmetric double well with barrier height $E_{\rm b}$ at the origin and two minima located at $x_i=\pm a$. The non-reactive modes are described as harmonic normal modes with frequencies $\omega_j$, whose equilibrium positions are shifted through coupling to the reactive coordinates. The corresponding coupling strengths are denoted by $\eta_{ij}$, and for simplicity we assume all such couplings to be identical, $\eta_{ij}=\eta_{\rm nor}$.

The cavity mode is characterized by a displaced harmonic oscillator with coordinate $x_{\mathrm{c}}$, momentum $p_{\mathrm{c}}$, and frequency $\omega_{\mathrm{c}}$. The light--matter coupling strength is determined by $\eta_{\mathrm{c}}=1/\sqrt{2\omega_{\mathrm{c}}\epsilon_0 V}$, where $\epsilon_0$ is the permittivity of the medium inside the cavity and $V$ is the cavity mode volume, together with the projection of the dipole operator
\begin{equation}
\label{dipole}
\vec{D}=\chi_1\sum_i\vec{\mu}(x_i)+\chi_2\sum_j\vec{\mu}(x_j)
\end{equation}
along the light polarization direction $\vec{e}$. The single-mode cavity assumption is valid for high-finesse cavities whose free spectral range significantly exceeds the Rabi splitting. Otherwise, the influence of neighboring cavity modes cannot be neglected.\cite{Ke_2025_JCP_p164703}

The system is coupled to a dissipative environment. Specifically, each system degree of freedom (DoF) is assumed to interact with an independent bosonic bath composed of an infinite set of harmonic oscillators with a continuous energy spectrum,
\begin{equation}
    H_{\mathrm{B}} = \sum_{\alpha}\sum_k \frac{P_{\alpha k}^2}{2}
    + \frac{1}{2}\omega_{\alpha k}^2\left(Q_{\alpha k}-\frac{c_{\alpha k}x_{\alpha}}{\omega_{\alpha k}^2}\right)^2.
\end{equation}
Here, $\alpha\in \{i,j,c\}$, where the indices $i=1,\ldots,N_{\rm mol}$ and $j=1,\ldots,N_{\rm nor}$ label the reactive and non-reactive coordinates, respectively.

In this work, the open quantum dynamics of the system are simulated using the HEOM formalism (see Ref.~\onlinecite{Tanimura_2020_JCP_p20901} and references therein). This approach relies on an exponential expansion of the bath correlation function, which fully encodes the influence of the Gaussian environment on the system dynamics,
\begin{equation}
\label{bathcorrelationfunction}
    C_{\alpha}(t)
    = \frac{1}{\pi} \int_{-\infty}^{\infty} \frac{e^{-i\omega t}}{1-e^{-\beta \omega}} J_{\alpha}(\omega)\mathrm{d}\omega
    = \sum_p \lambda_{\alpha }^2\eta_{\alpha p}e^{-\gamma_{\alpha p}t}.
\end{equation}
The spectral density
$J_{\alpha}(\omega)=\frac{\pi}{2}\sum_k \frac{c_{\alpha k}^2}{\omega_{\alpha k}}\delta(\omega -\omega_{\alpha k})$
characterizes the coupling-weighted distribution of bath modes. The reorganization energy,
$\lambda_{\alpha}^2=\frac{1}{\pi}\int_0^{\infty}\frac{J_{\alpha}(\omega)}{\omega}\mathrm{d}\omega$,
quantifies the overall coupling strength between the $\alpha$th system DoF and its bath.

In the twin space representation, the HEOM formalism yields a Schr\"odinger-like equation motion\cite{Borrelli_2019_JCP_p234102,Borrelli_2021_WCMS_p1539,Ke_2022_JCP_p194102}
\begin{equation}
\label{Schroedinger}
    \frac{\mathrm{d}|\Psi(t)\rangle}{\mathrm{d} t} = -i\mathcal{H} |\Psi(t)\rangle,
\end{equation}
for the extended wave function
\begin{equation}
|\Psi(t)\rangle=\sum_{\mathbf{v,n}}C_{\mathbf{v,n}}(t)
\prod_{\alpha}|v_{\alpha}\rangle \otimes |v'_{\alpha}\rangle
\otimes \prod_p|n_{\alpha p}\rangle,
\end{equation}
which is expanded in the system basis states $|v_{\alpha}\rangle$ and $|v'_{\alpha}\rangle$, as well as in the number states $|n_{\alpha p}\rangle$ of the bath pseudomodes stemming from the exponential expansion in \Eq{bathcorrelationfunction}. 

The non-Hermitian super-Hamiltonian $\mathcal{H}$ in twin space is given by 
\begin{equation}
\begin{split}
    \mathcal{H} = &\hat{H}_{\rm S}+\sum_{\alpha}\lambda_{\alpha}^2 \hat{x}_{\alpha}^2 -\tilde{H}_{\rm S}-\sum_{\alpha}\lambda_{\alpha}^2 \tilde{x}_{\alpha}^2 -i \sum_{\alpha}\sum_p  \gamma_{\alpha p} b_{\alpha p}^{+}b_{\alpha p}\\
    & +\sum_{\alpha}\sum_p \lambda_{\alpha}\left[(\hat{x}_{\alpha}-\tilde{x}_{\alpha})b_{\alpha p}     +(\eta_{\alpha p}\hat{x}_{\alpha}-\eta^*_{\alpha p}\tilde{x}_{\alpha})b_{\alpha p}^+ \right].
\end{split}
\end{equation}
Here, $\hat{x}_{\alpha}$ and $\tilde{x}_{\alpha}$ are superoperators acting in twin space, corresponding to the coordinate $x_{\alpha}$ in the forward and backward branches for the bra and ket wave functions of Hilbert space, respectively. Likewise, $\hat{H}_{\rm S}$ and $\tilde{H}_{\rm S}$ are the corresponding system super-Hamiltonians derived from $H_{\rm S}$ in \Eq{systemHamiltonian}. The operators $b_{\alpha p}^{+}$ and $b_{\alpha p}$ are the bosonic creation and annihilation operators of the bath pseudomodes, satisfying
$b_{\alpha p}^{+} |n_{\alpha p}\rangle = \sqrt{n_{\alpha p}+1}| n_{\alpha p}+1 \rangle$
and
$b_{\alpha p} |n_{\alpha p}\rangle = \sqrt{n_{\alpha p}}|n_{\alpha p}-1 \rangle$.

Finally, we employ a tree tensor network state (TTNS) decomposition of the high-rank coefficient tensor $C_{\mathbf{v,n}}(t)$ in the extended wave function $|\Psi(t)\rangle$, together with an efficient time-propagation scheme based on this ansatz and the time-dependent variational principle. Implementation details can be found in Refs.~\onlinecite{Ke_2023_JCP_p211102} and ~\onlinecite{Ke_J.Chem.Phys._2025_p64702}.

\section{\label{sec:results}Results}
To elucidate the physical mechanisms and connections underlying various cavity-induced modification phenomena of chemical reactivity, manifesting in some cases as rate enhancement and in others as rate suppression, we begin with a minimal model system consisting of a single reactive coordinate $(N_{\rm mol}=1)$ coupled to a single non-reactive mode $(N_{\rm nor}=1)$, both outside and inside an optical cavity. In this setup, we assume that the non-reactive mode is optically bright, while the reactive vibration is optically dark. This is achieved by setting $\chi_1=0$ and $\chi_2=1$ in \Eq{dipole}. The dipoles are aligned with the cavity polarization such that $\vec{\mu}(x_j)\cdot\vec{e}=x_j$. Despite its simplicity, this model allows for a detailed mechanistic analysis of the interplay among distinct energy exchange processes and competing energy transfer pathways, as well as how their cooperation and interference ultimately determine cavity-modified reaction dynamics. Building on the insights gained from this minimal system, we subsequently extend our analysis to the collective regime with $N_{\rm mol}>1$ and $N_{\rm nor}>1$ under different coupling scenarios, highlighting the increased complexity of reactive behavior in cavity-coupled molecular networks.

Throughout this work, the simulations are performed at room temperature ($T=300$~K). All bosonic baths are described by a Lorentzian–Debye spectral density function,
\begin{equation}
J_{\alpha}(\omega)=2\lambda_{\alpha}^2\frac{\omega \Omega_{\alpha}}{\omega^2+\Omega_{\alpha}^2}.
\end{equation}
Each bosonic bath is represented by four effective dissipative modes obtained from a Pad\'e pole decomposition,\cite{Hu_2010_JCP_p101106} and each environmental pseudomode is expanded in a local basis of dimension $d_{\rm e}=10$.

For the reactive potential energy surface, we adopt the parameters $E_{b}=2250\,\mathrm{cm}^{-1}$ and $a=44.4$~a.u., consistent with previous studies.\cite{Lindoy_2023_NC_p2733,Lindoy_2024_N_p2617,Ying_2023_JCP_p84104,Ying_2024_CM_p110,Vega_2025_JACS_p19727,Ke_J.Chem.Phys._2024_p224704,Ke_2024_JCP_p54104,Ke_J.Chem.Phys._2025_p64702} The lowest ten vibrational eigenstates $\{|v^{\rm mol}_k\rangle\}$ are retained for each reactive coordinate, corresponding to a local basis dimension $d_{\rm mol}=10$. Among the four lowest vibrational eigenstates below the reaction barrier, two distinct vibrational transitions play a dominant role in the cavity-modified reaction dynamics. However, for the solvent coupling parameters used throughout this work, $\lambda_{\rm mol}=100\,\mathrm{cm}^{-1}$ and $\Omega_{\rm mol}=200\,\mathrm{cm}^{-1}$, bath-induced broadening renders these transitions indistinguishable. As a result, they merge into a doubly degenerate band that is more conveniently described in terms of two diabatically localized transitions,
$|v^{\rm mol}_{L/R,0}\rangle \leftrightarrow |v^{\rm mol}_{L/R,1}\rangle$,
with a characteristic transition energy of approximately $\Delta E=1185\,\mathrm{cm}^{-1}$.\cite{Ke_2025_JCP_p164703} For the cavity mode and the non-reactive spectator modes, we retain the lowest six harmonic eigenstates, corresponding to $d_{\rm c}=6$ and $d_{\rm nor}=6$, respectively. 

At the initial moment, the reactive and non-reactive vibrational modes, the cavity mode, and all bosonic baths are assumed to be factorized and prepared in their respective thermal equilibrium states. Prior to time propagation, each reactive molecule is projected onto the reactant region in the left well ($x_i<0$) using the side operator $1-h_i$, where $h_i=\theta(x_i-x_i^{\ddagger})$ is the Heaviside step function projecting onto the product region in the right well ($x_i>0$). The time evolution is carried out with a time step of $\delta t=1$~fs, and the maximum bond dimension in the binary TTNS decomposition of $|\Psi(t)\rangle$ is set to $D_{\rm max}=40$. 

The dynamics are propagated up to tens of picoseconds, until a plateau time $t_{\rm plateau}$ is reached, indicating entry into the quasi-steady rate process. The forward reaction rate constant from the reactant to the product region is then computed as\cite{Miller_1983_JCP_p4889,Craig_2007_JCP_p144503,Chen_2009_JCP_p134505,Ke_2022_JCP_p34103}
\begin{equation}
k_i=\lim_{t\rightarrow t_{\rm plateau}} \frac{C_i(t)}{1-(1+1/K_i)P_i(t)}.
\end{equation}
Here, the numerator is the flux–side correlation function $C_i(t)=\langle {\rm II}|\tilde{F}_i|\Psi(t)\rangle$, where the flux operator is defined as $F_i=i[H_{\rm S}, h_i]$. The identity state
$|{\rm II}\rangle=|1_{\mathrm{sys}}\rangle\otimes |\mathbf{n=0}\rangle$,
with $|1_{\mathrm{sys}}\rangle=\otimes_s\sum_{v_s=v'_s}|v_s v'_s\rangle$, is the unit vector in twin space, while all environmental modes are initialized in their vacuum states. The time-dependent population probability in the product region is given by $P_i(t)=\langle {\rm II}|\hat{h}_i|\Psi(t)\rangle$. 
For the symmetric double well potential considered here, the equilibrium population probability satisfies $P_i^{\rm eq}=1/2$, yielding an equilibrium constant $K_i=P_i^{\rm eq}/(1-P_i^{\rm eq})=1$, both outside and inside the cavity.\cite{Ke_2025_JCP_p54109}  Owing to the homogeneity of the molecular ensemble, all molecules exhibit identical reaction kinetics. We therefore denote the reaction rates of an individual molecule outside and inside the cavity by $k_{\rm o}$ and $k_{\rm c}$, respectively.
\begin{figure}[H]
  \begin{minipage}[c]{0.45\textwidth}     
    \includegraphics[width=\textwidth]{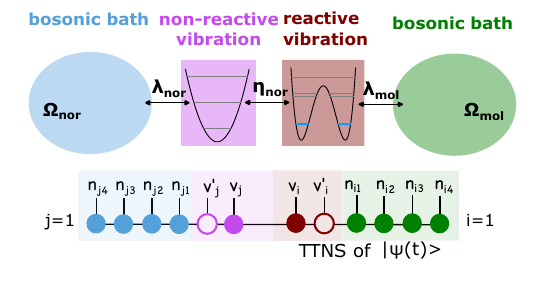}
  \end{minipage}
    \caption{Schematic illustration of a model system with a single reactive mode ($N_{\rm mol}=1$) interacting with a single non-reactive vibration ($N_{\rm nor}=1$ outside the cavity, together with a graphic representation of the TTNS decomposition of the extended wave function $|\Psi(t)\rangle$ for this scenario.
    } \label{fig:Nmol1Nnor1_model}
\end{figure}

\subsection{Single Reactive and Non-reactive Vibration Regime}
In this subsection, we analyze the simplest realization of our model to establish a clear physical baseline for understanding cavity-modified reaction dynamics. We begin by characterizing the reaction kinetics in the absence of the cavity field, which serves as a reference against which all subsequent cavity-induced effects will be assessed.
\begin{figure}[H]
  \begin{minipage}[c]{0.45\textwidth}   
  \raggedright a)  $\Omega_{\rm nor}=200\,\mathrm{cm}^{-1}$\
    \includegraphics[width=\textwidth]{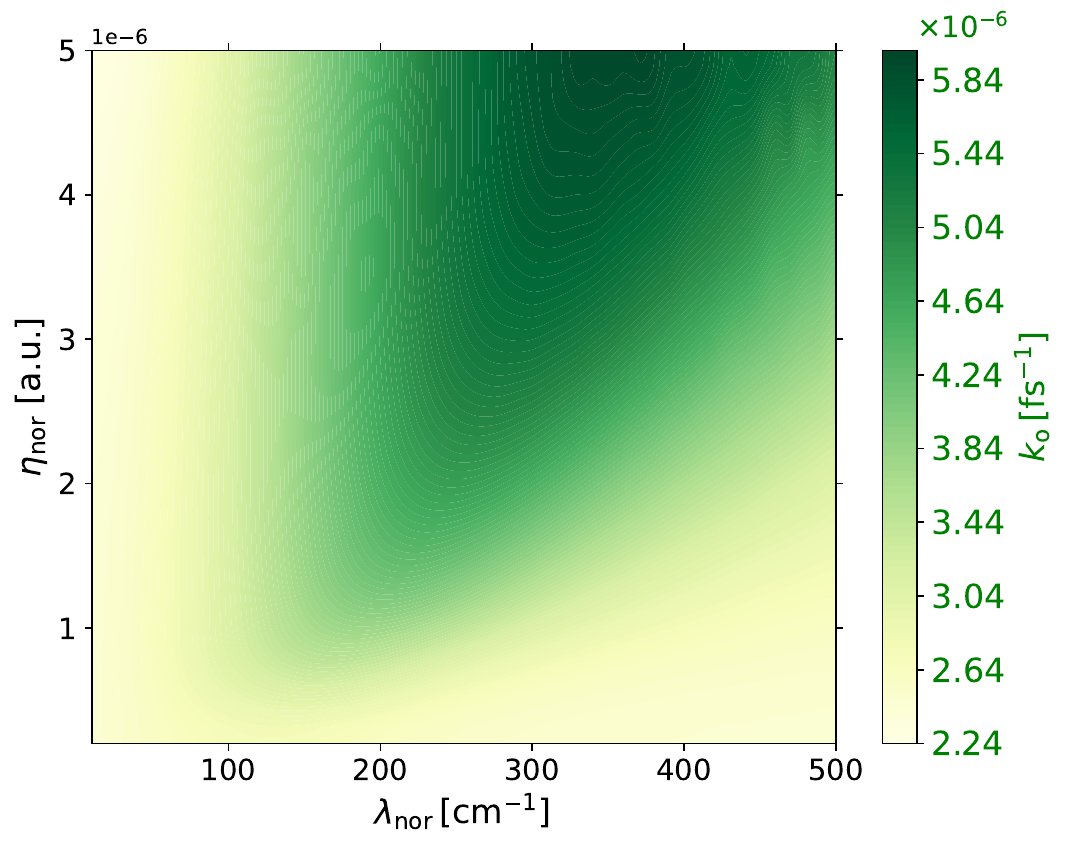}
  \end{minipage}
  \begin{minipage}[c]{0.45\textwidth}
  \raggedright b)  $\Omega_{\rm nor}=1000\,\mathrm{cm}^{-1}$
    \includegraphics[width=\textwidth]{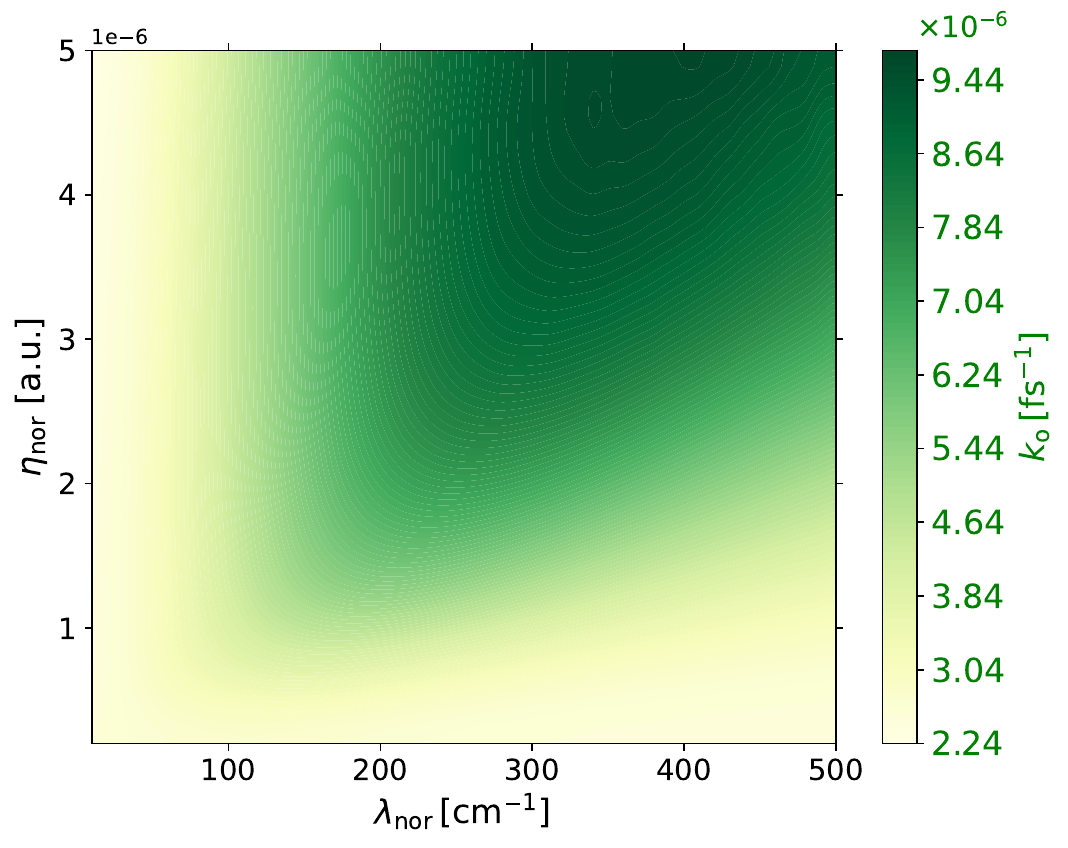}
  \end{minipage}
    \caption{Contour plots of the reaction rate $k_{\rm o}$ for the model illustrated in \Fig{fig:Nmol1Nnor1_model}, shown as functions of the coupling strength $\eta_{\rm nor}$ and the dissipation strength $\lambda_{\rm nor}$ for two different characterisitic bath frequencies $\Omega_{\rm nor}$. The frequency of the non-reactive vibrationa is fixed at $\omega_{\mathrm{nor}}=1185\,\mathrm{cm}^{-1}$. 
    } \label{fig:Nmol1Nnor1_rates}
\end{figure}

\subsubsection{Outside Cavity}
We consider a minimal scenario in which a single reactive molecule is coupled with strength $\eta_{\rm nor}$ to a single non-reactive spectator harmonic mode, with both DoFs embedded in their respective dissipative environments, as schematically illustrated in \Fig{fig:Nmol1Nnor1_model}. The non-reactive spectator mode may represent either an intramolecular vibration or a characteristic solvent coordinate. This cavity-free setup allows us to isolate the role of vibrational energy exchange and dissipation in governing the reaction rate.

We find that the reaction rate reaches its maximum when the frequency of the non-reactive mode is resonant with the reactive vibrational transition energy $\Delta E$, occurring at around $\omega_{\rm nor}\approx1185\,\mathrm{cm}^{-1}$. Under this resonant condition, the reaction rate is strongly influenced by the energy-exchange timescales between interacting subsystems--controlled by parameters such as $\eta_{\rm nor}$ and $\lambda_{\rm nor}$--as well as by the characteristic timescales of the bosonic baths, set by $\Omega_{\rm nor}$.

\Fig{fig:Nmol1Nnor1_rates} displays contour plots of the reaction rate $k_{\rm o}$ outside the cavity as functions of the coupling strength $\eta_{\rm nor}$ and the dissipative strength $\lambda_{\rm nor}$ of the non-reactive mode, for two different values of $\Omega_{\rm nor}$. Notably, the rate exhibits a pronounced light-cone-like structure along the diagonal direction, indicating a strong enhancement when both $\eta_{\rm nor}$ and $\lambda_{\rm nor}$ are increased simultaneously. Horizontal or vertical cuts at fixed $\eta_{\rm nor}$ or $\lambda_{\rm nor}$ reveal a turnover behavior as a function of the complementary parameter. In addition, the reaction rate is further enhanced for larger bath characteristic frequencies $\Omega_{\rm nor}$.

These observations highlight the important role of timescale matching among the various dynamical processes involved in the reaction pathway. Energy fluctuations originating in a bosonic bath can be transferred through the non-reactive and reactive modes and ultimately dissipated into another bath. Accelerating a single step within this energy-funneling channel does not necessarily enhance the reaction rate. Instead, the maximal rate emerges from a cooperative acceleration of all relevant energy exchange and dissipation processes. This cavity-free analysis therefore provides a crucial mechanistic reference for interpreting the cavity-induced modifications discussed in what follows.
 
\begin{figure}[H]
  \begin{minipage}[c]{0.45\textwidth}
    \includegraphics[width=\textwidth]{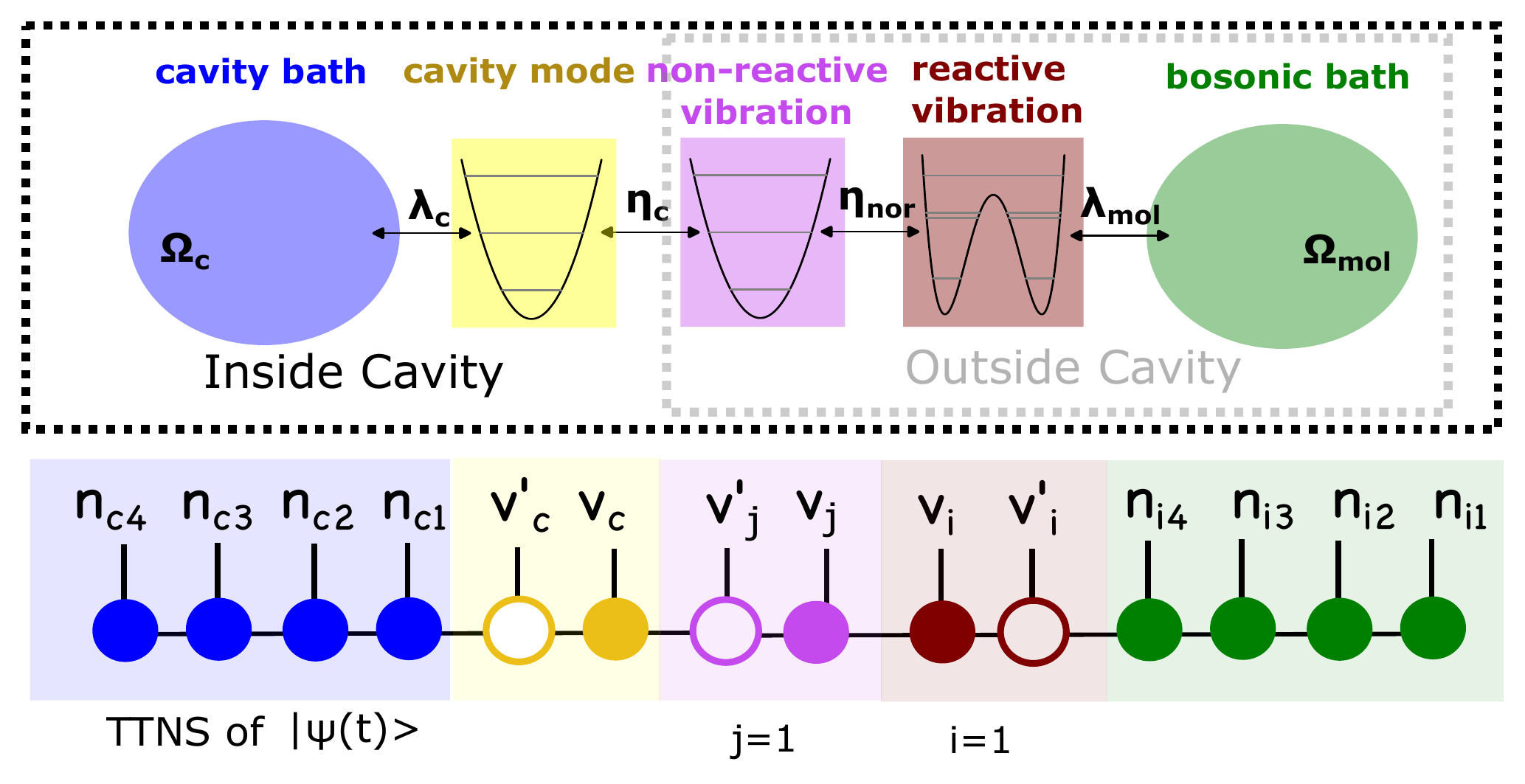}
  \end{minipage}
    \caption{
       Schematic illustration of a dissipation-free non-reactive vibrational mode bridging a single reactive molecule and a cavity mode, together with the TTNS representation of the extended wave function $|\Psi(t)\rangle$. 
    } \label{fig:Nmol1Nnor1_IC_model}
\end{figure}
\subsubsection{Inside Cavity}
We now turn to the influence of a discrete cavity mode on the reaction dynamics for the minimal single-molecule model introduced above. We assume that the non-reactive spectator vibration--rather than the reactive coordinate--is coupled to a single cavity mode. This configuration allows us to isolate how cavity-mediated energy flow indirectly reshapes the reaction dynamics.
\begin{figure}[H]
    \begin{minipage}[c]{0.45\textwidth}
      \raggedright a) $\eta_{\rm nor}=1\times 10^{-6}$~a.u. and $\eta_{\rm c}=0.00125$~a.u. 
    \includegraphics[width=\textwidth]{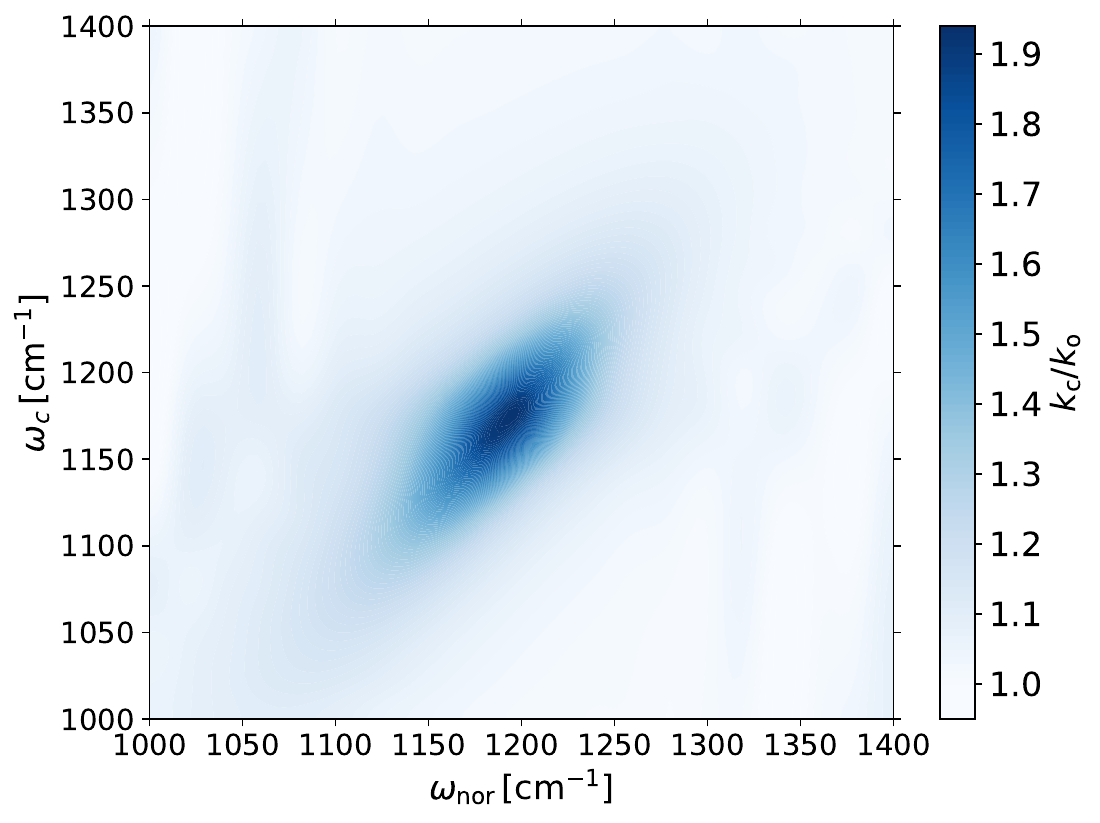}
  \end{minipage}
    \begin{minipage}[c]{0.45\textwidth}
      \raggedright b) $\eta_{\rm nor}=1\times 10^{-6}$~a.u. and $\eta_{\rm c}=0.005$~a.u. 
    \includegraphics[width=\textwidth]{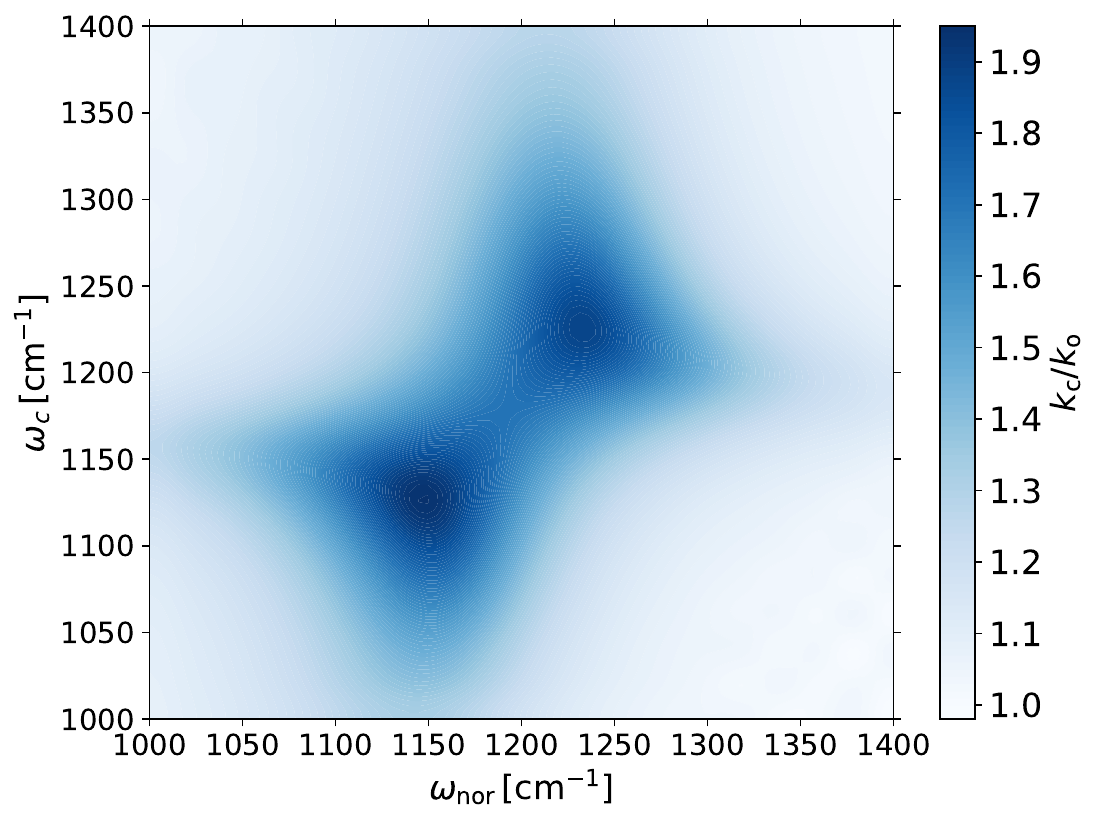}
  \end{minipage}
      \begin{minipage}[c]{0.45\textwidth}
      \raggedright c) $\eta_{\rm nor}=2\times 10^{-6}$~a.u. and $\eta_{\rm c}=0.00125$~a.u. 
    \includegraphics[width=\textwidth]{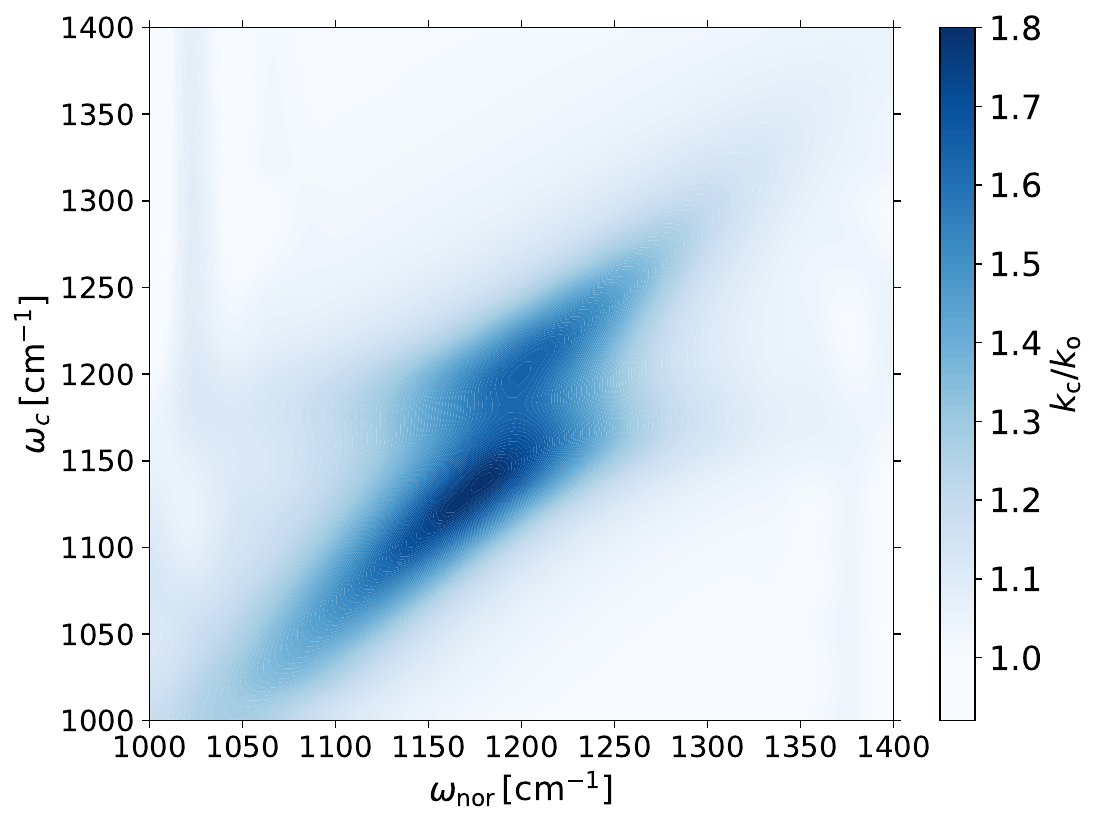}
  \end{minipage}
    \caption{
      Contour plots of the rate modification ratio $k_{\rm c}/k_{\rm o}$ as functions of the non-reactive vibrational frequency $\omega_{\rm nor}$ and the caivty frequency $\omega_{\rm c}$ for the model system illustrated in \Fig{fig:Nmol1Nnor1_IC_model}. Each panel corresponds to a different set of coupling strengths $\eta_{\rm c}$ and $\eta_{\rm nor}$. Other parameters are fixed at $\lambda_{\rm c}=200\,\mathrm{cm}^{-1}$, and $\Omega_{\rm c}=1000\,\mathrm{cm}^{-1}$.  
    } \label{fig:Nmol1Nnor1_IC_wcwr}
\end{figure}
\begin{figure}[H]
\centering
  \begin{minipage}[c]{0.45\textwidth}
  \raggedright a) $\eta_{\rm nor}=1\times10^{-6}$~a.u. \
    \includegraphics[width=\textwidth]{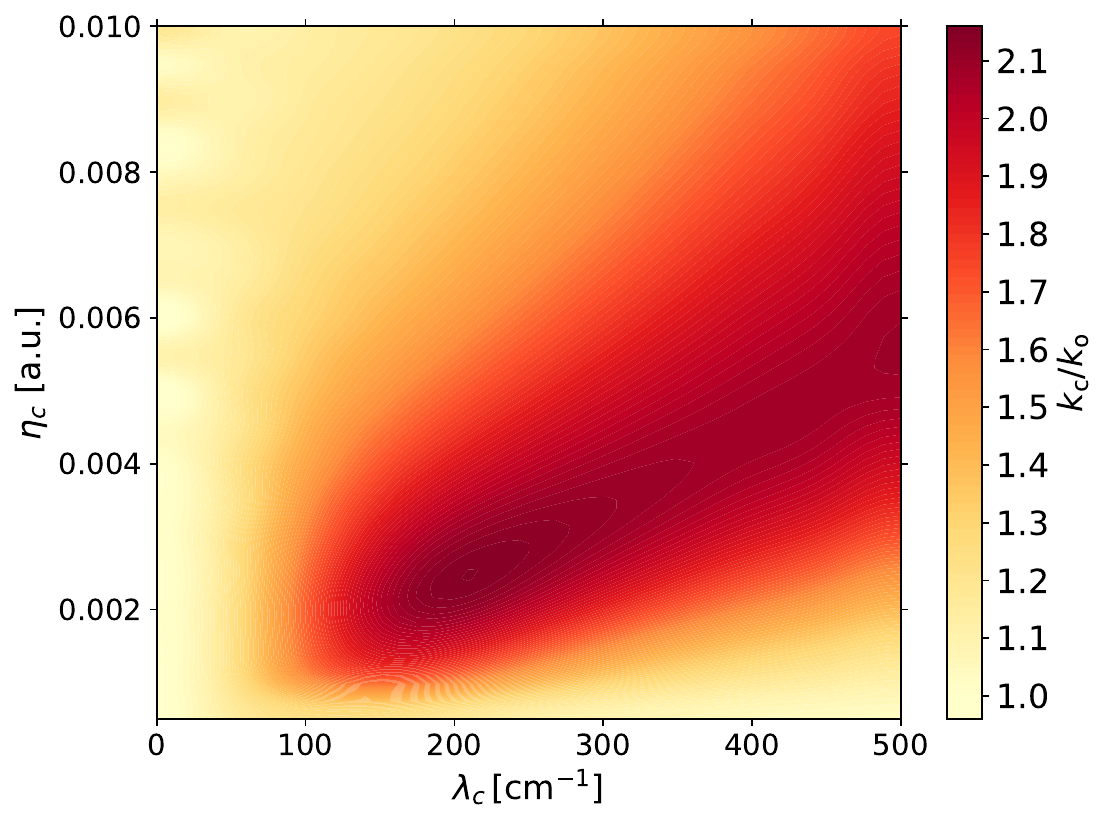}
  \end{minipage} 
 \begin{minipage}[c]{0.45\textwidth}
  \raggedright b) $\eta_{\rm nor}=2\times10^{-6}$~a.u.\
    \includegraphics[width=\textwidth]{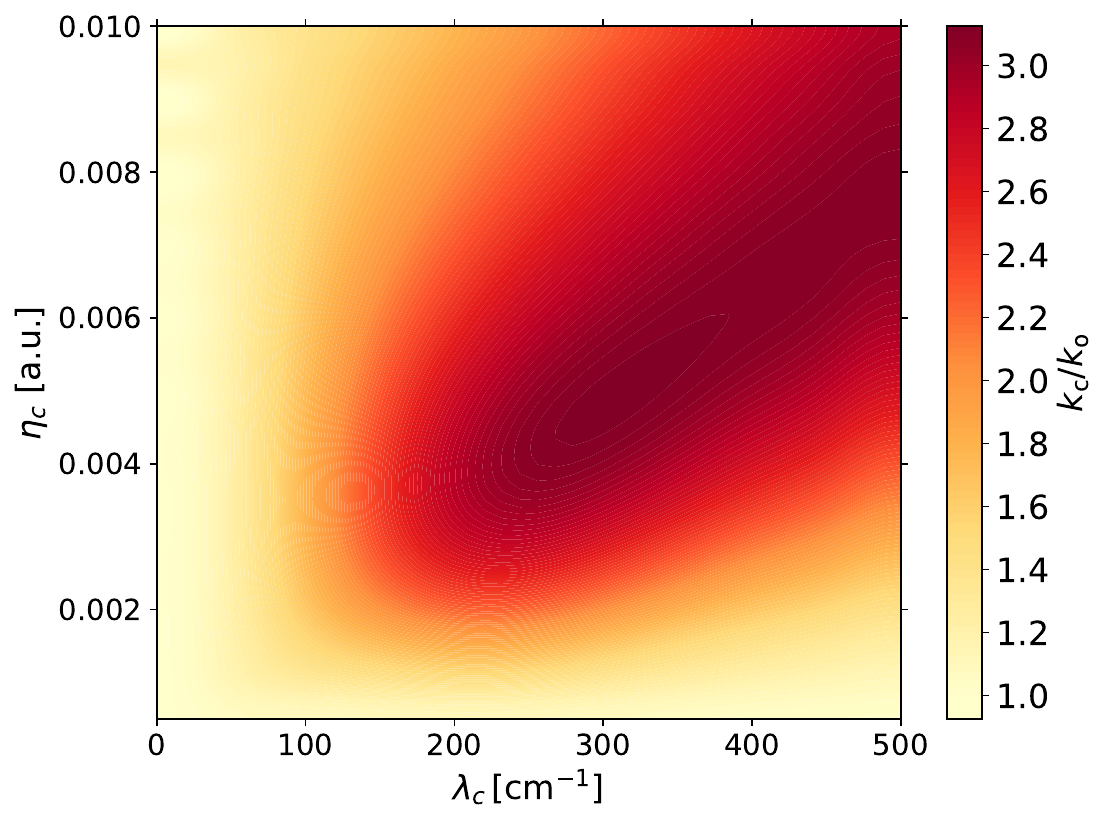}
  \end{minipage} 
    \begin{minipage}[c]{0.45\textwidth}
  \raggedright c) $\eta_{\rm nor}=5\times10^{-6}$~a.u.\
    \includegraphics[width=\textwidth]{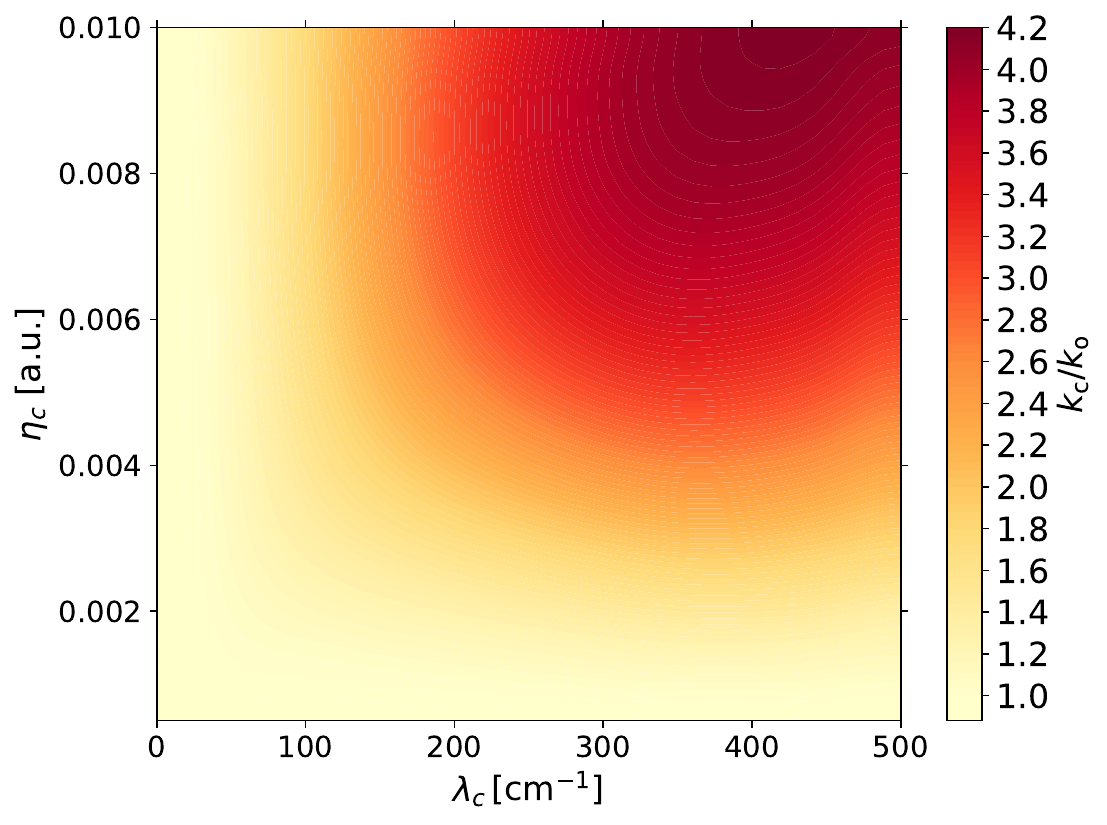}
  \end{minipage} 
    \caption{Contour plots of the rate ratio inside and outside the cavity, $k_{\rm c}/k_{\rm o}$, as functions of the light-matter coupling strength $\eta_{\rm c}$ and the cavity lossy strength $\lambda_{\rm c}$ for the model system illustrated in \Fig{fig:Nmol1Nnor1_IC_model}. Different values of the vibrational coupling strength $\eta_{\rm nor}$ are shown in different panels. The frequencies of the cavity mode and the non-reactive harmonic mode are fixed at $\omega_{\mathrm{c}}=\omega_{\mathrm{nor}}=1185\,\mathrm{cm}^{-1}$, and the cavity bath cutoff frequency is set to  $\Omega_{\mathrm{c}}=1000\,\mathrm{cm}^{-1}$. 
    } \label{fig:Nmol1Nnor1_IC_etaclambc}
\end{figure}

\begin{figure}[H]
  \begin{minipage}[c]{0.45\textwidth}
    \includegraphics[width=\textwidth]{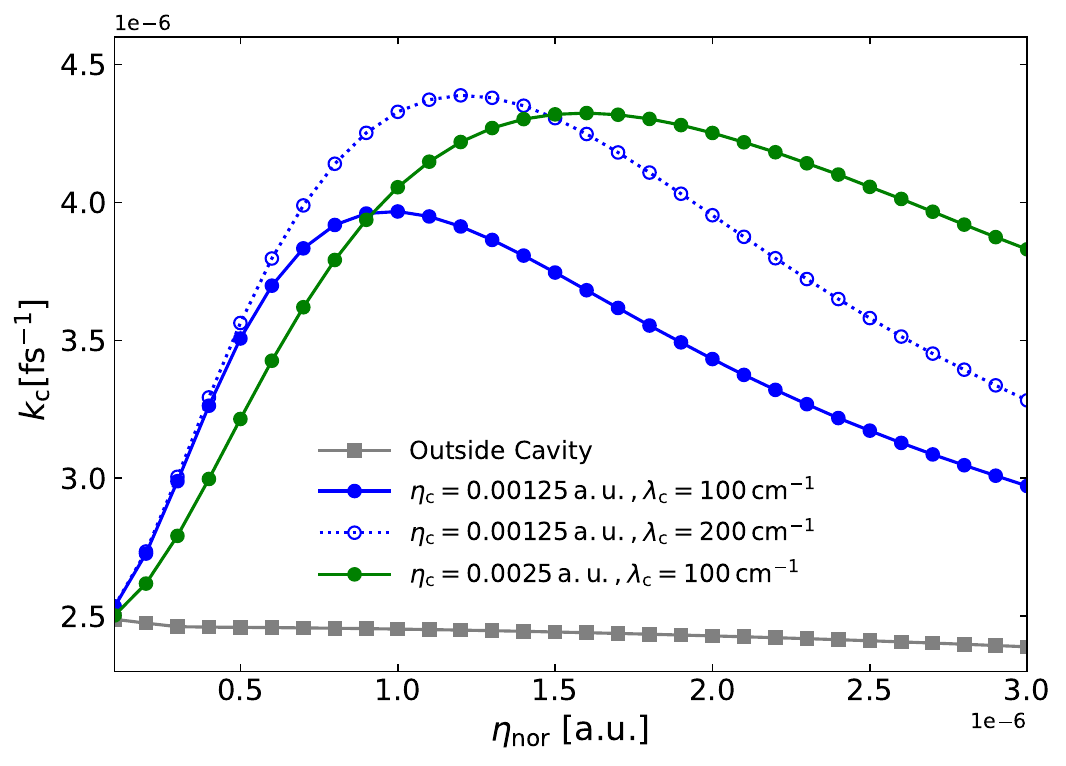}
  \end{minipage}
    \caption{
Reaction rates as a function of the vibrational coupling strength $\eta_{\rm nor}$ for several parameter sets of $\eta_{\rm c}$ and $\lambda_{\rm c}$. The frequencies of the cavity mode and the non-reactive vibrational mode are tuned close to resonance with the molecular vibrational transition, i.e. $\omega_{\rm c}= \omega_{\rm nor}=1185\,\mathrm{cm}^{-1}$, and the cavity bath cutoff frequency is fixed at $\Omega_{\rm c}=1000\,\mathrm{cm}^{-1}$.  For reference, the reaction rates outside the cavity $k_{\rm o}$ are shown as a gray squared line. 
    } \label{fig:Nmol1Nnor1_IC_etanor}
\end{figure}
\begin{table}
\centering
\begin{tabular}{||c || c | c | c | c ||} 
 \hline
$\eta_{\rm nor}$ [a.u.] & $5\times 10^{-7}$ & $1\times 10^{-6}$ & $2\times 10^{-6}$ & $5\times 10^{-6}$ \\ [0.5ex] 
 \hline\hline
 $\left(k_{\rm c}/k_{\rm o}\right)_{\rm max}$ & 1.5 & 2.2 & 3.1 & 4.2\\ 
  \hline
$\lambda_{\rm c}$ [$\mathrm{cm}^{-1}$] & 150 & 210 & 300 & 480 \\
 \hline
$\eta_{\rm c}$ [a.u.] & 0.00125 & 0.0025 & 0.005 & 0.01\\ [1ex] 
 \hline
\end{tabular}
\caption{Maximal rate enhancement $\left(k_{\rm c}/k_{\rm o}\right)_{\rm max}$ and the corresponding values of the cavity lossy strength $\lambda_{\rm c}$ and light-matter coupling strength $\eta_{\rm c}$, obtained for the minimal model with a single reactive mode $N_{\rm mol}=1$ and a dissipation-free non-reactive vibration $N_{\rm nor}=1$,  as illustrated in \Fig{fig:Nmol1Nnor1_IC_model}, for a given $\eta_{\rm nor}$. The frequencies of the cavity mode and the non-reactive mode are tuned in close resonance with the molecular vibrational transition, i.e. $\omega_{\rm c}= \omega_{\rm nor}=1185\,\mathrm{cm}^{-1}$, and the cavity bath cutoff frequency is fixed at $\Omega_{\rm c}=1000\,\mathrm{cm}^{-1}$.}
\label{table:1}
\end{table}

We first consider an idealized situation in which the non-reactive mode is decoupled from its dissipative bath, i.e., $\lambda_{\rm nor}=0$. In this limit, the non-reactive spectator mode acts as an effective bridge between the cavity mode and the reactive vibration. Energy fluctuations originating from the cavity bath can be transferred to the molecular solvent, or vice versa, through a sequential energy-transfer chainlet formed by the cavity mode, the non-reactive vibration, and the reactive coordinate, as illustrated schematically in \Fig{fig:Nmol1Nnor1_IC_model}. For this minimal system with $N_{\rm mol}=1$ and $N_{\rm nor}=1$, we find that the reaction rate is always enhanced inside the cavity. Notably, the location of the maximal enhancement becomes highly sensitive to the coupling parameters and must be understood in terms of state hybridization under strong coupling conditions.

To elucidate the role of vibrational and photonic resonances, \Fig{fig:Nmol1Nnor1_IC_wcwr} shows contour plots of the rate enhancement factor $k_{\rm c}/k_{\rm o}$ as functions of the non-reactive vibrational frequency $\omega_{\rm nor}$ and the cavity frequency $\omega_{\rm c}$ for three representative sets of coupling strengths $\eta_{\rm nor}$ and $\eta_{\rm c}$. Unless otherwise specified, the remaining parameters are fixed at $\lambda_{\rm c}=200\,\mathrm{cm}^{-1}$ and $\Omega_{\rm c}=1000\,\mathrm{cm}^{-1}$.

In the relatively weak coupling regime with $\eta_{\rm nor}=1\times10^{-6}$~a.u. and $\eta_{\rm c}=0.00125$~a.u., as shown in \Fig{fig:Nmol1Nnor1_IC_wcwr}~a), a narrow resonant peak emerges with a maximum enhancement factor $(k_{\rm c}/k_{\rm o})_{\rm max}=1.94$ at $\omega_{\rm nor}\approx1190\,\mathrm{cm}^{-1}$ and $\omega_{\rm c}\approx1170\,\mathrm{cm}^{-1}$, both in close resonance with the molecular vibrational transition. At this point, the quantized photon energy is most efficiently transmitted into the molecule via the resonant non-reactive vibration.

Upon increasing the light–matter coupling strength, the rate-enhancement landscape undergoes a qualitative transformation. As shown in \Fig{fig:Nmol1Nnor1_IC_wcwr}~b) for $\eta_{\rm c}=0.005$~a.u., the resonant peak broadens and evolves into a slanted avoided-crossing-like structure along the diagonal direction. Two local maxima appear at $(\omega_{\rm nor},\,\omega_{\rm c})=(1150,\,1125)\,\mathrm{cm}^{-1}$ with $(k_{\rm c}/k_{\rm o})_{\rm max}=1.94$, and at $(1230,\,1225)\,\mathrm{cm}^{-1}$ with $(k_{\rm c}/k_{\rm o})_{\rm max}=1.88$. As $\eta{\rm c}$ is further increased (see Supplementary Material), the separation between these two maxima grows, reflecting the progressively stronger hybridization between the cavity mode and the non-reactive vibrational mode. In this regime, the coupled system must be described in terms of upper and lower polaritonic states rather than bare cavity and vibrational excitations. 

Efficient rate enhancement occurs when either polariton branch becomes resonant with the molecular reactive vibrational transition, thereby enabling effective transmission of energy fluctuations from the cavity bath into the molecule and their subsequent dissipation into the solvent. For instance, when the upper polariton closely matches the reactive vibrational transition energy--corresponding to the peak at $\omega_{\rm nor}=1150\,\mathrm{cm}^{-1}$ and $\omega_{\rm c}=1125\,\mathrm{cm}^{-1}$--energy injected from the cavity bath is funneled through the upper polaritonic state into the molecular degrees of freedom. Analogously, at the higher-frequency maximum $(\omega_{\rm nor},\,\omega_{\rm c})=(1230,\,1225)\,\mathrm{cm}^{-1}$, the lower polariton becomes resonant with the molecular transition and mediates an efficient energy-transfer pathway.

In addition to cavity–vibration hybridization, strong coupling between the reactive and non-reactive vibrations also reshapes the resonant conditions perceived by the cavity. Upon increasing the vibrational coupling $\eta_{\rm nor}$, a different avoided-crossing silhouette emerges along the anti-diagonal direction of the $(\omega_{\rm nor},\,\omega_{\rm c})$ plane, as illustrated in \Fig{fig:Nmol1Nnor1_IC_wcwr}~c) for $\eta_{\rm nor}=2\times10^{-6}$~a.u. For a fixed value of $\omega_{\rm nor}$, a vertical cut through the contour plot--corresponding to the rate-modification profile $k_{\rm c}/k_{\rm o}$ as a function of the cavity frequency $\omega_{\rm c}$--reveals a clear splitting of the resonance peaks, with the separation between them increasing as $\eta_{\rm nor}$ is further enhanced (see Supplementary Material).

This behavior originates from the hybridization between the reactive and non-reactive vibrational modes, which generates two distinct hybrid vibrational excited states. As a consequence, the cavity mode can now resonantly couple to either of these hybridized vibrational transitions, giving rise to multiple resonant conditions for cavity-mediated energy transfer and, in turn, to the observed peak splitting in the rate-modification profile. This vibrational-mode hybridization thus introduces an additional layer of complexity into the cavity-induced rate landscape, further emphasizing that the optimal reaction dynamics needs to consider the collective interplay of all coupled degrees of freedom.

Beyond resonance conditions, maximal cavity-induced rate enhancement additionally requires a delicate balance among the relevant dynamical timescales governing light–matter coupling, vibrational energy exchange, and environmental dissipation, resembling the behavior observed outside the cavity. To illustrate this point, \Fig{fig:Nmol1Nnor1_IC_etaclambc} presents contour plots of $k_{\rm c}/k_{\rm o}$ as functions of the light–matter coupling strength $\eta_{\rm c}$ and the cavity loss strength $\lambda_{\rm c}$ for several values of $\eta_{\rm nor}$, with the frequencies fixed at $\omega_{\rm c}=\omega_{\rm nor}=1185\,\mathrm{cm}^{-1}$. 

For a given $\eta_{\rm nor}$, which sets the characteristic timescale for energy exchange between the non-reactive spectator mode and the reactive vibration, the maximal rate enhancement is achieved only at intermediate values of both $\eta_{\rm c}$ and $\lambda_{\rm c}$. As a representative example, when $\eta_{\rm nor}=1\times10^{-6}$~a.u., the maximum enhancement $(k_{\rm c}/k_{\rm o})_{\rm max}=2.2$ occurs at approximately $\eta_{\rm c}=0.0025$~a.u. and $\lambda_{\rm c}=210\,\mathrm{cm}^{-1}$. Deviations from this optimal regime, either toward weaker or stronger light-matter coupling or cavity dissipation, leads to a reduction of the rate enhancment. Conversely, for fixed values of $\eta_{\rm c}$ and $\lambda_{\rm c}$, the reaction rate exhibits a turnover as a function of $\eta_{\rm nor}$, as shown in Fig.~\ref{fig:Nmol1Nnor1_IC_etanor}. 

Taken together, these results demonstrate that cavity-induced rate enhancement relies on efficient energy transfer from the cavity bath, through the cavity mode and the non-reactive vibration, to the reactive coordinate and its surrounding solvent. Cavity-induced enhanced reactivity therefore emerges from the cooperative tuning of all relevant couplings rather than from the monotonic increase of any single parameter. Varying an individual coupling strength, such as the light–matter interaction, while holding all others fixed inevitably leads to a characteristic turnover behavior. Larger rate enhancements can thus be achieved only through the concerted increase of $\eta_{\rm nor}$, $\eta_{\rm c}$, and $\lambda_{\rm c}$, as summarized in Table~\ref{table:1}. Notably, the maximal rate enhancement is consistently accompanied by a narrowing of the frequency-dependent rate-modification profile as a function of $\omega_{\rm c}$ at fixed $\omega_{\rm nor}$, reflecting increasingly stringent resonance conditions at optimal coupling.

In short, rate modification and its optimization within a single reactive pathway constitute a strategic interplay of competition and cooperation among all energy-exchange processes involved. Most strikingly, the largest enhancements arise from the cooperative acceleration of these processes. However, once multiple reactive pathways are available, more intricate interference effects emerge. This is the case when dissipation of the non-reactive mode is explicitly included, as will be addressed below.

\begin{figure}[H]
  \begin{minipage}[c]{0.45\textwidth}
    \includegraphics[width=\textwidth]{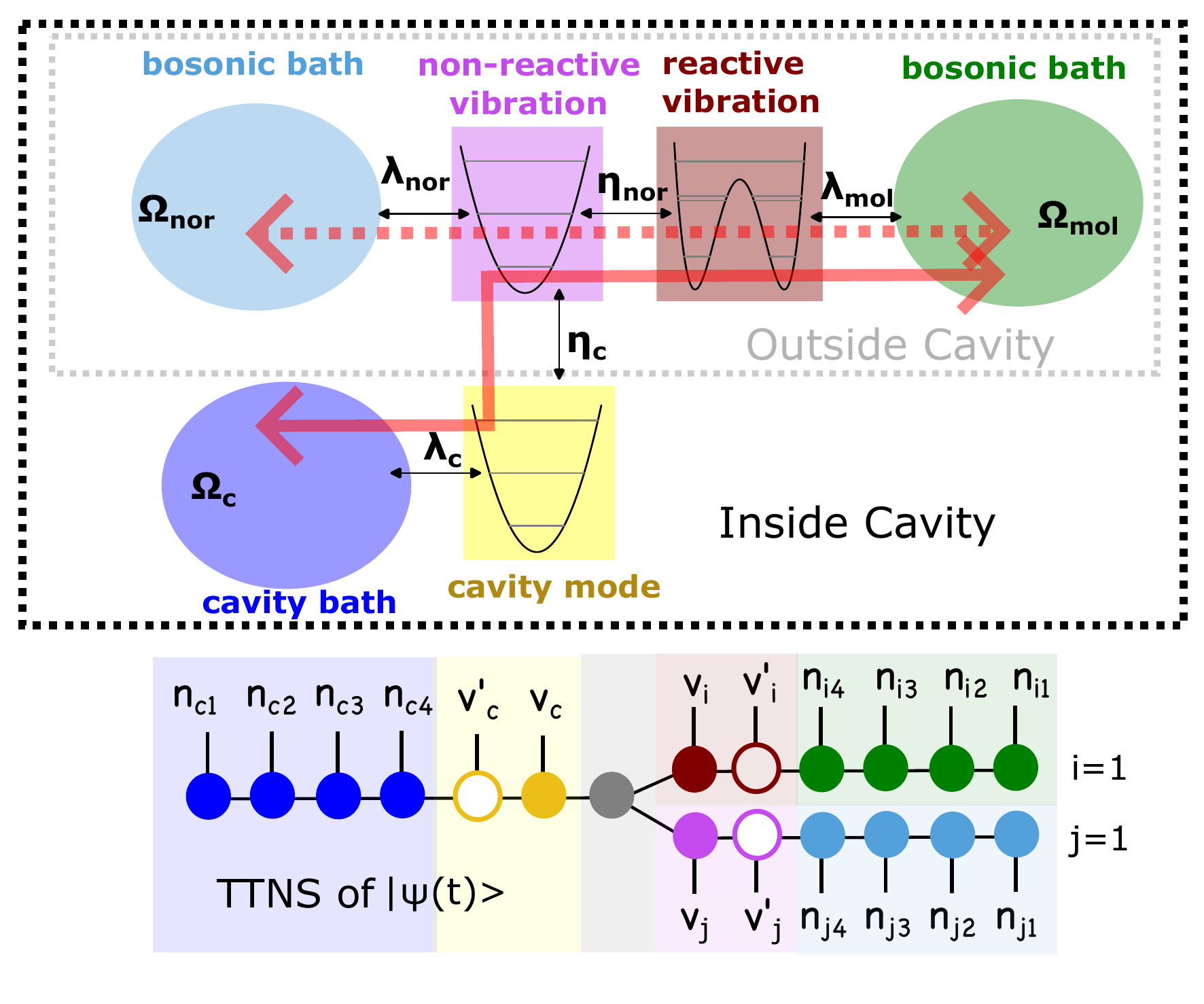}
  \end{minipage}
    \caption{
 Schematic illustration of a dissipative non-reactive vibrational mode bridging a single reactive molecule and a cavity mode, along with the TTNS representation of the extended wave function $|\Psi(t)\rangle$. 
    } \label{fig:Nmol1Nnor1_IC_lossy_model}
\end{figure}

\begin{figure}[H]
  \begin{minipage}[c]{0.5\textwidth}
    \includegraphics[width=\textwidth]{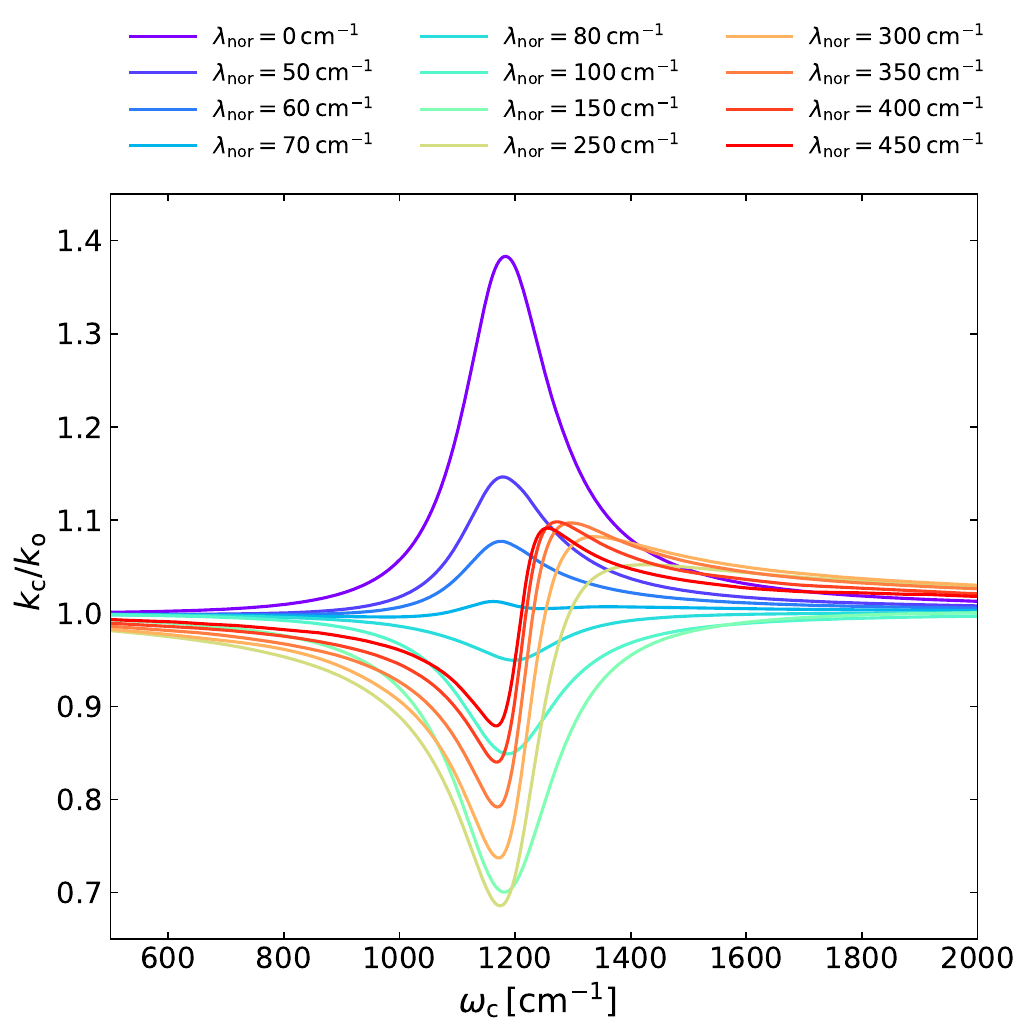}
  \end{minipage}
    \caption{Ratios of reaction rates $k_{\rm c}/k_{\rm o}$ as a function of the cavity frequency $\omega_{\rm c}$ for different dissipation strengths of the non-reactive vibrational mode $\lambda_{\rm nor}$. Other parameters are fixed at $\eta_{\rm nor}=1\times10^{-6}$~a.u., $\eta_{\rm c}=0.005$~a.u., $\omega_{\rm nor}=1185\,\mathrm{cm}^{-1}$, $\lambda_{\rm c}=100\,\mathrm{cm}^{-1}$, $\Omega_{\rm nor}=1000\,\mathrm{cm}^{-1}$, and $\Omega_{\rm c}=1000\,\mathrm{cm}^{-1}$.} 
 \label{fig:rateratios_fano1}
\end{figure}

We now relax the assumption of an isolated non-reactive vibration and explicitly introduce dissipation by coupling it to its own bath with finite strength $\lambda_{\rm nor}$, as schematically illustrated in \Fig{fig:Nmol1Nnor1_IC_lossy_model}. \Fig{fig:rateratios_fano1} shows the resulting rate-modification profiles $k_{\rm c}/k_{\rm o}$ as functions of the cavity frequency $\omega_{\rm c}$ for several values of $\lambda_{\rm nor}$, with all other parameters fixed as specified in the caption.


For weak dissipation, the rate-modification profile exhibits a resonant enhancement peak. As $\lambda_{\rm nor}$ increases, the profile gradually evolves into an asymmetric line shape featuring both a peak and a dip in the vicinity of the resonance. In this intermediate regime, the overall enhancement amplitude is reduced, rendering the asymmetry less apparent in Fig.~\ref{fig:rateratios_fano1}. Upon further increasing $\lambda_{\rm nor}$, the profile turns into a pronounced resonant suppression, with the depth of the dip growing monotonically with increasing dissipation. At even larger values of $\lambda_{\rm nor}$, the asymmetric line shape re-emerges, characterized by a dip on the low-frequency side and a peak on the high-frequency side of the resonant cavity frequency.

These qualitatively distinct rate modification profiles can be unified within the framework of Fano resonance, which is described by the characteristic lineshape formula
\begin{equation}
\label{Fanoformula}
    \sigma(\omega)= D^2 \frac{(q+\epsilon)^2}{1+\epsilon^2}.
\end{equation}
Here, $\epsilon=(\omega-\omega_{R})/(\Gamma_{R}/2)$ is the normalized detuning from the resonance frequency $\omega_{R}$, scaled by the half-width $\Gamma_R/2$ of the resonance. 
The dimensionless parameter $q=\cot(\delta)$ is the Fano asymmetry parameter, and $D=\sin(\delta)$, where $\delta$ denotes the relative phase shift between two interfering pathways. Depending on the value of $q$, the lineshape continuously interpolates between a symmetric Lorentzian peak, a symmetric anti-resonant dip, and an asymmetric profile featuring an adjacent peak and dip near resonance.  In the limiting cases $|q|\rightarrow \infty$, $\sigma(\omega)$ reduces to a symmetric Lorentzian peak with its maximum at $\omega=\omega_{R}$. In contrast, for $q=0$, the lineshap $\sigma(\omega)$ becomes a symmetric anti-resonance, exhibiting a minimum at $\omega=\omega_{R}$. For a finite value of $|q|$ between $0$ and $\infty$, the line shape displays both a maximum and a minimum located at $\omega_{\rm max}=\omega_{R}+\Gamma_{R}/(2q)$ and $\omega_{\rm min}=\omega_{R}-q\Gamma_{R}/2$, respectively. Representative intances of Fano line shapes for different values of $q$ can be found in Refs.~\onlinecite{Fano_1961_PR_p1866,Limonov_2017_NP_p543}.

The line shape formula in \Eq{Fanoformula} was first introduced by Ugo Fano,\cite{Fano_1935_INC_p154,Fano_1961_PR_p1866} as its name suggests, to explain the sharp asymmetric scattering profiles observed in Rydberg atomic spectra arising from autoionization. Since then, Fano resonances, characterized by their distinctive asymmetric line shapes, have been identified in a wide range of physical systems, including plasmonic metasurfaces,\cite{Miroshnichenko_2010_RMP_p2257,Rahmani_2013_LPR_p329} nanoscale photonic crystals,\cite{Limonov_2017_NP_p543,Limonov_2021_AOP_p703} as well as quantum impurity systems and tunneling junctions,\cite{Ujsaghy_2000_Prl_p2557,Zheng_2022_ACIE_p202210097} among many others.

At its core, the Fano resonance arises from quantum interference between two transition pathways. One pathway involves excitation into a localized discrete state, whose phase varies rapidly near resonance, while the other proceeds through a continuous band of states that is coupled to the discrete level and exhibits a slowly varying phase. The Fano parameter $q$ quantifies the relative weight of these two pathways and is given by the cotangent of the phase shift $\delta$ between them. Constructive interference between the pathways results in resonant enhancement, whereas destructive interference leads to resonant suppression. When both contributions are present, these two effects coexist, producing a characteristic asymmetric spectral profile with a closely spaced maximum and minimum on opposite sides of the resonance frequency.

\begin{figure}[H]
  \begin{minipage}[c]{0.5\textwidth}
    \includegraphics[width=\textwidth]{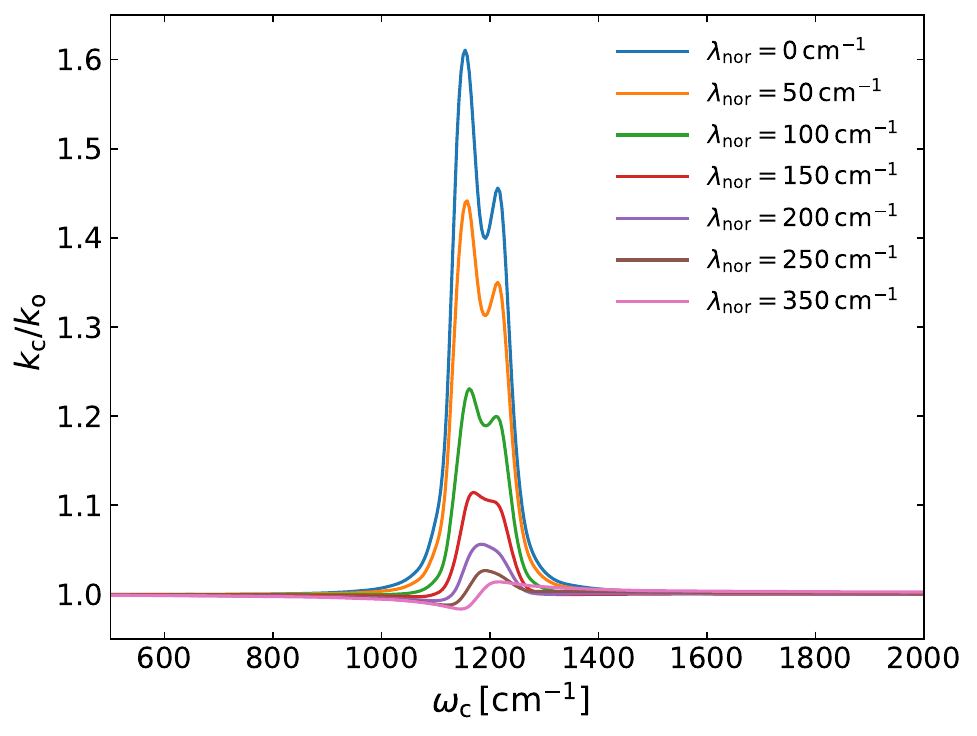}
  \end{minipage}
    \caption{
 Ratios of reaction rates $k_{\rm c}/k_{\rm o}$ as a function of the cavity frequency $\omega_{\rm c}$ for different dissipation strengths of the non-reactive vibrational mode $\lambda_{\rm nor}$. Other parameters are fixed at $\eta_{\rm nor}=2\times10^{-6}$~a.u., $\eta_{\rm c}=0.00125$~a.u., $\omega_{\rm nor}=1185\,\mathrm{cm}^{-1}$, $\lambda_{\rm c}=100\,\mathrm{cm}^{-1}$, $\Omega_{\rm nor}=1000\,\mathrm{cm}^{-1}$, and $\Omega_{\rm c}=1000\,\mathrm{cm}^{-1}$. 
    } \label{fig:rateratios_fano2}
\end{figure}

In the context of cooperative polaritonic chemistry, or more specifically for the present cavity-coupled minimal model system illustrated in \Fig{fig:Nmol1Nnor1_IC_lossy_model}, the conditions required for the emergence of a Fano resonance are naturally fulfilled. The discrete cavity mode is side-coupled to a non-reactive spectator vibration, which itself is dissipative due to its coupling to a continuous bosonic bath. Together, these DoFs form an effective T-shaped configuration that supports the interference of two distinct energy-transfer pathways.

The first pathway, which already exists outside the cavity, transfers energy fluctuations from the bosonic bath associated with the non-reactive mode through the reactive vibration and into the surrounding solvent. This pathway is depicted as a broad double-headed dotted arrow in \Fig{fig:Nmol1Nnor1_IC_lossy_model}. The introduction of the cavity mode enables a second pathway, shown as a broad double-headed solid arrow, which connects the cavity bath to the solvent via the cavity mode, the non-reactive vibration, and finally the reactive coordinate.

When the second pathway is in phase with the first, vibrational transitions in the non-reactive mode are reinforced. The resulting enhancement of vibrational excitation in the non-reactive mode, followed by spontaneous emission, can stimulate excitation of the reactive vibration through their mutual coupling, ultimately leading to an increased reaction rate. In contrast, when the two pathways are out of phase, vibrational excitation of the non-reactive mode triggered by its continuous bath is partially or completely cancelled by energy transfer into the cavity mode, with the excess energy subsequently dissipated into the cavity bath. Under these conditions, the reaction rate inside the cavity is suppressed.

In the absence of dissipation of the non-reactive mode ($\lambda_{\rm nor}=0$), corresponding to the regime explored earlier, the reaction dynamics inside the cavity is governed exclusively by the second pathway. This limit corresponds to $|q|\rightarrow\infty$ and results in purely resonant rate enhancement. As $\lambda_{\rm nor}$ is increased to intermediate values, destructive interference between the two pathways becomes operative, leading to resonant rate suppression, as previously reported in quantum dynamical simulations.\cite{Lindoy_2023_NC_p2733,Vega_2025_JACS_p19727}. Upon further increasing $\lambda_{\rm nor}$, a larger continuum of bath states participates in the dynamics,\cite{Miroshnichenko_2010_RMP_p2257} giving rise to the re-emergence of asymmetric rate modification profiles characteristic of finite-$q$ Fano interference.

Finally, in the strong-coupling regime, vibrational hybridization becomes intertwined with Fano interference, leading to more intricate rate modification profiles. To illustrate this interplay, we increase the coupling strength between the reactive and non-reactive vibrational modes to $\eta_{\rm nor}=2\times10^{-6}$~a.u. and simultaneously reduce the light–matter coupling to $\eta_{\rm c}=0.00125$~a.u. The resulting rate modification profiles for various values of $\lambda_{\rm nor}$ are shown in \Fig{fig:rateratios_fano2}.

When $\lambda_{\rm nor}$ is small, the peak in the rate modification profile splits into a doublet. As explained before, this splitting originates from the hybridization between the reactive and non-reactive vibrational modes, which generates lower and higher hybridized vibrational excited states. In this regime, energy transfer from the cavity bath to the solvent is dominant, and under these conditions, it is most efficient when the cavity frequency is resonant with the transition to either of these hybridized states.

As $\lambda_{\rm nor}$ is increased, two concurrent effects emerge. First, dissipation induces spectral broadening, causing the initially well-resolved doublet to gradually merge. Second, stronger coupling to the dissipative bath boosts Fano interference between two reactive pathways: one mediated by the discrete cavity mode, and the other involving the continuum of more bath states surrounding the hybridized vibrational modes. This interference increases the asymmetry of the rate modification line shape, as shown in \Fig{fig:rateratios_fano2}.

In summary, even this minimal cavity-coupled model system, consisting of a single reactive and a single non-reactive mode interacting with a single cavity mode, already reveals the intricate nature of cavity-induced reaction dynamics. These processes are dictated by a complex interplay of multiple factors, including mode hybridization, the balancing of dynamical timescales, and interference between distinct energy transfer pathways. Consequently, cavity-induced rate modifications are highly sensitive to structural and environmental parameters. Even slight variations can switch the cavity effect from resonant rate enhancement to suppression, or give rise to Fano-type asymmetric features within a narrow frequency window--phenomena that have not yet been observed experimentally. Building on these insights, in the next section, we extend our analysis to the collective regime, moving beyond the single-molecule and single–non-reactive-mode assumption ($N_{\rm mol}=N_{\rm nor}=1$), to explore how collective interactions further enrich cavity-modified chemical reactivity.

\subsection{Collective Regime}
While single-molecule vibrational strong coupling has been achieved in a carefully engineered nanophotonic platform,\cite{Chikkaraddy_2016_N_p127} the vast majority of polaritonic chemistry experiments are conducted in the collective regime, where a large number of molecules are simultaneously coupled to a common optical cavity mode. In this setting, molecular vibrations interact not only with the cavity field but also indirectly with one another through their shared coupling to the confined electromagnetic environment, potentially leading to qualitative changes in reaction dynamics. Understanding how cavity-modified reactivity evolves from the single-molecule limit to the collective regime is therefore essential for developing a realistic and predictive theoretical framework for polariton-assisted chemistry.

\begin{figure}
  \begin{minipage}[c]{0.45\textwidth}   
  \raggedright a)  
    \includegraphics[width=\textwidth]{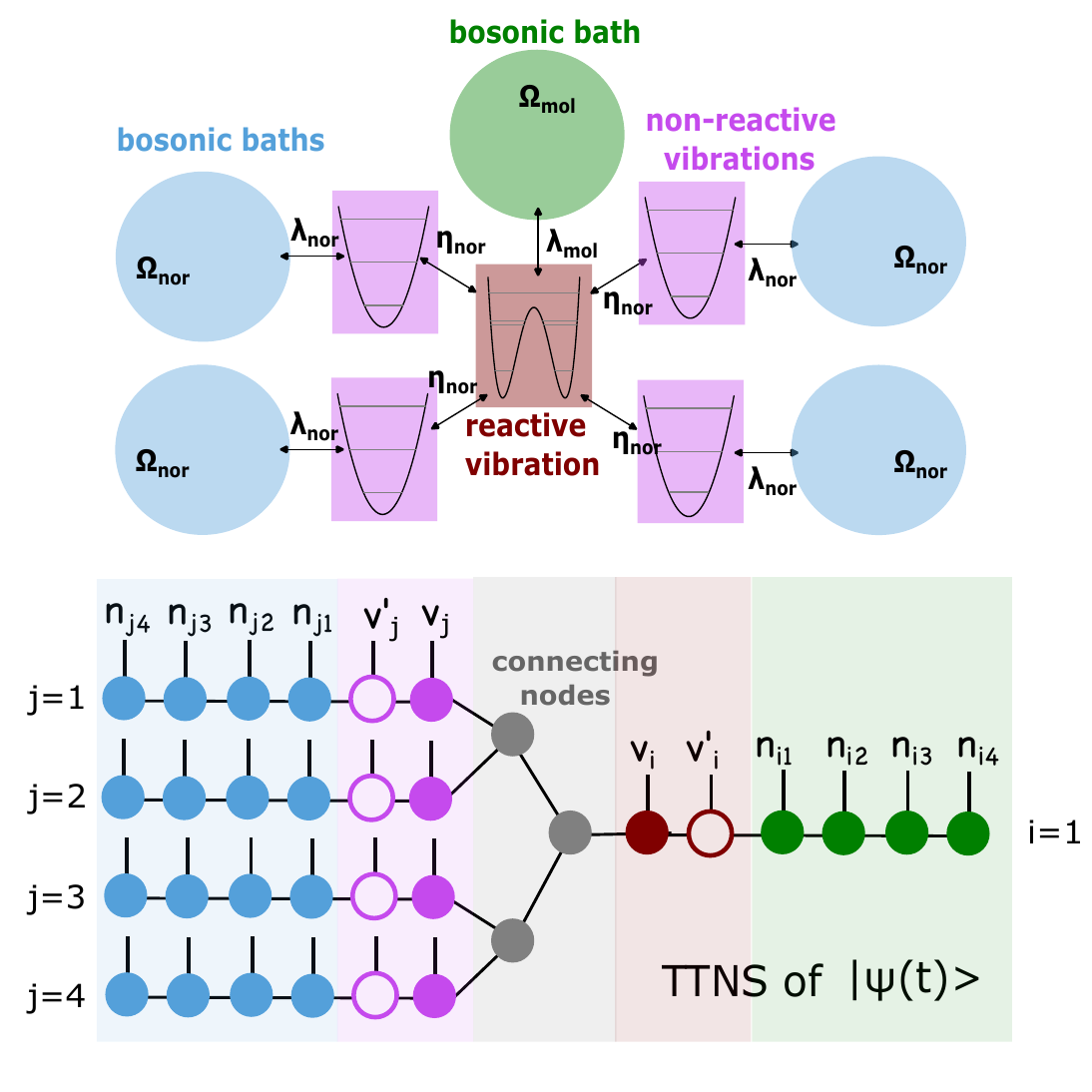}
  \end{minipage}
  \begin{minipage}[c]{0.45\textwidth}   
  \raggedright b)  
    \includegraphics[width=\textwidth]{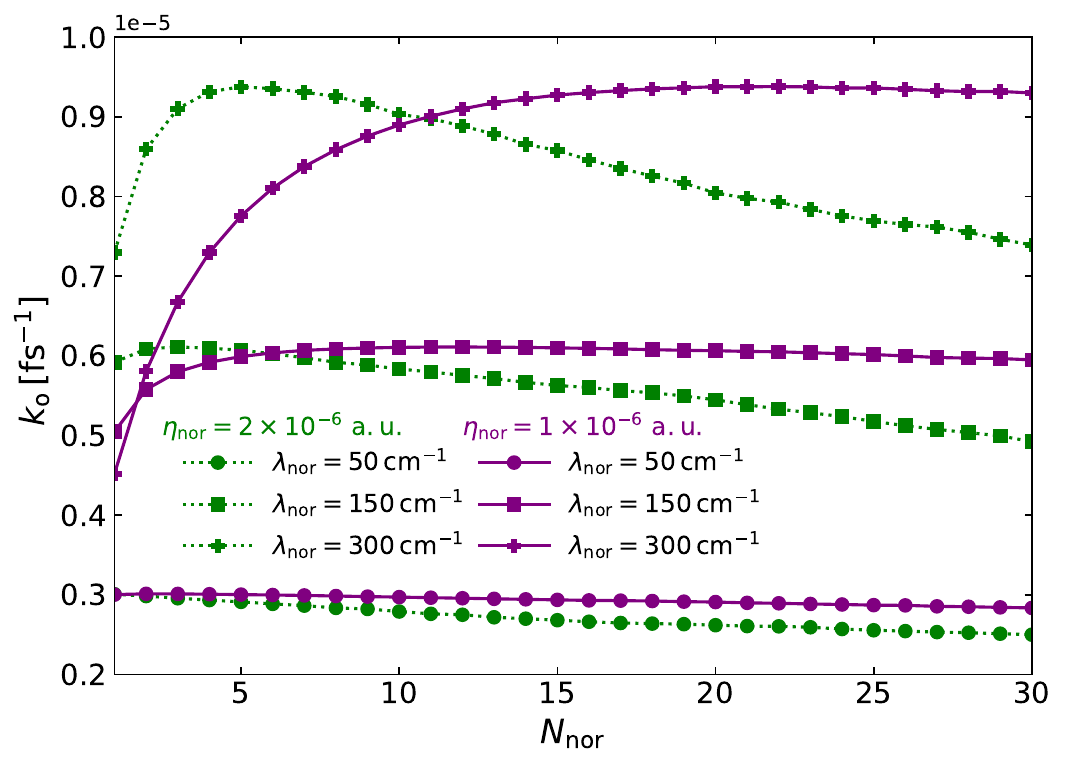}
  \end{minipage}
    \caption{a) Schematic illustration of a model system with a single reactive mode ($N_{\rm mol}=1$) interacting with four non-reactive vibrational modes ($N_{\rm nor}=4$ outside the cavity, along with the graphic representation of the TTNS decomposition of the extended wave function $|\Psi(t)\rangle$ in this scenario. b) Reaction rates outside the cavity $k_{\rm o}$ as a function of the number of non-reactive spectator modes $N_{\rm nor}$.  The frequencies of all non-reactive vibrational modes are fixed at $\omega_{\rm nor}=1185\,\mathrm{cm}^{-1}$, and we set $\Omega_{\rm nor}=1000\,\mathrm{cm}^{-1}$.
    } \label{fig:Nmol1NnorN_OC}
\end{figure}
\subsubsection{Outside Cavity}
The non-reactive spectator modes may represent either internal molecular vibrations or characteristic solvent motions. In the latter case, a reactive molecule is typically coupled to multiple solvent molecules whose absorption bands may overlap with those of the reactive vibration, a scenario that is particularly relevant in studies of cooperative polaritonic chemistry. We therefore begin by examining the influence of collective coupling to multiple non-reactive spectator modes outside the optical cavity. As an illustrative example, the case with $N_{\rm nor}=4$ is shown schematically in \Fig{fig:Nmol1NnorN_OC}~a).

\Fig{fig:Nmol1NnorN_OC}~b) displays the reaction rate as a function of the number of non-reactive modes $N_{\rm nor}$ under resonant conditions, with $\omega_{\rm nor}=1185\,\mathrm{cm}^{-1}$. Notably, increasing the number of coupled spectator modes does not monotonically enhance the reaction rate. Instead, an optimal number of non-reactive modes exists that maximizes the rate, and this optimal value depends sensitively on the system parameters, as illustrated in \Fig{fig:Nmol1NnorN_OC}~b).

Stronger dissipation of the non-reactive modes, characterized by larger $\lambda_{\rm nor}$, favors larger-scale aggregation to achieve maximal rate enhancement. For example, when $\eta_{\rm nor}=1\times10^{-6}$~a.u., the optimal number of modes increases from $N_{\rm nor}=2$ for $\lambda_{\rm nor}=50\,\mathrm{cm}^{-1}$, to $N_{\rm nor}=12$ for $\lambda_{\rm nor}=150\,\mathrm{cm}^{-1}$, and further to $N_{\rm nor}=20$ for $\lambda_{\rm nor}=300\,\mathrm{cm}^{-1}$. Beyond the turnover point, the reaction rate decreases gradually with further increases in $N_{\rm nor}$. In contrast, increasing the vibrational coupling strength $\eta_{\rm nor}$ reduces the value of $N_{\rm nor}$ at which the maximal rate is attained and accelerates the post-maximum decay of the rate. In particular, when $\eta_{\rm nor}=2\times10^{-6}$~a.u. and $\lambda_{\rm nor}=50\,\mathrm{cm}^{-1}$, the non-reactive modes act predominantly as rate-suppressing channels.

The inclusion of additional non-reactive modes introduces more dissipation channels and, consequently, more available energy-transfer pathways. However, the resulting reaction rate is not a trivial classical sum of contributions from individual pathways. Rather, it emerges from their quantum-mechanical coherent superposition, implying that interference effects may play a crucial role in determining the overall chemical response. Under certain parameter regimes, the additional pathways introduced by increasing $N_{\rm nor}$ interfere constructively with existing channels, leading to cooperative rate enhancement. Conversely, when the newly introduced pathways interfere destructively with others, the net energy flow toward the reactive coordinate is reduced, resulting in rate suppression. The emergence of an optimal number of non-reactive modes that maximizes the reaction rate therefore reflects a balance point at which constructive interference among collective vibrational pathways is most efficient, before destructive interference starts acting up.

\begin{figure*}
\centering
  \begin{minipage}[c]{0.8\textwidth}
  \raggedright a) \
    \includegraphics[width=\textwidth]{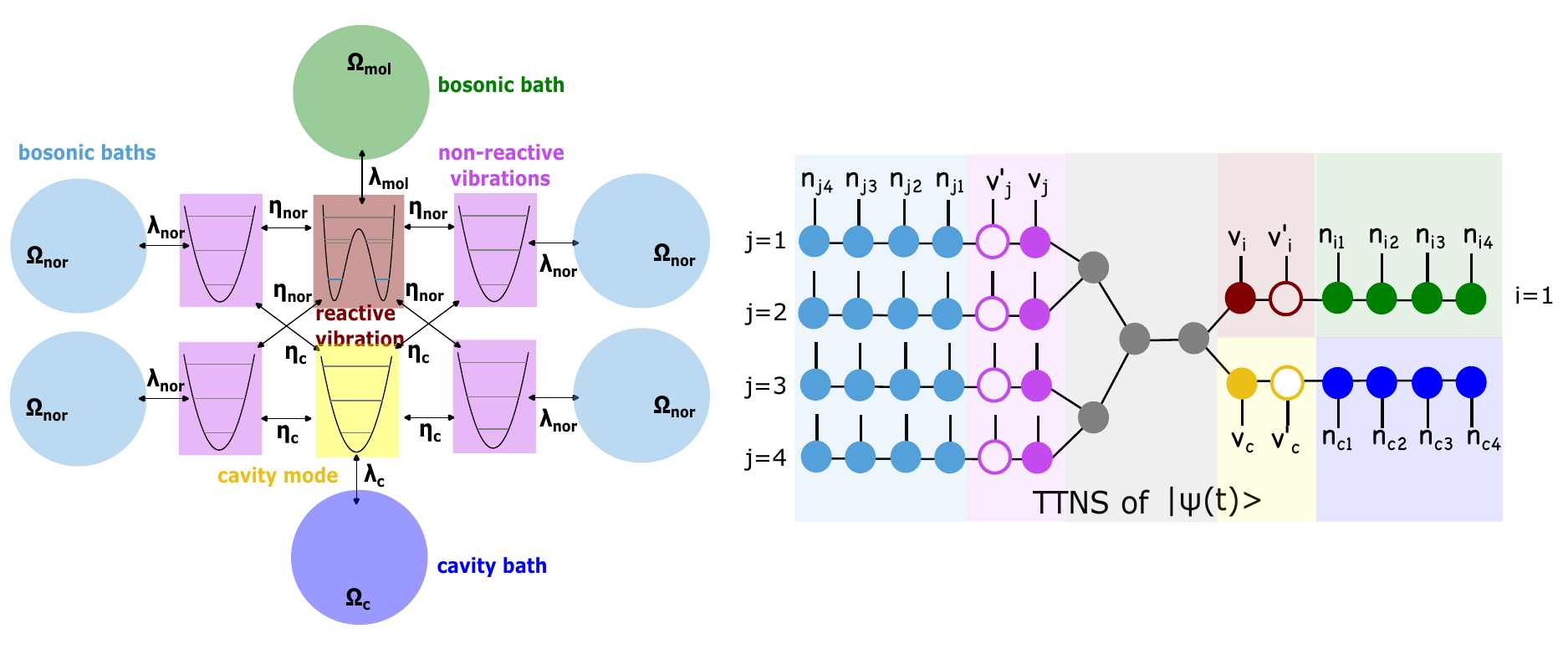}
  \end{minipage} 
  \begin{minipage}[c]{0.45\textwidth}
  \raggedright b) $\eta_{\rm c}=0.00125\,\mathrm{a.u.}$ \
    \includegraphics[width=\textwidth]{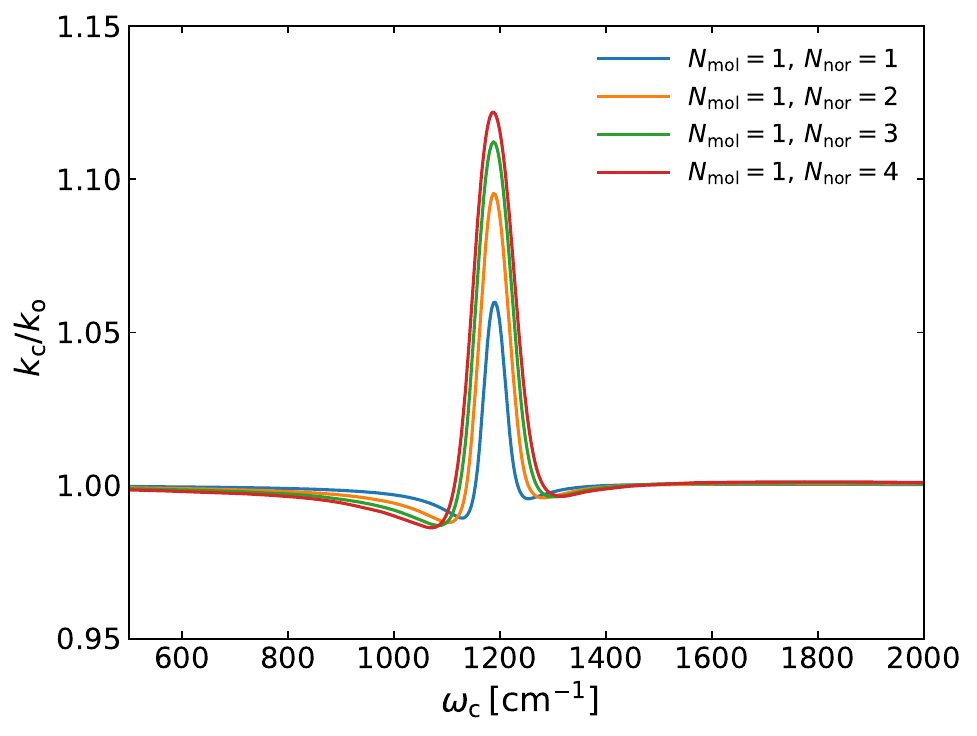}
  \end{minipage} 
 \begin{minipage}[c]{0.45\textwidth}
  \raggedright c) $\eta_{\rm c}=0.005\,\mathrm{a.u.}$\
    \includegraphics[width=\textwidth]{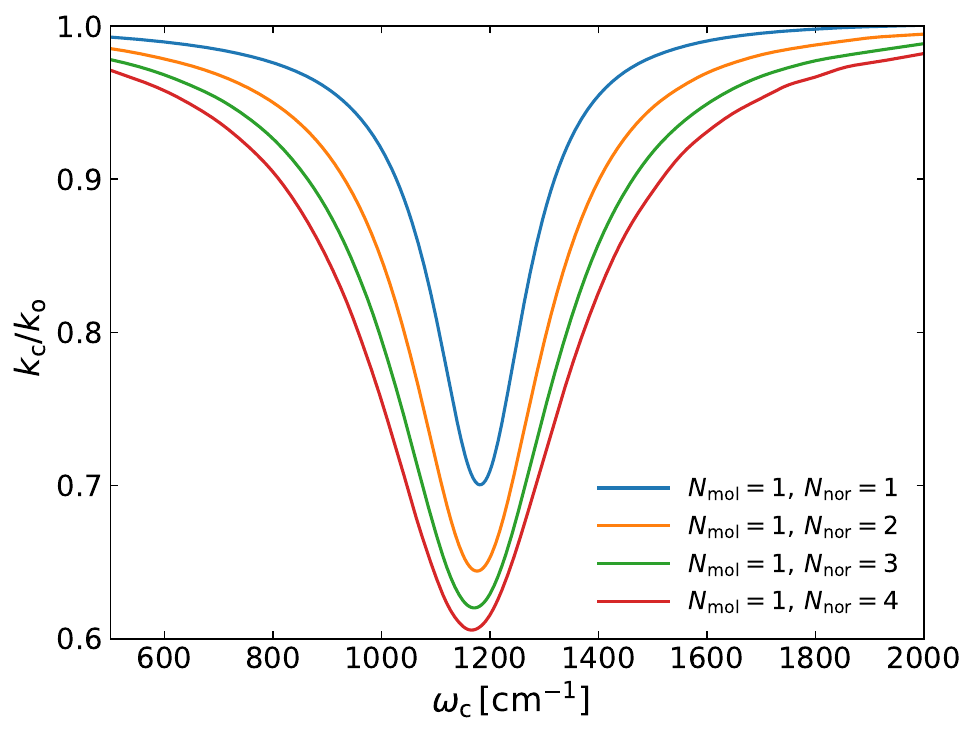}
  \end{minipage} 
    \caption{a) Schematic illustration of a model system with a single reactive mode ($N_{\rm mol}=1$) interacting with four non-reactive vibrational modes ($N_{\rm nor}=4$ inside the optical cavity, along with the graphic representation of the TTNS decomposition of the extended wave function $|\Psi(t)\rangle$ in this scenario. b) and c) Rate modification profiles as a function of the cavity frequency $\omega_{\rm c}$ for $N_{\rm mol}=1$ and $N_{\rm nor}$ ranging from 1 to 4, for two different cavity lossy strengths $\lambda_{\rm c}$. Other parameters are 
    fix at $\lambda_{\rm c}=100\,\mathrm{cm}^{-1}$, $\Omega_{\rm c}=1000\,\mathrm{cm}^{-1}$, and $\eta_{\rm nor}=1\times 10^{-6}$~a.u., $\lambda_{\rm nor}=150\,\mathrm{cm}^{-1}$, $\Omega_{\rm nor}=1000\,\mathrm{cm}^{-1}$, and $\omega_{\rm nor}=1185\,\mathrm{cm}^{-1}$.} \label{fig:rateratiosaggregation_Nmol1Nnor1-4_dissipative}
\end{figure*}

\begin{figure}
\centering
  \begin{minipage}[c]{0.45\textwidth}
  \raggedright a) \
    \includegraphics[width=\textwidth]{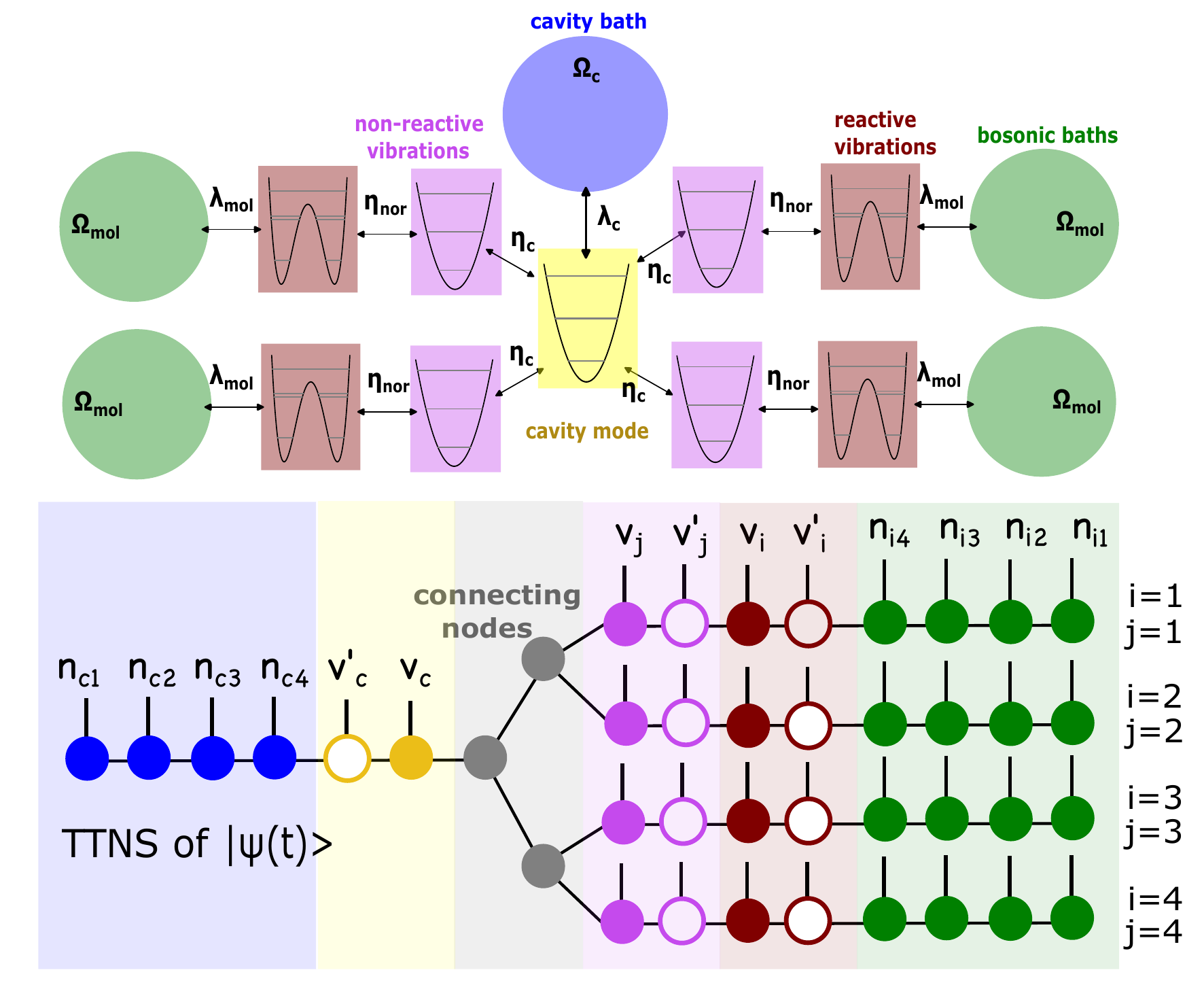}
  \end{minipage} 
  \begin{minipage}[c]{0.45\textwidth}
  \raggedright b) $\lambda_{\rm c}=50\,\mathrm{cm}^{-1}$ \
    \includegraphics[width=\textwidth]{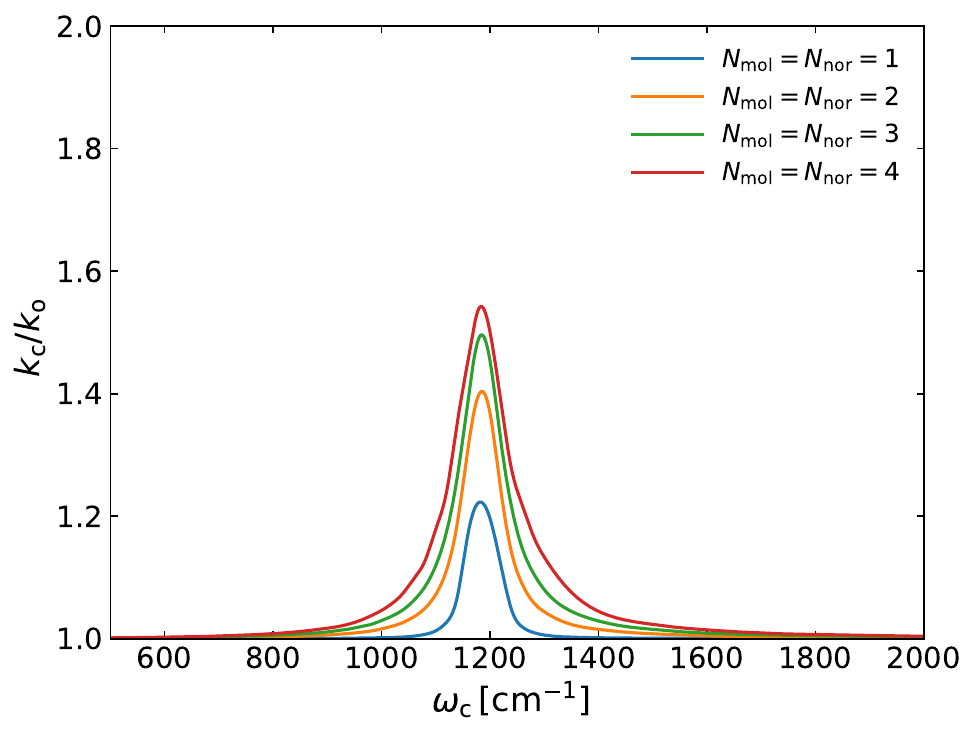}
  \end{minipage} 
 \begin{minipage}[c]{0.45\textwidth}
  \raggedright c) $\lambda_{\rm c}=200\,\mathrm{cm}^{-1}$\
    \includegraphics[width=\textwidth]{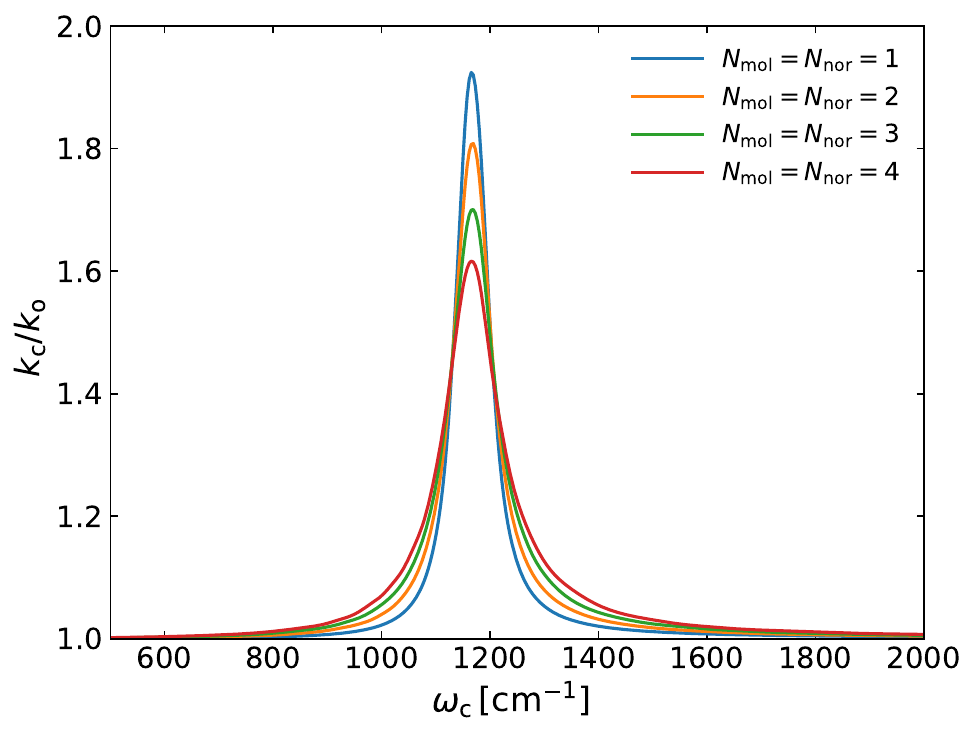}
  \end{minipage} 
    \caption{a) Schematic illustration of a model system with four reactive molecules ($N_{\rm mol}=1$) collectively coupled to a single cavity mode via dissipation-free non-reactive vibrational modes ($N_{\rm nor}=4$ and $\lambda_{\rm nor}=0$, along with the graphic representation of the TTNS decomposition of the extended wave function $|\Psi(t)\rangle$ in this scenario. b) and c) Rate modification profiles as a function of the cavity frequency $\omega_{\rm c}$ for $N_{\rm mol}=N_{\rm nor}$ ranging from 1 to 4.  Each panel corresponds to a different cavity lossy strength $\lambda_{\rm c}$. Other parameters are 
    fix $\eta_{\rm c}=0.00125$~a.u. and $\eta_{\rm nor}=1\times 10^{-6}$~a.u., and $\omega_{\rm nor}=1185\,\mathrm{cm}^{-1}$.} \label{fig:rateratiosaggregation_NmolNnor1-4_dissipationfree}
\end{figure}
\begin{figure}
\centering
  \begin{minipage}[c]{0.4\textwidth}
  \raggedright a) 
    \includegraphics[width=\textwidth]{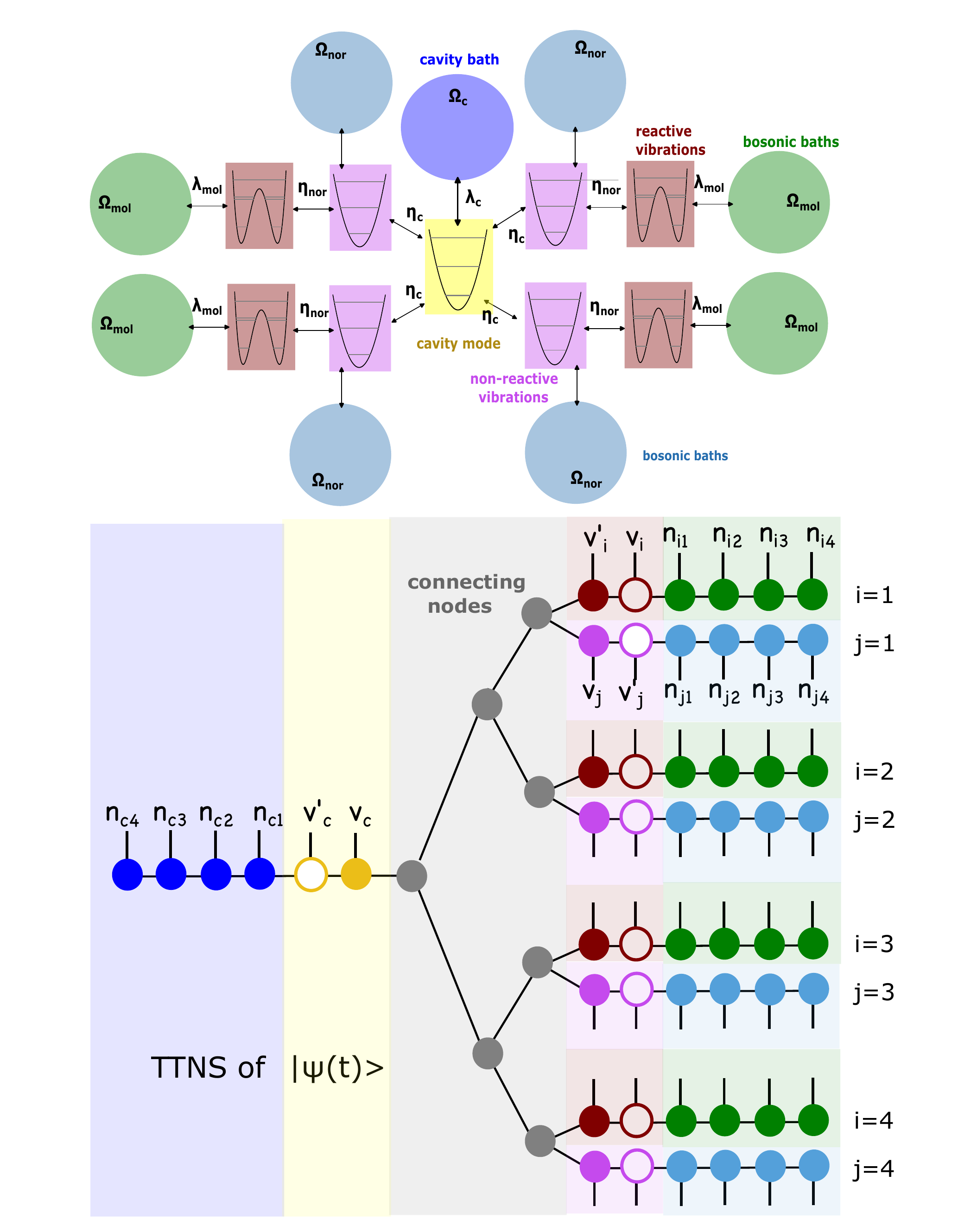}
  \end{minipage} 
  \begin{minipage}[c]{0.45\textwidth}
  \raggedright b) 
    \includegraphics[width=\textwidth]{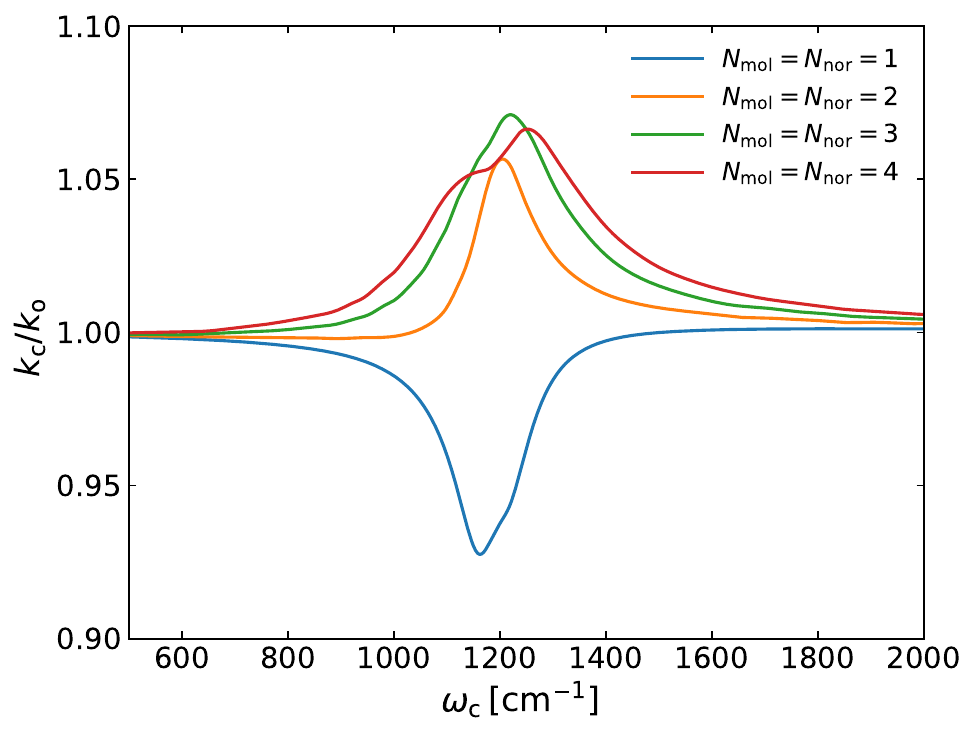}
  \end{minipage} 
    \caption{a) Schematic illustration of a model system with four reactive molecules ($N_{\rm mol}=4$) collectively coupled to a single cavity mode via dissipative non-reactive vibrational modes ($N_{\rm nor}=4$), along with the graphic representation of the TTNS decomposition of the extended wave function $|\Psi(t)\rangle$ in this scenario. b) Rate modification profiles as a function of the cavity frequency $\omega_{\rm c}$ for $N_{\rm mol}=N_{\rm nor}$ ranging from 1 to 4. Other parameters are 
    fix at $\eta_{\rm c}=0.0025\,\mathrm{a.u.}$, $\lambda_{\rm c}=100\,\mathrm{cm}^{-1}$, $\Omega_{\rm c}=1000\,\mathrm{cm}^{-1}$, and $\eta_{\rm nor}=1\times 10^{-6}$~a.u., $\lambda_{\rm nor}=150\,\mathrm{cm}^{-1}$, $\Omega_{\rm nor}=1000\,\mathrm{cm}^{-1}$, and $\omega_{\rm nor}=1185\,\mathrm{cm}^{-1}$.} \label{fig:rateratiosaggregation_NmolNnor1-4_dissipative}
\end{figure}
\subsubsection{Inside Cavity}
Next, we explore reaction dynamics on three different molecular network configurations, comprising multiple identical molecules interconnected via the cavity mode and non-reactive vibrational modes, in order to gain preliminary insights into quantum collective effects induced by the cavity on chemical reactivities. 

The first network configuration is a straightforward extension of the model system discussed above, where a single cavity mode is collectively coupled to several non-reactive spectator modes surrounding and interacting with a single reactive molecule, as schematically illustrated in \Fig{fig:rateratiosaggregation_Nmol1Nnor1-4_dissipative}~a), along with an optimal tree tensor network state decomposition of the extended wavefunction. \Fig{fig:rateratiosaggregation_Nmol1Nnor1-4_dissipative}~b) and c) show the rate modification profiles for the reactive molecule as a function of $\omega_{\rm c}$ for an increasing number of non-reactive dissipative harmonic oscillators $N_{\rm nor}$ ranging from 1 to 4, under two different light-matter coupling strengths. While the Fano lineshape formula is not directly applicable to systems with more than two interfering pathways,\cite{Miroshnichenko_2010_RMP_p2257} characteristic Fano-like features, such as asymmetric rate modification lineshapes and anti-resonances, still appear in the collective regime, implying the participation of interference effects. When $\eta_{\rm c}=0.00125$~a.u., reaction rates are enhanced when the cavity frequency is blue-detuned from $\Delta E$ and suppressed for red-detuned $\omega_{\rm c}$. Increasing $N_{\rm nor}$ amplifies the asymmetry, producing a much more pronounced enhancement peak in the blue-detuned region. With stronger light-matter coupling, $\eta_{\rm c}=0.005$~a.u., the cavity triggers destructive interference, resulting in resonant rate suppression, which is further reinforced by the ensemble of non-reactive vibrational modes bridging the cavity and the reactive molecule.

The second and third network scenarios correspond to cavity-linked molecular networks where the cavity couples to non-reactive modes representing intramolecular vibrations interacting with the inherent reaction coordinate. 

First, we consider dissipation-free non-reactive modes, as illustrated in \Fig{fig:rateratiosaggregation_NmolNnor1-4_dissipationfree}~a), which is the collective extension of the minimal model in \Fig{fig:Nmol1Nnor1_IC_model}. As in the single-molecule case, reaction rates remain enhanced inside the cavity. However, depending on the system parameters, aggregation can produce qualitatively different outcomes. \Fig{fig:rateratiosaggregation_NmolNnor1-4_dissipationfree}~b) and c) show rate modification profiles as a function of $\omega_{\rm c}$ for $N_{\rm mol}=N_{\rm nor} \in \{1,2,3,4\}$ under two different cavity loss strengths $\lambda_{\rm c}$, with fixed $\eta_{\rm c}=0.00125$~a.u., $\eta_{\rm nor}=1\times 10^{-6}$~a.u., and $\omega_{\rm nor}=1185\,\mathrm{cm}^{-1}$. In a weakly lossy cavity ($\lambda_{\rm c}=50\,\mathrm{cm}^{-1}$), collective coupling  of the non-reactive vibrations to the cavity enhances the reaction rates of individual molecules. By contrast, in a strongly lossy cavity ($\lambda_{\rm c}=200\,\mathrm{cm}^{-1}$), the formation of a cavity-mediated molecular network can instead undermine the resonant rate enhancement for an embedded reactive molecule in the network. These findings are consistent with with our previous work,\cite{Ke_J.Chem.Phys._2025_p64702} where reactive molecular modes were directly and collectively coupled to the cavity mode. 

Reaction dynamics becomes extremely complicated in the collective regime when the mediating non-reactive modes are dissipative, as schematically illustrated in \Fig{fig:rateratiosaggregation_NmolNnor1-4_dissipative}~a). An example is presented in \Fig{fig:rateratiosaggregation_NmolNnor1-4_dissipative}~b), displaying reaction rate modification profiles versus $\omega_{\rm c}$ in this scenario for varying $N_{\rm mol}$. Here, the aggregation effect fundamentally alters the rate modification trends. In the single-molecule limit, the reaction rates are suppressed inside the cavity; however, once molecules are correlated via the cavity mode, individual molecules experience rate enhancement. Furthermore, the involvement of additional molecules not only broadens the resonant width of the enhancement peaks but can also induce peak splitting. These observations highlight the emergent many-body interference effects, which can be viewed a collective analog of the Fano interference discussed in the single-molecule case. Each added molecule or non-reactive mode introduces new energy-transfer pathways that coherently superimpose, modifying the lineshape asymmetry or producing multi-peak structures. In this sense, the collective regime realizes effectively a high-dimensional generalization of the Fano phenomenon, where quantum interference among multiple coherent pathways governs the cavity-modified chemical reactivity.

Whether a simple, universal expression can capture cavity-induced rate modification in the collective regime remains an open question to us. Achieving a general and deeper understanding will likely require extensive case-by-case studies. The preliminary results presented here are intended to emphasize and raise awareness of the inherent complexity and extreme sensitivity of cavity-modified reaction dynamics. The kinetic behavior is not merely shaped by the static architecture of a molecular network linked by a cavity.  More importantly, the resulting rate modification emerges from an intricate and dynamical interplay of all available energy transfer steps and pathways built upon the aggregated molecular network. This situation is more akin to analyzing traffic flows in a network of roads with numerous intersections and roundabouts, which is considerably more complex than mapping the roads themselves. Crucially, quantum dynamical cooperative effects, such as interference, may be central to the observed modifications, highlighting the nontrivial collective quantum nature of the reaction dynamics. 

We should also note that the current study does not aim to directly explain the state-of-the-art polaritonic chemistry experiments or resolve any ongoing controversies. It remains unclear whether the key observations reported here, based on the few-molecule models, are directly transferable to the thermodynamic limit, which is relevant for most experiments where millions of molecules are involved and collectively coupled to discrete cavity modes. Nevertheless, recent experimental efforts to achieve vibrational strong coupling with a downsized number of molecules\cite{Chikkaraddy_2016_N_p127} may provide a manipulable platform to test these theoretical predictions and thus offer further insights into the underlying mechanisms and the potential roles of quantum resonant superposition and interference in cavity-modified condensed-phase chemical reactions.

\section{\label{sec:conclusion}Conclusion}
To summarize, we have performed a systematic and numerically exact quantum dynamical investigation of cavity-modified cooperative chemical reaction processes in condensed-phase environments, with a particular focus on scenarios in which IR-active non-reactive vibrational modes mediate energy transfer between the cavity field and optically dark reactive coordinates.

Our results and analyses are organized according to a hierarchical progression in model complexity, allowing us to disentangle, step by step, the key physical factors and underlying mechanisms responsible for cavity-induced rate modifications. Beginning with a minimal model consisting of a single reaction coordinate coupled to a single non-reactive vibrational mode, both outside and inside the cavity, we demonstrate that once the cavity becomes an integral element of an energy-transfer chain, optimal rate enhancement is achieved only when two conditions are simultaneously satisfied. First, the appropriate resonance condition must be met, which requires a full account of mode hybridization in regimes where any pair of subsystems becomes strongly coupled; properly treating such hybridization is essential for accurately describing the resulting reaction dynamics. Second, the characteristic timescales of all participating energy-exchange processes are well balanced. This finding underscores that cavity-induced reactivity cannot be optimized by simply strengthening a single interaction, such as the light–matter coupling, in isolation.

Moreover, when the cavity introduces an additional reaction pathway that coherently interferes with pathways already present outside the cavity, the overall rate modification is governed by quantum interference between cavity-assisted and non-cavity channels. Depending on the relative phases and strengths of these distinct pathways, the reaction rate profile as a function of the cavity frequency $\omega_{\rm c}$ may exhibit resonant enhancement, resonant suppression, or even asymmetric Fano-type line shapes in which enhancement peaks and suppression dips coexist within a narrow resonance window. This interference-based picture provides a unifying framework for understanding the diverse cavity-induced reactivity modification trends observed even in minimal model systems.

Extending the analysis to the collective regime with multiple reactive and non-reactive modes further reveals the pronounced complexity and extreme sensitivity of cavity-modified reaction dynamics. Collective coupling through the cavity effectively binds molecules into an interconnected network, within which small variations in molecular composition, environmental parameters, or system size can qualitatively alter the resulting reaction rate modifications. Our preliminary results imply that cavity-induced chemical reactivity in molecular ensembles is inherently a collective quantum dynamical phenomenon. Looking ahead, further advances will require continued methodological optimization and algorithmic acceleration, with the goal of achieving a deeper and more unified theoretical understanding of these collective quantum effects. Identifying robust design principles for cavity-controlled chemistry therefore remains an important direction for future research.

\begin{acknowledgments}
The author thanks the Swiss National Science Foundation for the award of a research fellowship (Grant No. TMPFP2\_224947). 
\end{acknowledgments}

\section*{Supplementary information}
See the Supplementary Material for additional results presented as contour plots of the rate modification ratio $k_{\rm c}/k_{\rm o}$ as functions of the cavity frequency $\omega_{\rm c}$ and the non-reactive mode frequency $\omega_{\rm nor}$, for various choices of the light–matter coupling strength $\eta_{\rm c}$ and the vibrational coupling strength $\eta_{\rm nor}$.

\section*{Data Availability Statement}
The data and code that support the findings of this work are available from the corresponding author upon reasonable request.

%\bibliography{polaritonfano_2026}% Produces the bibliography via BibTeX.

\begin{thebibliography}{59}%
\makeatletter
\providecommand \@ifxundefined [1]{%
 \@ifx{#1\undefined}
}%
\providecommand \@ifnum [1]{%
 \ifnum #1\expandafter \@firstoftwo
 \else \expandafter \@secondoftwo
 \fi
}%
\providecommand \@ifx [1]{%
 \ifx #1\expandafter \@firstoftwo
 \else \expandafter \@secondoftwo
 \fi
}%
\providecommand \natexlab [1]{#1}%
\providecommand \enquote  [1]{``#1''}%
\providecommand \bibnamefont  [1]{#1}%
\providecommand \bibfnamefont [1]{#1}%
\providecommand \citenamefont [1]{#1}%
\providecommand \href@noop [0]{\@secondoftwo}%
\providecommand \href [0]{\begingroup \@sanitize@url \@href}%
\providecommand \@href[1]{\@@startlink{#1}\@@href}%
\providecommand \@@href[1]{\endgroup#1\@@endlink}%
\providecommand \@sanitize@url [0]{\catcode `\\12\catcode `\$12\catcode
  `\&12\catcode `\#12\catcode `\^12\catcode `\_12\catcode `\%12\relax}%
\providecommand \@@startlink[1]{}%
\providecommand \@@endlink[0]{}%
\providecommand \url  [0]{\begingroup\@sanitize@url \@url }%
\providecommand \@url [1]{\endgroup\@href {#1}{\urlprefix }}%
\providecommand \urlprefix  [0]{URL }%
\providecommand \Eprint [0]{\href }%
\providecommand \doibase [0]{http://dx.doi.org/}%
\providecommand \selectlanguage [0]{\@gobble}%
\providecommand \bibinfo  [0]{\@secondoftwo}%
\providecommand \bibfield  [0]{\@secondoftwo}%
\providecommand \translation [1]{[#1]}%
\providecommand \BibitemOpen [0]{}%
\providecommand \bibitemStop [0]{}%
\providecommand \bibitemNoStop [0]{.\EOS\space}%
\providecommand \EOS [0]{\spacefactor3000\relax}%
\providecommand \BibitemShut  [1]{\csname bibitem#1\endcsname}%
\let\auto@bib@innerbib\@empty
%</preamble>
\bibitem [{\citenamefont {Ebbesen}(2016)}]{Ebbesen_2016_ACR_p2403}%
  \BibitemOpen
  \bibfield  {author} {\bibinfo {author} {\bibfnamefont {T.~W.}\ \bibnamefont
  {Ebbesen}},\ }\bibfield  {title} {\enquote {\bibinfo {title} {Hybrid
  light--matter states in a molecular and material science perspective},}\
  }\href@noop {} {\bibfield  {journal} {\bibinfo  {journal} {Acc. Chem. Res.}\
  }\textbf {\bibinfo {volume} {49}},\ \bibinfo {pages} {2403--2412} (\bibinfo
  {year} {2016})}\BibitemShut {NoStop}%
\bibitem [{\citenamefont {Kasprzak}\ \emph {et~al.}(2006)\citenamefont
  {Kasprzak}, \citenamefont {Richard}, \citenamefont {Kundermann},
  \citenamefont {Baas}, \citenamefont {Jeambrun}, \citenamefont {Keeling},
  \citenamefont {Marchetti}, \citenamefont {Szyma{\'n}ska}, \citenamefont
  {Andr{\'e}}, \citenamefont {Staehli} \emph
  {et~al.}}]{Kasprzak_Nat_2006_p409}%
  \BibitemOpen
  \bibfield  {author} {\bibinfo {author} {\bibfnamefont {J.}~\bibnamefont
  {Kasprzak}}, \bibinfo {author} {\bibfnamefont {M.}~\bibnamefont {Richard}},
  \bibinfo {author} {\bibfnamefont {S.}~\bibnamefont {Kundermann}}, \bibinfo
  {author} {\bibfnamefont {A.}~\bibnamefont {Baas}}, \bibinfo {author}
  {\bibfnamefont {P.}~\bibnamefont {Jeambrun}}, \bibinfo {author}
  {\bibfnamefont {J.~M.~J.}\ \bibnamefont {Keeling}}, \bibinfo {author}
  {\bibfnamefont {F.}~\bibnamefont {Marchetti}}, \bibinfo {author}
  {\bibfnamefont {M.}~\bibnamefont {Szyma{\'n}ska}}, \bibinfo {author}
  {\bibfnamefont {R.}~\bibnamefont {Andr{\'e}}}, \bibinfo {author}
  {\bibfnamefont {J.~a.}\ \bibnamefont {Staehli}},  \emph {et~al.},\ }\bibfield
   {title} {\enquote {\bibinfo {title} {Bose--einstein condensation of exciton
  polaritons},}\ }\href@noop {} {\bibfield  {journal} {\bibinfo  {journal}
  {Nat}\ }\textbf {\bibinfo {volume} {443}},\ \bibinfo {pages} {409--414}
  (\bibinfo {year} {2006})}\BibitemShut {NoStop}%
\bibitem [{\citenamefont {Aberra~Guebrou}\ \emph {et~al.}(2012)\citenamefont
  {Aberra~Guebrou}, \citenamefont {Symonds}, \citenamefont {Homeyer},
  \citenamefont {Plenet}, \citenamefont {Gartstein}, \citenamefont
  {Agranovich},\ and\ \citenamefont
  {Bellessa}}]{AberraGuebrou_2012_PRL_p66401}%
  \BibitemOpen
  \bibfield  {author} {\bibinfo {author} {\bibfnamefont {S.}~\bibnamefont
  {Aberra~Guebrou}}, \bibinfo {author} {\bibfnamefont {C.}~\bibnamefont
  {Symonds}}, \bibinfo {author} {\bibfnamefont {E.}~\bibnamefont {Homeyer}},
  \bibinfo {author} {\bibfnamefont {J.}~\bibnamefont {Plenet}}, \bibinfo
  {author} {\bibfnamefont {Y.~N.}\ \bibnamefont {Gartstein}}, \bibinfo {author}
  {\bibfnamefont {V.~M.}\ \bibnamefont {Agranovich}}, \ and\ \bibinfo {author}
  {\bibfnamefont {J.}~\bibnamefont {Bellessa}},\ }\bibfield  {title} {\enquote
  {\bibinfo {title} {Coherent emission from a disordered organic
  semiconductor<? format?> induced by strong coupling with surface plasmons},}\
  }\href@noop {} {\bibfield  {journal} {\bibinfo  {journal} {Phys. Rev. Lett.}\
  }\textbf {\bibinfo {volume} {108}},\ \bibinfo {pages} {066401} (\bibinfo
  {year} {2012})}\BibitemShut {NoStop}%
\bibitem [{\citenamefont {Coles}\ \emph {et~al.}(2014)\citenamefont {Coles},
  \citenamefont {Somaschi}, \citenamefont {Michetti}, \citenamefont {Clark},
  \citenamefont {Lagoudakis}, \citenamefont {Savvidis},\ and\ \citenamefont
  {Lidzey}}]{Coles_Nat.Mater._2014_p712}%
  \BibitemOpen
  \bibfield  {author} {\bibinfo {author} {\bibfnamefont {D.~M.}\ \bibnamefont
  {Coles}}, \bibinfo {author} {\bibfnamefont {N.}~\bibnamefont {Somaschi}},
  \bibinfo {author} {\bibfnamefont {P.}~\bibnamefont {Michetti}}, \bibinfo
  {author} {\bibfnamefont {C.}~\bibnamefont {Clark}}, \bibinfo {author}
  {\bibfnamefont {P.~G.}\ \bibnamefont {Lagoudakis}}, \bibinfo {author}
  {\bibfnamefont {P.~G.}\ \bibnamefont {Savvidis}}, \ and\ \bibinfo {author}
  {\bibfnamefont {D.~G.}\ \bibnamefont {Lidzey}},\ }\bibfield  {title}
  {\enquote {\bibinfo {title} {Polariton-mediated energy transfer between
  organic dyes in a strongly coupled optical microcavity},}\ }\href@noop {}
  {\bibfield  {journal} {\bibinfo  {journal} {Nat. Mater.}\ }\textbf {\bibinfo
  {volume} {13}},\ \bibinfo {pages} {712--719} (\bibinfo {year}
  {2014})}\BibitemShut {NoStop}%
\bibitem [{\citenamefont {Orgiu}\ \emph {et~al.}(2015)\citenamefont {Orgiu},
  \citenamefont {George}, \citenamefont {Hutchison}, \citenamefont {Devaux},
  \citenamefont {Dayen}, \citenamefont {Doudin}, \citenamefont {Stellacci},
  \citenamefont {Genet}, \citenamefont {Schachenmayer}, \citenamefont {Genes}
  \emph {et~al.}}]{Orgiu_2015_NM_p1123}%
  \BibitemOpen
  \bibfield  {author} {\bibinfo {author} {\bibfnamefont {E.}~\bibnamefont
  {Orgiu}}, \bibinfo {author} {\bibfnamefont {J.}~\bibnamefont {George}},
  \bibinfo {author} {\bibfnamefont {J.}~\bibnamefont {Hutchison}}, \bibinfo
  {author} {\bibfnamefont {E.}~\bibnamefont {Devaux}}, \bibinfo {author}
  {\bibfnamefont {J.}~\bibnamefont {Dayen}}, \bibinfo {author} {\bibfnamefont
  {B.}~\bibnamefont {Doudin}}, \bibinfo {author} {\bibfnamefont
  {F.}~\bibnamefont {Stellacci}}, \bibinfo {author} {\bibfnamefont
  {C.}~\bibnamefont {Genet}}, \bibinfo {author} {\bibfnamefont
  {J.}~\bibnamefont {Schachenmayer}}, \bibinfo {author} {\bibfnamefont
  {C.}~\bibnamefont {Genes}},  \emph {et~al.},\ }\bibfield  {title} {\enquote
  {\bibinfo {title} {Conductivity in organic semiconductors hybridized with the
  vacuum field},}\ }\href@noop {} {\bibfield  {journal} {\bibinfo  {journal}
  {Nat. Mater.}\ }\textbf {\bibinfo {volume} {14}},\ \bibinfo {pages}
  {1123--1129} (\bibinfo {year} {2015})}\BibitemShut {NoStop}%
\bibitem [{\citenamefont {Hagenm{\"u}ller}\ \emph {et~al.}(2017)\citenamefont
  {Hagenm{\"u}ller}, \citenamefont {Schachenmayer}, \citenamefont {Sch{\"u}tz},
  \citenamefont {Genes},\ and\ \citenamefont
  {Pupillo}}]{Hagenmueller_2017_PRL_p223601}%
  \BibitemOpen
  \bibfield  {author} {\bibinfo {author} {\bibfnamefont {D.}~\bibnamefont
  {Hagenm{\"u}ller}}, \bibinfo {author} {\bibfnamefont {J.}~\bibnamefont
  {Schachenmayer}}, \bibinfo {author} {\bibfnamefont {S.}~\bibnamefont
  {Sch{\"u}tz}}, \bibinfo {author} {\bibfnamefont {C.}~\bibnamefont {Genes}}, \
  and\ \bibinfo {author} {\bibfnamefont {G.}~\bibnamefont {Pupillo}},\
  }\bibfield  {title} {\enquote {\bibinfo {title} {Cavity-enhanced transport of
  charge},}\ }\href@noop {} {\bibfield  {journal} {\bibinfo  {journal} {Phys.
  Rev. Lett.}\ }\textbf {\bibinfo {volume} {119}},\ \bibinfo {pages} {223601}
  (\bibinfo {year} {2017})}\BibitemShut {NoStop}%
\bibitem [{\citenamefont {Hsu}, \citenamefont {Ding},\ and\ \citenamefont
  {Schatz}(2017)}]{Hsu_2017_JPCL_p2357}%
  \BibitemOpen
  \bibfield  {author} {\bibinfo {author} {\bibfnamefont {L.-Y.}\ \bibnamefont
  {Hsu}}, \bibinfo {author} {\bibfnamefont {W.}~\bibnamefont {Ding}}, \ and\
  \bibinfo {author} {\bibfnamefont {G.~C.}\ \bibnamefont {Schatz}},\ }\bibfield
   {title} {\enquote {\bibinfo {title} {Plasmon-coupled resonance energy
  transfer},}\ }\href@noop {} {\bibfield  {journal} {\bibinfo  {journal} {J.
  Phys. Chem. Lett.}\ }\textbf {\bibinfo {volume} {8}},\ \bibinfo {pages}
  {2357--2367} (\bibinfo {year} {2017})}\BibitemShut {NoStop}%
\bibitem [{\citenamefont {Baieva}\ \emph {et~al.}(2017)\citenamefont {Baieva},
  \citenamefont {Hakamaa}, \citenamefont {Groenhof}, \citenamefont {Heikkila},\
  and\ \citenamefont {Toppari}}]{Baieva_2017_AP_p28}%
  \BibitemOpen
  \bibfield  {author} {\bibinfo {author} {\bibfnamefont {S.}~\bibnamefont
  {Baieva}}, \bibinfo {author} {\bibfnamefont {O.}~\bibnamefont {Hakamaa}},
  \bibinfo {author} {\bibfnamefont {G.}~\bibnamefont {Groenhof}}, \bibinfo
  {author} {\bibfnamefont {T.~T.}\ \bibnamefont {Heikkila}}, \ and\ \bibinfo
  {author} {\bibfnamefont {J.~J.}\ \bibnamefont {Toppari}},\ }\bibfield
  {title} {\enquote {\bibinfo {title} {Dynamics of strongly coupled modes
  between surface plasmon polaritons and photoactive molecules: the effect of
  the stokes shift},}\ }\href@noop {} {\bibfield  {journal} {\bibinfo
  {journal} {ACS Photonics}\ }\textbf {\bibinfo {volume} {4}},\ \bibinfo
  {pages} {28--37} (\bibinfo {year} {2017})}\BibitemShut {NoStop}%
\bibitem [{\citenamefont {Mony}\ \emph {et~al.}(2021)\citenamefont {Mony},
  \citenamefont {Climent}, \citenamefont {Petersen}, \citenamefont
  {Moth-Poulsen}, \citenamefont {Feist},\ and\ \citenamefont
  {B{\"o}rjesson}}]{Mony_2021_AFM_p2010737}%
  \BibitemOpen
  \bibfield  {author} {\bibinfo {author} {\bibfnamefont {J.}~\bibnamefont
  {Mony}}, \bibinfo {author} {\bibfnamefont {C.}~\bibnamefont {Climent}},
  \bibinfo {author} {\bibfnamefont {A.~U.}\ \bibnamefont {Petersen}}, \bibinfo
  {author} {\bibfnamefont {K.}~\bibnamefont {Moth-Poulsen}}, \bibinfo {author}
  {\bibfnamefont {J.}~\bibnamefont {Feist}}, \ and\ \bibinfo {author}
  {\bibfnamefont {K.}~\bibnamefont {B{\"o}rjesson}},\ }\bibfield  {title}
  {\enquote {\bibinfo {title} {Photoisomerization efficiency of a solar thermal
  fuel in the strong coupling regime},}\ }\href@noop {} {\bibfield  {journal}
  {\bibinfo  {journal} {Adv. Funct. Mater.}\ }\textbf {\bibinfo {volume}
  {31}},\ \bibinfo {pages} {2010737} (\bibinfo {year} {2021})}\BibitemShut
  {NoStop}%
\bibitem [{\citenamefont {Hayashi}, \citenamefont {Fukushima},\ and\
  \citenamefont {Murakoshi}(2024)}]{Hayashi_2024_JCP_p181101}%
  \BibitemOpen
  \bibfield  {author} {\bibinfo {author} {\bibfnamefont {T.}~\bibnamefont
  {Hayashi}}, \bibinfo {author} {\bibfnamefont {T.}~\bibnamefont {Fukushima}},
  \ and\ \bibinfo {author} {\bibfnamefont {K.}~\bibnamefont {Murakoshi}},\
  }\bibfield  {title} {\enquote {\bibinfo {title} {Role of cavity strong
  coupling on single electron transfer reaction rate at electrode--electrolyte
  interface},}\ }\href@noop {} {\bibfield  {journal} {\bibinfo  {journal} {J.
  Chem. Phys.}\ }\textbf {\bibinfo {volume} {161}},\ \bibinfo {pages} {181101}
  (\bibinfo {year} {2024})}\BibitemShut {NoStop}%
\bibitem [{\citenamefont {Sandik}\ \emph {et~al.}(2025)\citenamefont {Sandik},
  \citenamefont {Feist}, \citenamefont {Garc{\'\i}a-Vidal},\ and\ \citenamefont
  {Schwartz}}]{Sandik_2025_NM_p344}%
  \BibitemOpen
  \bibfield  {author} {\bibinfo {author} {\bibfnamefont {G.}~\bibnamefont
  {Sandik}}, \bibinfo {author} {\bibfnamefont {J.}~\bibnamefont {Feist}},
  \bibinfo {author} {\bibfnamefont {F.~J.}\ \bibnamefont {Garc{\'\i}a-Vidal}},
  \ and\ \bibinfo {author} {\bibfnamefont {T.}~\bibnamefont {Schwartz}},\
  }\bibfield  {title} {\enquote {\bibinfo {title} {Cavity-enhanced energy
  transport in molecular systems},}\ }\href@noop {} {\bibfield  {journal}
  {\bibinfo  {journal} {Nat. Mater.}\ }\textbf {\bibinfo {volume} {24}},\
  \bibinfo {pages} {344--355} (\bibinfo {year} {2025})}\BibitemShut {NoStop}%
\bibitem [{\citenamefont {Thomas}\ \emph {et~al.}(2016)\citenamefont {Thomas},
  \citenamefont {George}, \citenamefont {Shalabney}, \citenamefont {Dryzhakov},
  \citenamefont {Varma}, \citenamefont {Moran}, \citenamefont {Chervy},
  \citenamefont {Zhong}, \citenamefont {Devaux}, \citenamefont {Genet} \emph
  {et~al.}}]{Thomas_2016_ACE_p11634}%
  \BibitemOpen
  \bibfield  {author} {\bibinfo {author} {\bibfnamefont {A.}~\bibnamefont
  {Thomas}}, \bibinfo {author} {\bibfnamefont {J.}~\bibnamefont {George}},
  \bibinfo {author} {\bibfnamefont {A.}~\bibnamefont {Shalabney}}, \bibinfo
  {author} {\bibfnamefont {M.}~\bibnamefont {Dryzhakov}}, \bibinfo {author}
  {\bibfnamefont {S.~J.}\ \bibnamefont {Varma}}, \bibinfo {author}
  {\bibfnamefont {J.}~\bibnamefont {Moran}}, \bibinfo {author} {\bibfnamefont
  {T.}~\bibnamefont {Chervy}}, \bibinfo {author} {\bibfnamefont
  {X.}~\bibnamefont {Zhong}}, \bibinfo {author} {\bibfnamefont
  {E.}~\bibnamefont {Devaux}}, \bibinfo {author} {\bibfnamefont
  {C.}~\bibnamefont {Genet}},  \emph {et~al.},\ }\bibfield  {title} {\enquote
  {\bibinfo {title} {Ground-state chemical reactivity under vibrational
  coupling to the vacuum electromagnetic field},}\ }\href@noop {} {\bibfield
  {journal} {\bibinfo  {journal} {Angew. Chem-ger. Edit.}\ }\textbf {\bibinfo
  {volume} {128}},\ \bibinfo {pages} {11634--11638} (\bibinfo {year}
  {2016})}\BibitemShut {NoStop}%
\bibitem [{\citenamefont {Vergauwe}\ \emph {et~al.}(2019)\citenamefont
  {Vergauwe}, \citenamefont {Thomas}, \citenamefont {Nagarajan}, \citenamefont
  {Shalabney}, \citenamefont {George}, \citenamefont {Chervy}, \citenamefont
  {Seidel}, \citenamefont {Devaux}, \citenamefont {Torbeev},\ and\
  \citenamefont {Ebbesen}}]{Vergauwe_2019_ACIE_p15324}%
  \BibitemOpen
  \bibfield  {author} {\bibinfo {author} {\bibfnamefont {R.~M.}\ \bibnamefont
  {Vergauwe}}, \bibinfo {author} {\bibfnamefont {A.}~\bibnamefont {Thomas}},
  \bibinfo {author} {\bibfnamefont {K.}~\bibnamefont {Nagarajan}}, \bibinfo
  {author} {\bibfnamefont {A.}~\bibnamefont {Shalabney}}, \bibinfo {author}
  {\bibfnamefont {J.}~\bibnamefont {George}}, \bibinfo {author} {\bibfnamefont
  {T.}~\bibnamefont {Chervy}}, \bibinfo {author} {\bibfnamefont
  {M.}~\bibnamefont {Seidel}}, \bibinfo {author} {\bibfnamefont
  {E.}~\bibnamefont {Devaux}}, \bibinfo {author} {\bibfnamefont
  {V.}~\bibnamefont {Torbeev}}, \ and\ \bibinfo {author} {\bibfnamefont
  {T.~W.}\ \bibnamefont {Ebbesen}},\ }\bibfield  {title} {\enquote {\bibinfo
  {title} {Modification of enzyme activity by vibrational strong coupling of
  water},}\ }\href@noop {} {\bibfield  {journal} {\bibinfo  {journal} {Angew.
  Chem. - Int. Ed.}\ }\textbf {\bibinfo {volume} {58}},\ \bibinfo {pages}
  {15324--15328} (\bibinfo {year} {2019})}\BibitemShut {NoStop}%
\bibitem [{\citenamefont {Thomas}\ \emph {et~al.}(2019)\citenamefont {Thomas},
  \citenamefont {Lethuillier-Karl}, \citenamefont {Nagarajan}, \citenamefont
  {Vergauwe}, \citenamefont {George}, \citenamefont {Chervy}, \citenamefont
  {Shalabney}, \citenamefont {Devaux}, \citenamefont {Genet}, \citenamefont
  {Moran} \emph {et~al.}}]{Thomas_2019_S_p615}%
  \BibitemOpen
  \bibfield  {author} {\bibinfo {author} {\bibfnamefont {A.}~\bibnamefont
  {Thomas}}, \bibinfo {author} {\bibfnamefont {L.}~\bibnamefont
  {Lethuillier-Karl}}, \bibinfo {author} {\bibfnamefont {K.}~\bibnamefont
  {Nagarajan}}, \bibinfo {author} {\bibfnamefont {R.~M.}\ \bibnamefont
  {Vergauwe}}, \bibinfo {author} {\bibfnamefont {J.}~\bibnamefont {George}},
  \bibinfo {author} {\bibfnamefont {T.}~\bibnamefont {Chervy}}, \bibinfo
  {author} {\bibfnamefont {A.}~\bibnamefont {Shalabney}}, \bibinfo {author}
  {\bibfnamefont {E.}~\bibnamefont {Devaux}}, \bibinfo {author} {\bibfnamefont
  {C.}~\bibnamefont {Genet}}, \bibinfo {author} {\bibfnamefont
  {J.}~\bibnamefont {Moran}},  \emph {et~al.},\ }\bibfield  {title} {\enquote
  {\bibinfo {title} {Tilting a ground-state reactivity landscape by vibrational
  strong coupling},}\ }\href@noop {} {\bibfield  {journal} {\bibinfo  {journal}
  {Science}\ }\textbf {\bibinfo {volume} {363}},\ \bibinfo {pages} {615--619}
  (\bibinfo {year} {2019})}\BibitemShut {NoStop}%
\bibitem [{\citenamefont {Hirai}\ \emph {et~al.}(2020)\citenamefont {Hirai},
  \citenamefont {Takeda}, \citenamefont {Hutchison},\ and\ \citenamefont
  {Uji-i}}]{Hirai_2020_ACE_p5370}%
  \BibitemOpen
  \bibfield  {author} {\bibinfo {author} {\bibfnamefont {K.}~\bibnamefont
  {Hirai}}, \bibinfo {author} {\bibfnamefont {R.}~\bibnamefont {Takeda}},
  \bibinfo {author} {\bibfnamefont {J.~A.}\ \bibnamefont {Hutchison}}, \ and\
  \bibinfo {author} {\bibfnamefont {H.}~\bibnamefont {Uji-i}},\ }\bibfield
  {title} {\enquote {\bibinfo {title} {Modulation of {P}rins cyclization by
  vibrational strong coupling},}\ }\href@noop {} {\bibfield  {journal}
  {\bibinfo  {journal} {Angew. Chem-ger. Edit.}\ }\textbf {\bibinfo {volume}
  {132}},\ \bibinfo {pages} {5370--5373} (\bibinfo {year} {2020})}\BibitemShut
  {NoStop}%
\bibitem [{\citenamefont {Pang}\ \emph {et~al.}(2020)\citenamefont {Pang},
  \citenamefont {Thomas}, \citenamefont {Nagarajan}, \citenamefont {Vergauwe},
  \citenamefont {Joseph}, \citenamefont {Patrahau}, \citenamefont {Wang},
  \citenamefont {Genet},\ and\ \citenamefont
  {Ebbesen}}]{Pang_2020_ACIE_p10436}%
  \BibitemOpen
  \bibfield  {author} {\bibinfo {author} {\bibfnamefont {Y.}~\bibnamefont
  {Pang}}, \bibinfo {author} {\bibfnamefont {A.}~\bibnamefont {Thomas}},
  \bibinfo {author} {\bibfnamefont {K.}~\bibnamefont {Nagarajan}}, \bibinfo
  {author} {\bibfnamefont {R.~M.}\ \bibnamefont {Vergauwe}}, \bibinfo {author}
  {\bibfnamefont {K.}~\bibnamefont {Joseph}}, \bibinfo {author} {\bibfnamefont
  {B.}~\bibnamefont {Patrahau}}, \bibinfo {author} {\bibfnamefont
  {K.}~\bibnamefont {Wang}}, \bibinfo {author} {\bibfnamefont {C.}~\bibnamefont
  {Genet}}, \ and\ \bibinfo {author} {\bibfnamefont {T.~W.}\ \bibnamefont
  {Ebbesen}},\ }\bibfield  {title} {\enquote {\bibinfo {title} {On the role of
  symmetry in vibrational strong coupling: the case of charge-transfer
  complexation},}\ }\href@noop {} {\bibfield  {journal} {\bibinfo  {journal}
  {Angew. Chem. - Int. Ed.}\ }\textbf {\bibinfo {volume} {59}},\ \bibinfo
  {pages} {10436--10440} (\bibinfo {year} {2020})}\BibitemShut {NoStop}%
\bibitem [{\citenamefont {Sau}\ \emph {et~al.}(2021)\citenamefont {Sau},
  \citenamefont {Nagarajan}, \citenamefont {Patrahau}, \citenamefont
  {Lethuillier-Karl}, \citenamefont {Vergauwe}, \citenamefont {Thomas},
  \citenamefont {Moran}, \citenamefont {Genet},\ and\ \citenamefont
  {Ebbesen}}]{Sau_2021_ACIE_p5712}%
  \BibitemOpen
  \bibfield  {author} {\bibinfo {author} {\bibfnamefont {A.}~\bibnamefont
  {Sau}}, \bibinfo {author} {\bibfnamefont {K.}~\bibnamefont {Nagarajan}},
  \bibinfo {author} {\bibfnamefont {B.}~\bibnamefont {Patrahau}}, \bibinfo
  {author} {\bibfnamefont {L.}~\bibnamefont {Lethuillier-Karl}}, \bibinfo
  {author} {\bibfnamefont {R.~M.}\ \bibnamefont {Vergauwe}}, \bibinfo {author}
  {\bibfnamefont {A.}~\bibnamefont {Thomas}}, \bibinfo {author} {\bibfnamefont
  {J.}~\bibnamefont {Moran}}, \bibinfo {author} {\bibfnamefont
  {C.}~\bibnamefont {Genet}}, \ and\ \bibinfo {author} {\bibfnamefont {T.~W.}\
  \bibnamefont {Ebbesen}},\ }\bibfield  {title} {\enquote {\bibinfo {title}
  {Modifying {W}oodward--{H}offmann stereoselectivity under vibrational strong
  coupling},}\ }\href@noop {} {\bibfield  {journal} {\bibinfo  {journal}
  {Angew. Chem. - Int. Ed.}\ }\textbf {\bibinfo {volume} {60}},\ \bibinfo
  {pages} {5712--5717} (\bibinfo {year} {2021})}\BibitemShut {NoStop}%
\bibitem [{\citenamefont {Ahn}\ \emph {et~al.}(2023)\citenamefont {Ahn},
  \citenamefont {Triana}, \citenamefont {Recabal}, \citenamefont {Herrera},\
  and\ \citenamefont {Simpkins}}]{Ahn_Sci_2023_p1165}%
  \BibitemOpen
  \bibfield  {author} {\bibinfo {author} {\bibfnamefont {W.}~\bibnamefont
  {Ahn}}, \bibinfo {author} {\bibfnamefont {J.~F.}\ \bibnamefont {Triana}},
  \bibinfo {author} {\bibfnamefont {F.}~\bibnamefont {Recabal}}, \bibinfo
  {author} {\bibfnamefont {F.}~\bibnamefont {Herrera}}, \ and\ \bibinfo
  {author} {\bibfnamefont {B.~S.}\ \bibnamefont {Simpkins}},\ }\bibfield
  {title} {\enquote {\bibinfo {title} {Modification of ground-state chemical
  reactivity via light--matter coherence in infrared cavities},}\ }\href@noop
  {} {\bibfield  {journal} {\bibinfo  {journal} {Science}\ }\textbf {\bibinfo
  {volume} {380}},\ \bibinfo {pages} {1165--1168} (\bibinfo {year}
  {2023})}\BibitemShut {NoStop}%
\bibitem [{\citenamefont {Patrahau}\ \emph {et~al.}(2024)\citenamefont
  {Patrahau}, \citenamefont {Piejko}, \citenamefont {Mayer}, \citenamefont
  {Antheaume}, \citenamefont {Sangchai}, \citenamefont {Ragazzon},
  \citenamefont {Jayachandran}, \citenamefont {Devaux}, \citenamefont {Genet},
  \citenamefont {Moran} \emph
  {et~al.}}]{Patrahau_Angew.Chem.Int.Ed._2024_p202401368}%
  \BibitemOpen
  \bibfield  {author} {\bibinfo {author} {\bibfnamefont {B.}~\bibnamefont
  {Patrahau}}, \bibinfo {author} {\bibfnamefont {M.}~\bibnamefont {Piejko}},
  \bibinfo {author} {\bibfnamefont {R.~J.}\ \bibnamefont {Mayer}}, \bibinfo
  {author} {\bibfnamefont {C.}~\bibnamefont {Antheaume}}, \bibinfo {author}
  {\bibfnamefont {T.}~\bibnamefont {Sangchai}}, \bibinfo {author}
  {\bibfnamefont {G.}~\bibnamefont {Ragazzon}}, \bibinfo {author}
  {\bibfnamefont {A.}~\bibnamefont {Jayachandran}}, \bibinfo {author}
  {\bibfnamefont {E.}~\bibnamefont {Devaux}}, \bibinfo {author} {\bibfnamefont
  {C.}~\bibnamefont {Genet}}, \bibinfo {author} {\bibfnamefont
  {J.}~\bibnamefont {Moran}},  \emph {et~al.},\ }\bibfield  {title} {\enquote
  {\bibinfo {title} {Direct observation of polaritonic chemistry by nuclear
  magnetic resonance spectroscopy},}\ }\href@noop {} {\bibfield  {journal}
  {\bibinfo  {journal} {Angew. Chem. - Int. Ed.}\ }\textbf {\bibinfo {volume}
  {63}},\ \bibinfo {pages} {e202401368} (\bibinfo {year} {2024})}\BibitemShut
  {NoStop}%
\bibitem [{\citenamefont {Verdelli}\ \emph {et~al.}(2024)\citenamefont
  {Verdelli}, \citenamefont {Wei}, \citenamefont {Joseph}, \citenamefont
  {Abdelkhalik}, \citenamefont {Goudarzi}, \citenamefont {Askes}, \citenamefont
  {Baldi}, \citenamefont {Meijer},\ and\ \citenamefont
  {Gomez~Rivas}}]{Verdelli_2024_ACIE_p202409528}%
  \BibitemOpen
  \bibfield  {author} {\bibinfo {author} {\bibfnamefont {F.}~\bibnamefont
  {Verdelli}}, \bibinfo {author} {\bibfnamefont {Y.-C.}\ \bibnamefont {Wei}},
  \bibinfo {author} {\bibfnamefont {K.}~\bibnamefont {Joseph}}, \bibinfo
  {author} {\bibfnamefont {M.~S.}\ \bibnamefont {Abdelkhalik}}, \bibinfo
  {author} {\bibfnamefont {M.}~\bibnamefont {Goudarzi}}, \bibinfo {author}
  {\bibfnamefont {S.~H.}\ \bibnamefont {Askes}}, \bibinfo {author}
  {\bibfnamefont {A.}~\bibnamefont {Baldi}}, \bibinfo {author} {\bibfnamefont
  {E.}~\bibnamefont {Meijer}}, \ and\ \bibinfo {author} {\bibfnamefont
  {J.}~\bibnamefont {Gomez~Rivas}},\ }\bibfield  {title} {\enquote {\bibinfo
  {title} {Polaritonic chemistry enabled by non-local metasurfaces},}\
  }\href@noop {} {\bibfield  {journal} {\bibinfo  {journal} {Angew. Chem. Int.
  Ed.}\ }\textbf {\bibinfo {volume} {63}},\ \bibinfo {pages} {e202409528}
  (\bibinfo {year} {2024})}\BibitemShut {NoStop}%
\bibitem [{\citenamefont {Mahato}\ \emph {et~al.}(2025)\citenamefont {Mahato},
  \citenamefont {Mony}, \citenamefont {Baliyan}, \citenamefont {Biswas},\ and\
  \citenamefont {Thomas}}]{Mahato_2025_ACIE_p202424247}%
  \BibitemOpen
  \bibfield  {author} {\bibinfo {author} {\bibfnamefont {M.~K.}\ \bibnamefont
  {Mahato}}, \bibinfo {author} {\bibfnamefont {K.~S.}\ \bibnamefont {Mony}},
  \bibinfo {author} {\bibfnamefont {H.}~\bibnamefont {Baliyan}}, \bibinfo
  {author} {\bibfnamefont {S.}~\bibnamefont {Biswas}}, \ and\ \bibinfo {author}
  {\bibfnamefont {A.}~\bibnamefont {Thomas}},\ }\bibfield  {title} {\enquote
  {\bibinfo {title} {Exploring excited state proton transfer in thin films
  under vibrational strong coupling},}\ }\href@noop {} {\bibfield  {journal}
  {\bibinfo  {journal} {Angew. Chem. Int. Ed.}\ }\textbf {\bibinfo {volume}
  {137}},\ \bibinfo {pages} {e202424247} (\bibinfo {year} {2025})}\BibitemShut
  {NoStop}%
\bibitem [{\citenamefont {Lather}\ \emph {et~al.}(2019)\citenamefont {Lather},
  \citenamefont {Bhatt}, \citenamefont {Thomas}, \citenamefont {Ebbesen},\ and\
  \citenamefont {George}}]{Lather_2019_ACIE_p10635}%
  \BibitemOpen
  \bibfield  {author} {\bibinfo {author} {\bibfnamefont {J.}~\bibnamefont
  {Lather}}, \bibinfo {author} {\bibfnamefont {P.}~\bibnamefont {Bhatt}},
  \bibinfo {author} {\bibfnamefont {A.}~\bibnamefont {Thomas}}, \bibinfo
  {author} {\bibfnamefont {T.~W.}\ \bibnamefont {Ebbesen}}, \ and\ \bibinfo
  {author} {\bibfnamefont {J.}~\bibnamefont {George}},\ }\bibfield  {title}
  {\enquote {\bibinfo {title} {Cavity catalysis by cooperative vibrational
  strong coupling of reactant and solvent molecules},}\ }\href@noop {}
  {\bibfield  {journal} {\bibinfo  {journal} {Angew. Chem. - Int. Ed.}\
  }\textbf {\bibinfo {volume} {58}},\ \bibinfo {pages} {10635--10638} (\bibinfo
  {year} {2019})}\BibitemShut {NoStop}%
\bibitem [{\citenamefont {Lian}\ \emph {et~al.}(2025)\citenamefont {Lian},
  \citenamefont {Song}, \citenamefont {Si}, \citenamefont {Zhao}, \citenamefont
  {Chen}, \citenamefont {Xu},\ and\ \citenamefont
  {Zhang}}]{Lian_2025_AP_p3557}%
  \BibitemOpen
  \bibfield  {author} {\bibinfo {author} {\bibfnamefont {J.}~\bibnamefont
  {Lian}}, \bibinfo {author} {\bibfnamefont {Y.}~\bibnamefont {Song}}, \bibinfo
  {author} {\bibfnamefont {Q.}~\bibnamefont {Si}}, \bibinfo {author}
  {\bibfnamefont {X.}~\bibnamefont {Zhao}}, \bibinfo {author} {\bibfnamefont
  {L.}~\bibnamefont {Chen}}, \bibinfo {author} {\bibfnamefont {R.}~\bibnamefont
  {Xu}}, \ and\ \bibinfo {author} {\bibfnamefont {F.}~\bibnamefont {Zhang}},\
  }\bibfield  {title} {\enquote {\bibinfo {title} {Continuous-flow fabry-pérot
  cavity for enhanced catalysis via cooperative vibrational strong coupling},}\
  }\href@noop {} {\bibfield  {journal} {\bibinfo  {journal} {ACS Photonics}\
  }\textbf {\bibinfo {volume} {12}},\ \bibinfo {pages} {3557--3564} (\bibinfo
  {year} {2025})}\BibitemShut {NoStop}%
\bibitem [{\citenamefont {Wiesehan}\ and\ \citenamefont
  {Xiong}(2021)}]{Wiesehan_2021_JCP_p241103}%
  \BibitemOpen
  \bibfield  {author} {\bibinfo {author} {\bibfnamefont {G.~D.}\ \bibnamefont
  {Wiesehan}}\ and\ \bibinfo {author} {\bibfnamefont {W.}~\bibnamefont
  {Xiong}},\ }\bibfield  {title} {\enquote {\bibinfo {title} {Negligible rate
  enhancement from reported cooperative vibrational strong coupling
  catalysis},}\ }\href@noop {} {\bibfield  {journal} {\bibinfo  {journal} {J.
  Chem. Phys.}\ }\textbf {\bibinfo {volume} {155}},\ \bibinfo {pages} {241103}
  (\bibinfo {year} {2021})}\BibitemShut {NoStop}%
\bibitem [{\citenamefont {Hiura}, \citenamefont {Shalabney},\ and\
  \citenamefont {George}(2021)}]{Hiura_2019__p}%
  \BibitemOpen
  \bibfield  {author} {\bibinfo {author} {\bibfnamefont {H.}~\bibnamefont
  {Hiura}}, \bibinfo {author} {\bibfnamefont {A.}~\bibnamefont {Shalabney}}, \
  and\ \bibinfo {author} {\bibfnamefont {J.}~\bibnamefont {George}},\
  }\bibfield  {title} {\enquote {\bibinfo {title} {Vacuum-field catalysis:
  Accelerated reactions by vibrational ultra strong coupling},}\ }\href@noop {}
  {\bibfield  {journal} {\bibinfo  {journal} {chemRxiv:7234721.v5}\ } (\bibinfo
  {year} {2021})}\BibitemShut {NoStop}%
\bibitem [{\citenamefont {Imperatore}, \citenamefont {Asbury},\ and\
  \citenamefont {Giebink}(2021)}]{Imperatore_2021_JCP_p191103}%
  \BibitemOpen
  \bibfield  {author} {\bibinfo {author} {\bibfnamefont {M.~V.}\ \bibnamefont
  {Imperatore}}, \bibinfo {author} {\bibfnamefont {J.~B.}\ \bibnamefont
  {Asbury}}, \ and\ \bibinfo {author} {\bibfnamefont {N.~C.}\ \bibnamefont
  {Giebink}},\ }\bibfield  {title} {\enquote {\bibinfo {title} {Reproducibility
  of cavity-enhanced chemical reaction rates in the vibrational strong coupling
  regime},}\ }\href@noop {} {\bibfield  {journal} {\bibinfo  {journal} {J.
  Chem. Phys.}\ }\textbf {\bibinfo {volume} {154}},\ \bibinfo {pages} {191103}
  (\bibinfo {year} {2021})}\BibitemShut {NoStop}%
\bibitem [{\citenamefont {Wang}\ \emph {et~al.}(2024)\citenamefont {Wang},
  \citenamefont {Rosenmann}, \citenamefont {Muntean},\ and\ \citenamefont
  {Ma}}]{Wang_2024_Ap_p263}%
  \BibitemOpen
  \bibfield  {author} {\bibinfo {author} {\bibfnamefont {Y.}~\bibnamefont
  {Wang}}, \bibinfo {author} {\bibfnamefont {D.}~\bibnamefont {Rosenmann}},
  \bibinfo {author} {\bibfnamefont {J.~V.}\ \bibnamefont {Muntean}}, \ and\
  \bibinfo {author} {\bibfnamefont {X.}~\bibnamefont {Ma}},\ }\bibfield
  {title} {\enquote {\bibinfo {title} {Polaritonic bright and dark states
  collectively affect the reactivity of a hydrolysis reaction},}\ }\href@noop
  {} {\bibfield  {journal} {\bibinfo  {journal} {ACS photonics}\ }\textbf
  {\bibinfo {volume} {12}},\ \bibinfo {pages} {263--270} (\bibinfo {year}
  {2024})}\BibitemShut {NoStop}%
\bibitem [{\citenamefont {Lindoy}, \citenamefont {Mandal},\ and\ \citenamefont
  {Reichman}(2023)}]{Lindoy_2023_NC_p2733}%
  \BibitemOpen
  \bibfield  {author} {\bibinfo {author} {\bibfnamefont {L.~P.}\ \bibnamefont
  {Lindoy}}, \bibinfo {author} {\bibfnamefont {A.}~\bibnamefont {Mandal}}, \
  and\ \bibinfo {author} {\bibfnamefont {D.~R.}\ \bibnamefont {Reichman}},\
  }\bibfield  {title} {\enquote {\bibinfo {title} {Quantum dynamical effects of
  vibrational strong coupling in chemical reactivity},}\ }\href@noop {}
  {\bibfield  {journal} {\bibinfo  {journal} {Nat. Commun.}\ }\textbf {\bibinfo
  {volume} {14}},\ \bibinfo {pages} {2733} (\bibinfo {year}
  {2023})}\BibitemShut {NoStop}%
\bibitem [{\citenamefont {Vega}, \citenamefont {Ying},\ and\ \citenamefont
  {Huo}(2025)}]{Vega_2025_JACS_p19727}%
  \BibitemOpen
  \bibfield  {author} {\bibinfo {author} {\bibfnamefont {S.~M.}\ \bibnamefont
  {Vega}}, \bibinfo {author} {\bibfnamefont {W.}~\bibnamefont {Ying}}, \ and\
  \bibinfo {author} {\bibfnamefont {P.}~\bibnamefont {Huo}},\ }\bibfield
  {title} {\enquote {\bibinfo {title} {Theoretical insights into the resonant
  suppression effect in vibrational polariton chemistry},}\ }\href@noop {}
  {\bibfield  {journal} {\bibinfo  {journal} {J. Am. Chem. Soc.}\ }\textbf
  {\bibinfo {volume} {147}},\ \bibinfo {pages} {19727--19737} (\bibinfo {year}
  {2025})}\BibitemShut {NoStop}%
\bibitem [{\citenamefont {Ke}(2023)}]{Ke_2023_JCP_p211102}%
  \BibitemOpen
  \bibfield  {author} {\bibinfo {author} {\bibfnamefont {Y.}~\bibnamefont
  {Ke}},\ }\bibfield  {title} {\enquote {\bibinfo {title} {Tree tensor network
  state approach for solving hierarchical equations of motion},}\ }\href@noop
  {} {\bibfield  {journal} {\bibinfo  {journal} {J. Chem. Phys.}\ }\textbf
  {\bibinfo {volume} {158}},\ \bibinfo {pages} {211102} (\bibinfo {year}
  {2023})}\BibitemShut {NoStop}%
\bibitem [{\citenamefont {Flick}\ \emph {et~al.}(2017)\citenamefont {Flick},
  \citenamefont {Ruggenthaler}, \citenamefont {Appel},\ and\ \citenamefont
  {Rubio}}]{Flick_2017_PotNAoS_p3026}%
  \BibitemOpen
  \bibfield  {author} {\bibinfo {author} {\bibfnamefont {J.}~\bibnamefont
  {Flick}}, \bibinfo {author} {\bibfnamefont {M.}~\bibnamefont {Ruggenthaler}},
  \bibinfo {author} {\bibfnamefont {H.}~\bibnamefont {Appel}}, \ and\ \bibinfo
  {author} {\bibfnamefont {A.}~\bibnamefont {Rubio}},\ }\bibfield  {title}
  {\enquote {\bibinfo {title} {Atoms and molecules in cavities, from weak to
  strong coupling in quantum-electrodynamics ({QED}) chemistry},}\ }\href@noop
  {} {\bibfield  {journal} {\bibinfo  {journal} {Proc. Natl. Acad. Sci. USA}\
  }\textbf {\bibinfo {volume} {114}},\ \bibinfo {pages} {3026--3034} (\bibinfo
  {year} {2017})}\BibitemShut {NoStop}%
\bibitem [{\citenamefont {Rokaj}\ \emph {et~al.}(2018)\citenamefont {Rokaj},
  \citenamefont {Welakuh}, \citenamefont {Ruggenthaler},\ and\ \citenamefont
  {Rubio}}]{Rokaj_2018_JPBAMOP_p34005}%
  \BibitemOpen
  \bibfield  {author} {\bibinfo {author} {\bibfnamefont {V.}~\bibnamefont
  {Rokaj}}, \bibinfo {author} {\bibfnamefont {D.~M.}\ \bibnamefont {Welakuh}},
  \bibinfo {author} {\bibfnamefont {M.}~\bibnamefont {Ruggenthaler}}, \ and\
  \bibinfo {author} {\bibfnamefont {A.}~\bibnamefont {Rubio}},\ }\bibfield
  {title} {\enquote {\bibinfo {title} {Light--matter interaction in the
  long-wavelength limit: no ground-state without dipole self-energy},}\
  }\href@noop {} {\bibfield  {journal} {\bibinfo  {journal} {J. Phys. B: At.
  Mol. Opt. Phys.}\ }\textbf {\bibinfo {volume} {51}},\ \bibinfo {pages}
  {034005} (\bibinfo {year} {2018})}\BibitemShut {NoStop}%
\bibitem [{\citenamefont {Mandal}\ \emph {et~al.}(2023)\citenamefont {Mandal},
  \citenamefont {Taylor}, \citenamefont {Weight}, \citenamefont {Koessler},
  \citenamefont {Li},\ and\ \citenamefont {Huo}}]{Mandal_2023_CR_p9786}%
  \BibitemOpen
  \bibfield  {author} {\bibinfo {author} {\bibfnamefont {A.}~\bibnamefont
  {Mandal}}, \bibinfo {author} {\bibfnamefont {M.~A.}\ \bibnamefont {Taylor}},
  \bibinfo {author} {\bibfnamefont {B.~M.}\ \bibnamefont {Weight}}, \bibinfo
  {author} {\bibfnamefont {E.~R.}\ \bibnamefont {Koessler}}, \bibinfo {author}
  {\bibfnamefont {X.}~\bibnamefont {Li}}, \ and\ \bibinfo {author}
  {\bibfnamefont {P.}~\bibnamefont {Huo}},\ }\bibfield  {title} {\enquote
  {\bibinfo {title} {Theoretical advances in polariton chemistry and molecular
  cavity quantum electrodynamics},}\ }\href@noop {} {\bibfield  {journal}
  {\bibinfo  {journal} {Chem. Rev.}\ }\textbf {\bibinfo {volume} {123}},\
  \bibinfo {pages} {9786--9879} (\bibinfo {year} {2023})}\BibitemShut {NoStop}%
\bibitem [{\citenamefont {Ke}\ and\ \citenamefont
  {Assan}(2025)}]{Ke_2025_JCP_p164703}%
  \BibitemOpen
  \bibfield  {author} {\bibinfo {author} {\bibfnamefont {Y.}~\bibnamefont
  {Ke}}\ and\ \bibinfo {author} {\bibfnamefont {J.}~\bibnamefont {Assan}},\
  }\bibfield  {title} {\enquote {\bibinfo {title} {Harnessing multi-mode
  optical structure for chemical reactivity},}\ }\href@noop {} {\bibfield
  {journal} {\bibinfo  {journal} {J. Chem. Phys.}\ }\textbf {\bibinfo {volume}
  {163}},\ \bibinfo {pages} {164703} (\bibinfo {year} {2025})}\BibitemShut
  {NoStop}%
\bibitem [{\citenamefont {Tanimura}(2020)}]{Tanimura_2020_JCP_p20901}%
  \BibitemOpen
  \bibfield  {author} {\bibinfo {author} {\bibfnamefont {Y.}~\bibnamefont
  {Tanimura}},\ }\bibfield  {title} {\enquote {\bibinfo {title} {Numerically
  “exact” approach to open quantum dynamics: The hierarchical equations of
  motion ({HEOM})},}\ }\href@noop {} {\bibfield  {journal} {\bibinfo  {journal}
  {J. Chem. Phys.}\ }\textbf {\bibinfo {volume} {153}},\ \bibinfo {pages}
  {020901} (\bibinfo {year} {2020})}\BibitemShut {NoStop}%
\bibitem [{\citenamefont {Borrelli}(2019)}]{Borrelli_2019_JCP_p234102}%
  \BibitemOpen
  \bibfield  {author} {\bibinfo {author} {\bibfnamefont {R.}~\bibnamefont
  {Borrelli}},\ }\bibfield  {title} {\enquote {\bibinfo {title} {Density matrix
  dynamics in twin-formulation: An efficient methodology based on tensor-train
  representation of reduced equations of motion},}\ }\href@noop {} {\bibfield
  {journal} {\bibinfo  {journal} {J. Chem. Phys.}\ }\textbf {\bibinfo {volume}
  {150}},\ \bibinfo {pages} {234102} (\bibinfo {year} {2019})}\BibitemShut
  {NoStop}%
\bibitem [{\citenamefont {Borrelli}\ and\ \citenamefont
  {Gelin}(2021)}]{Borrelli_2021_WCMS_p1539}%
  \BibitemOpen
  \bibfield  {author} {\bibinfo {author} {\bibfnamefont {R.}~\bibnamefont
  {Borrelli}}\ and\ \bibinfo {author} {\bibfnamefont {M.~F.}\ \bibnamefont
  {Gelin}},\ }\bibfield  {title} {\enquote {\bibinfo {title} {Finite
  temperature quantum dynamics of complex systems: Integrating thermo-field
  theories and tensor-train methods},}\ }\href@noop {} {\bibfield  {journal}
  {\bibinfo  {journal} {WIREs Comput Mol Sci}\ ,\ \bibinfo {pages} {e1539}}
  (\bibinfo {year} {2021})}\BibitemShut {NoStop}%
\bibitem [{\citenamefont {Ke}, \citenamefont {Borrelli},\ and\ \citenamefont
  {Thoss}(2022)}]{Ke_2022_JCP_p194102}%
  \BibitemOpen
  \bibfield  {author} {\bibinfo {author} {\bibfnamefont {Y.}~\bibnamefont
  {Ke}}, \bibinfo {author} {\bibfnamefont {R.}~\bibnamefont {Borrelli}}, \ and\
  \bibinfo {author} {\bibfnamefont {M.}~\bibnamefont {Thoss}},\ }\bibfield
  {title} {\enquote {\bibinfo {title} {Hierarchical equations of motion
  approach to hybrid fermionic and bosonic environments: Matrix product state
  formulation in twin space},}\ }\href@noop {} {\bibfield  {journal} {\bibinfo
  {journal} {J. Chem. Phys.}\ }\textbf {\bibinfo {volume} {156}},\ \bibinfo
  {pages} {194102} (\bibinfo {year} {2022})}\BibitemShut {NoStop}%
\bibitem [{\citenamefont
  {Ke}(2025{\natexlab{a}})}]{Ke_J.Chem.Phys._2025_p64702}%
  \BibitemOpen
  \bibfield  {author} {\bibinfo {author} {\bibfnamefont {Y.}~\bibnamefont
  {Ke}},\ }\bibfield  {title} {\enquote {\bibinfo {title} {Stochastic resonance
  in vibrational polariton chemistry},}\ }\href@noop {} {\bibfield  {journal}
  {\bibinfo  {journal} {J. Chem. Phys.}\ }\textbf {\bibinfo {volume} {162}},\
  \bibinfo {pages} {064702} (\bibinfo {year} {2025}{\natexlab{a}})}\BibitemShut
  {NoStop}%
\bibitem [{\citenamefont {Hu}, \citenamefont {Xu},\ and\ \citenamefont
  {Yan}(2010)}]{Hu_2010_JCP_p101106}%
  \BibitemOpen
  \bibfield  {author} {\bibinfo {author} {\bibfnamefont {J.}~\bibnamefont
  {Hu}}, \bibinfo {author} {\bibfnamefont {R.-X.}\ \bibnamefont {Xu}}, \ and\
  \bibinfo {author} {\bibfnamefont {Y.}~\bibnamefont {Yan}},\ }\bibfield
  {title} {\enquote {\bibinfo {title} {Communication: Pad{\'e} spectrum
  decomposition of {F}ermi function and {B}ose function},}\ }\href {\doibase
  10.1063/1.3484491} {\bibfield  {journal} {\bibinfo  {journal} {J. Chem.
  Phys.}\ }\textbf {\bibinfo {volume} {133}},\ \bibinfo {pages} {101106}
  (\bibinfo {year} {2010})}\BibitemShut {NoStop}%
\bibitem [{\citenamefont {Lindoy}, \citenamefont {Mandal},\ and\ \citenamefont
  {Reichman}(2024)}]{Lindoy_2024_N_p2617}%
  \BibitemOpen
  \bibfield  {author} {\bibinfo {author} {\bibfnamefont {L.~P.}\ \bibnamefont
  {Lindoy}}, \bibinfo {author} {\bibfnamefont {A.}~\bibnamefont {Mandal}}, \
  and\ \bibinfo {author} {\bibfnamefont {D.~R.}\ \bibnamefont {Reichman}},\
  }\bibfield  {title} {\enquote {\bibinfo {title} {Investigating the collective
  nature of cavity-modified chemical kinetics under vibrational strong
  coupling},}\ }\href@noop {} {\bibfield  {journal} {\bibinfo  {journal}
  {Nanophotonics}\ }\textbf {\bibinfo {volume} {13}},\ \bibinfo {pages}
  {2617--2633} (\bibinfo {year} {2024})}\BibitemShut {NoStop}%
\bibitem [{\citenamefont {Ying}\ and\ \citenamefont
  {Huo}(2023)}]{Ying_2023_JCP_p84104}%
  \BibitemOpen
  \bibfield  {author} {\bibinfo {author} {\bibfnamefont {W.}~\bibnamefont
  {Ying}}\ and\ \bibinfo {author} {\bibfnamefont {P.}~\bibnamefont {Huo}},\
  }\bibfield  {title} {\enquote {\bibinfo {title} {{Resonance theory and
  quantum dynamics simulations of vibrational polariton chemistry}},}\
  }\href@noop {} {\bibfield  {journal} {\bibinfo  {journal} {J. Chem. Phys.}\
  }\textbf {\bibinfo {volume} {159}},\ \bibinfo {pages} {084104} (\bibinfo
  {year} {2023})}\BibitemShut {NoStop}%
\bibitem [{\citenamefont {Ying}\ and\ \citenamefont
  {Huo}(2024)}]{Ying_2024_CM_p110}%
  \BibitemOpen
  \bibfield  {author} {\bibinfo {author} {\bibfnamefont {W.}~\bibnamefont
  {Ying}}\ and\ \bibinfo {author} {\bibfnamefont {P.}~\bibnamefont {Huo}},\
  }\bibfield  {title} {\enquote {\bibinfo {title} {Resonance theory of
  vibrational strong coupling enhanced polariton chemistry and the role of
  photonic mode lifetime},}\ }\href@noop {} {\bibfield  {journal} {\bibinfo
  {journal} {Commun. Mater.}\ }\textbf {\bibinfo {volume} {5}},\ \bibinfo
  {pages} {110} (\bibinfo {year} {2024})}\BibitemShut {NoStop}%
\bibitem [{\citenamefont {Ke}\ and\ \citenamefont
  {Richardson}(2024{\natexlab{a}})}]{Ke_J.Chem.Phys._2024_p224704}%
  \BibitemOpen
  \bibfield  {author} {\bibinfo {author} {\bibfnamefont {Y.}~\bibnamefont
  {Ke}}\ and\ \bibinfo {author} {\bibfnamefont {J.~O.}\ \bibnamefont
  {Richardson}},\ }\bibfield  {title} {\enquote {\bibinfo {title} {Insights
  into the mechanisms of optical cavity-modified ground-state chemical
  reactions},}\ }\href@noop {} {\bibfield  {journal} {\bibinfo  {journal} {J.
  Chem. Phys.}\ }\textbf {\bibinfo {volume} {160}},\ \bibinfo {pages} {224704}
  (\bibinfo {year} {2024}{\natexlab{a}})}\BibitemShut {NoStop}%
\bibitem [{\citenamefont {Ke}\ and\ \citenamefont
  {Richardson}(2024{\natexlab{b}})}]{Ke_2024_JCP_p54104}%
  \BibitemOpen
  \bibfield  {author} {\bibinfo {author} {\bibfnamefont {Y.}~\bibnamefont
  {Ke}}\ and\ \bibinfo {author} {\bibfnamefont {J.~O.}\ \bibnamefont
  {Richardson}},\ }\bibfield  {title} {\enquote {\bibinfo {title} {Quantum
  nature of reactivity modification in vibrational polariton chemistry},}\
  }\href@noop {} {\bibfield  {journal} {\bibinfo  {journal} {J. Chem. Phys.}\
  }\textbf {\bibinfo {volume} {161}},\ \bibinfo {pages} {054104} (\bibinfo
  {year} {2024}{\natexlab{b}})}\BibitemShut {NoStop}%
\bibitem [{\citenamefont {Miller}, \citenamefont {Schwartz},\ and\
  \citenamefont {Tromp}(1983)}]{Miller_1983_JCP_p4889}%
  \BibitemOpen
  \bibfield  {author} {\bibinfo {author} {\bibfnamefont {W.~H.}\ \bibnamefont
  {Miller}}, \bibinfo {author} {\bibfnamefont {S.~D.}\ \bibnamefont
  {Schwartz}}, \ and\ \bibinfo {author} {\bibfnamefont {J.~W.}\ \bibnamefont
  {Tromp}},\ }\bibfield  {title} {\enquote {\bibinfo {title} {Quantum
  mechanical rate constants for bimolecular reactions},}\ }\href@noop {}
  {\bibfield  {journal} {\bibinfo  {journal} {J. Chem. Phys.}\ }\textbf
  {\bibinfo {volume} {79}},\ \bibinfo {pages} {4889--4898} (\bibinfo {year}
  {1983})}\BibitemShut {NoStop}%
\bibitem [{\citenamefont {Craig}, \citenamefont {Thoss},\ and\ \citenamefont
  {Wang}(2007)}]{Craig_2007_JCP_p144503}%
  \BibitemOpen
  \bibfield  {author} {\bibinfo {author} {\bibfnamefont {I.~R.}\ \bibnamefont
  {Craig}}, \bibinfo {author} {\bibfnamefont {M.}~\bibnamefont {Thoss}}, \ and\
  \bibinfo {author} {\bibfnamefont {H.}~\bibnamefont {Wang}},\ }\bibfield
  {title} {\enquote {\bibinfo {title} {Proton transfer reactions in model
  condensed-phase environments: Accurate quantum dynamics using the multilayer
  multiconfiguration time-dependent {H}artree approach},}\ }\href@noop {}
  {\bibfield  {journal} {\bibinfo  {journal} {J. Chem. Phys.}\ }\textbf
  {\bibinfo {volume} {127}},\ \bibinfo {pages} {144503} (\bibinfo {year}
  {2007})}\BibitemShut {NoStop}%
\bibitem [{\citenamefont {Chen}\ and\ \citenamefont
  {Shi}(2009)}]{Chen_2009_JCP_p134505}%
  \BibitemOpen
  \bibfield  {author} {\bibinfo {author} {\bibfnamefont {L.}~\bibnamefont
  {Chen}}\ and\ \bibinfo {author} {\bibfnamefont {Q.}~\bibnamefont {Shi}},\
  }\bibfield  {title} {\enquote {\bibinfo {title} {Quantum rate dynamics for
  proton transfer reactions in condensed phase: The exact hierarchical
  equations of motion approach},}\ }\href@noop {} {\bibfield  {journal}
  {\bibinfo  {journal} {J. Chem. Phys.}\ }\textbf {\bibinfo {volume} {130}},\
  \bibinfo {pages} {134505} (\bibinfo {year} {2009})}\BibitemShut {NoStop}%
\bibitem [{\citenamefont {Ke}\ \emph {et~al.}(2022)\citenamefont {Ke},
  \citenamefont {Kaspar}, \citenamefont {Erpenbeck}, \citenamefont {Peskin},\
  and\ \citenamefont {Thoss}}]{Ke_2022_JCP_p34103}%
  \BibitemOpen
  \bibfield  {author} {\bibinfo {author} {\bibfnamefont {Y.}~\bibnamefont
  {Ke}}, \bibinfo {author} {\bibfnamefont {C.}~\bibnamefont {Kaspar}}, \bibinfo
  {author} {\bibfnamefont {A.}~\bibnamefont {Erpenbeck}}, \bibinfo {author}
  {\bibfnamefont {U.}~\bibnamefont {Peskin}}, \ and\ \bibinfo {author}
  {\bibfnamefont {M.}~\bibnamefont {Thoss}},\ }\bibfield  {title} {\enquote
  {\bibinfo {title} {Nonequilibrium reaction rate theory: Formulation and
  implementation within the hierarchical equations of motion approach},}\
  }\href@noop {} {\bibfield  {journal} {\bibinfo  {journal} {J. Chem. Phys.}\
  }\textbf {\bibinfo {volume} {157}},\ \bibinfo {pages} {034103} (\bibinfo
  {year} {2022})}\BibitemShut {NoStop}%
\bibitem [{\citenamefont {Ke}(2025{\natexlab{b}})}]{Ke_2025_JCP_p54109}%
  \BibitemOpen
  \bibfield  {author} {\bibinfo {author} {\bibfnamefont {Y.}~\bibnamefont
  {Ke}},\ }\bibfield  {title} {\enquote {\bibinfo {title} {Non-equilibrium
  origin of cavity-induced resonant modifications of chemical reactivities},}\
  }\href@noop {} {\bibfield  {journal} {\bibinfo  {journal} {J. Chem. Phys.}\
  }\textbf {\bibinfo {volume} {163}},\ \bibinfo {pages} {054109} (\bibinfo
  {year} {2025}{\natexlab{b}})}\BibitemShut {NoStop}%
\bibitem [{\citenamefont {Fano}(1961)}]{Fano_1961_PR_p1866}%
  \BibitemOpen
  \bibfield  {author} {\bibinfo {author} {\bibfnamefont {U.}~\bibnamefont
  {Fano}},\ }\bibfield  {title} {\enquote {\bibinfo {title} {Effects of
  configuration interaction on intensities and phase shifts},}\ }\href@noop {}
  {\bibfield  {journal} {\bibinfo  {journal} {Phys. Rev.}\ }\textbf {\bibinfo
  {volume} {124}},\ \bibinfo {pages} {1866} (\bibinfo {year}
  {1961})}\BibitemShut {NoStop}%
\bibitem [{\citenamefont {Limonov}\ \emph {et~al.}(2017)\citenamefont
  {Limonov}, \citenamefont {Rybin}, \citenamefont {Poddubny},\ and\
  \citenamefont {Kivshar}}]{Limonov_2017_NP_p543}%
  \BibitemOpen
  \bibfield  {author} {\bibinfo {author} {\bibfnamefont {M.~F.}\ \bibnamefont
  {Limonov}}, \bibinfo {author} {\bibfnamefont {M.~V.}\ \bibnamefont {Rybin}},
  \bibinfo {author} {\bibfnamefont {A.~N.}\ \bibnamefont {Poddubny}}, \ and\
  \bibinfo {author} {\bibfnamefont {Y.~S.}\ \bibnamefont {Kivshar}},\
  }\bibfield  {title} {\enquote {\bibinfo {title} {Fano resonances in
  photonics},}\ }\href@noop {} {\bibfield  {journal} {\bibinfo  {journal} {Nat.
  Photon.}\ }\textbf {\bibinfo {volume} {11}},\ \bibinfo {pages} {543--554}
  (\bibinfo {year} {2017})}\BibitemShut {NoStop}%
\bibitem [{\citenamefont {Fano}(1935)}]{Fano_1935_INC_p154}%
  \BibitemOpen
  \bibfield  {author} {\bibinfo {author} {\bibfnamefont {U.}~\bibnamefont
  {Fano}},\ }\bibfield  {title} {\enquote {\bibinfo {title} {Sullo spettro di
  assorbimento dei gas nobili presso il limite dello spettro d’arco},}\
  }\href@noop {} {\bibfield  {journal} {\bibinfo  {journal} {Il Nuovo Cimento}\
  }\textbf {\bibinfo {volume} {12}},\ \bibinfo {pages} {154--161} (\bibinfo
  {year} {1935})}\BibitemShut {NoStop}%
\bibitem [{\citenamefont {Miroshnichenko}, \citenamefont {Flach},\ and\
  \citenamefont {Kivshar}(2010)}]{Miroshnichenko_2010_RMP_p2257}%
  \BibitemOpen
  \bibfield  {author} {\bibinfo {author} {\bibfnamefont {A.~E.}\ \bibnamefont
  {Miroshnichenko}}, \bibinfo {author} {\bibfnamefont {S.}~\bibnamefont
  {Flach}}, \ and\ \bibinfo {author} {\bibfnamefont {Y.~S.}\ \bibnamefont
  {Kivshar}},\ }\bibfield  {title} {\enquote {\bibinfo {title} {Fano resonances
  in nanoscale structures},}\ }\href@noop {} {\bibfield  {journal} {\bibinfo
  {journal} {Rev. Mod. Phys.}\ }\textbf {\bibinfo {volume} {82}},\ \bibinfo
  {pages} {2257--2298} (\bibinfo {year} {2010})}\BibitemShut {NoStop}%
\bibitem [{\citenamefont {Rahmani}, \citenamefont {Luk'yanchuk},\ and\
  \citenamefont {Hong}(2013)}]{Rahmani_2013_LPR_p329}%
  \BibitemOpen
  \bibfield  {author} {\bibinfo {author} {\bibfnamefont {M.}~\bibnamefont
  {Rahmani}}, \bibinfo {author} {\bibfnamefont {B.}~\bibnamefont
  {Luk'yanchuk}}, \ and\ \bibinfo {author} {\bibfnamefont {M.}~\bibnamefont
  {Hong}},\ }\bibfield  {title} {\enquote {\bibinfo {title} {Fano resonance in
  novel plasmonic nanostructures},}\ }\href@noop {} {\bibfield  {journal}
  {\bibinfo  {journal} {Laser Photonics Rev.}\ }\textbf {\bibinfo {volume}
  {7}},\ \bibinfo {pages} {329--349} (\bibinfo {year} {2013})}\BibitemShut
  {NoStop}%
\bibitem [{\citenamefont {Limonov}(2021)}]{Limonov_2021_AOP_p703}%
  \BibitemOpen
  \bibfield  {author} {\bibinfo {author} {\bibfnamefont {M.~F.}\ \bibnamefont
  {Limonov}},\ }\bibfield  {title} {\enquote {\bibinfo {title} {Fano resonance
  for applications},}\ }\href@noop {} {\bibfield  {journal} {\bibinfo
  {journal} {Adv. Opt. Photonics.}\ }\textbf {\bibinfo {volume} {13}},\
  \bibinfo {pages} {703--771} (\bibinfo {year} {2021})}\BibitemShut {NoStop}%
\bibitem [{\citenamefont {{\'U}js{\'a}ghy}\ \emph {et~al.}(2000)\citenamefont
  {{\'U}js{\'a}ghy}, \citenamefont {Kroha}, \citenamefont {Szunyogh},\ and\
  \citenamefont {Zawadowski}}]{Ujsaghy_2000_Prl_p2557}%
  \BibitemOpen
  \bibfield  {author} {\bibinfo {author} {\bibfnamefont {O.}~\bibnamefont
  {{\'U}js{\'a}ghy}}, \bibinfo {author} {\bibfnamefont {J.}~\bibnamefont
  {Kroha}}, \bibinfo {author} {\bibfnamefont {L.}~\bibnamefont {Szunyogh}}, \
  and\ \bibinfo {author} {\bibfnamefont {A.}~\bibnamefont {Zawadowski}},\
  }\bibfield  {title} {\enquote {\bibinfo {title} {Theory of the fano resonance
  in the stm tunneling density of states due to a single kondo impurity},}\
  }\href@noop {} {\bibfield  {journal} {\bibinfo  {journal} {Physical review
  letters}\ }\textbf {\bibinfo {volume} {85}},\ \bibinfo {pages} {2557}
  (\bibinfo {year} {2000})}\BibitemShut {NoStop}%
\bibitem [{\citenamefont {Zheng}\ \emph {et~al.}(2022)\citenamefont {Zheng},
  \citenamefont {Duan}, \citenamefont {Zhou}, \citenamefont {Li}, \citenamefont
  {Zhou}, \citenamefont {Wang}, \citenamefont {Chen}, \citenamefont {Zhu},
  \citenamefont {Li}, \citenamefont {Bai} \emph
  {et~al.}}]{Zheng_2022_ACIE_p202210097}%
  \BibitemOpen
  \bibfield  {author} {\bibinfo {author} {\bibfnamefont {Y.}~\bibnamefont
  {Zheng}}, \bibinfo {author} {\bibfnamefont {P.}~\bibnamefont {Duan}},
  \bibinfo {author} {\bibfnamefont {Y.}~\bibnamefont {Zhou}}, \bibinfo {author}
  {\bibfnamefont {C.}~\bibnamefont {Li}}, \bibinfo {author} {\bibfnamefont
  {D.}~\bibnamefont {Zhou}}, \bibinfo {author} {\bibfnamefont {Y.}~\bibnamefont
  {Wang}}, \bibinfo {author} {\bibfnamefont {L.-C.}\ \bibnamefont {Chen}},
  \bibinfo {author} {\bibfnamefont {Z.}~\bibnamefont {Zhu}}, \bibinfo {author}
  {\bibfnamefont {X.}~\bibnamefont {Li}}, \bibinfo {author} {\bibfnamefont
  {J.}~\bibnamefont {Bai}},  \emph {et~al.},\ }\bibfield  {title} {\enquote
  {\bibinfo {title} {Fano resonance in single-molecule junctions},}\
  }\href@noop {} {\bibfield  {journal} {\bibinfo  {journal} {Angew. Chem. Int.
  Ed.}\ }\textbf {\bibinfo {volume} {134}},\ \bibinfo {pages} {e202210097}
  (\bibinfo {year} {2022})}\BibitemShut {NoStop}%
\bibitem [{\citenamefont {Chikkaraddy}\ \emph {et~al.}(2016)\citenamefont
  {Chikkaraddy}, \citenamefont {De~Nijs}, \citenamefont {Benz}, \citenamefont
  {Barrow}, \citenamefont {Scherman}, \citenamefont {Rosta}, \citenamefont
  {Demetriadou}, \citenamefont {Fox}, \citenamefont {Hess},\ and\ \citenamefont
  {Baumberg}}]{Chikkaraddy_2016_N_p127}%
  \BibitemOpen
  \bibfield  {author} {\bibinfo {author} {\bibfnamefont {R.}~\bibnamefont
  {Chikkaraddy}}, \bibinfo {author} {\bibfnamefont {B.}~\bibnamefont
  {De~Nijs}}, \bibinfo {author} {\bibfnamefont {F.}~\bibnamefont {Benz}},
  \bibinfo {author} {\bibfnamefont {S.~J.}\ \bibnamefont {Barrow}}, \bibinfo
  {author} {\bibfnamefont {O.~A.}\ \bibnamefont {Scherman}}, \bibinfo {author}
  {\bibfnamefont {E.}~\bibnamefont {Rosta}}, \bibinfo {author} {\bibfnamefont
  {A.}~\bibnamefont {Demetriadou}}, \bibinfo {author} {\bibfnamefont
  {P.}~\bibnamefont {Fox}}, \bibinfo {author} {\bibfnamefont {O.}~\bibnamefont
  {Hess}}, \ and\ \bibinfo {author} {\bibfnamefont {J.~J.}\ \bibnamefont
  {Baumberg}},\ }\bibfield  {title} {\enquote {\bibinfo {title}
  {Single-molecule strong coupling at room temperature in plasmonic
  nanocavities},}\ }\href@noop {} {\bibfield  {journal} {\bibinfo  {journal}
  {Nature}\ }\textbf {\bibinfo {volume} {535}},\ \bibinfo {pages} {127--130}
  (\bibinfo {year} {2016})}\BibitemShut {NoStop}%
\end{thebibliography}
%
\end{document}